\documentclass[11pt]{article}
\usepackage[utf8]{inputenc}
\usepackage{lipsum} 
\usepackage{fullpage}
\usepackage{xcolor}

\usepackage{epstopdf}
\usepackage[caption=false]{subfig}

\usepackage{authblk}
\usepackage{graphicx}
\usepackage{amssymb}
\usepackage{amsmath}
\usepackage{bm}

\usepackage[numbers,sort&compress]{natbib}
\bibpunct[, ]{[}{]}{,}{n}{,}{,}

\newcommand{\ind}{\stackrel{\rm ind}{\sim}}

\usepackage{url}

\usepackage{color}
\usepackage[normalem]{ulem} 
\usepackage{tikz}
\usetikzlibrary{shapes,arrows,fit}
\usetikzlibrary{bayesnet}
\usepackage{multirow}
\usepackage{pifont}

\usepackage{xr}
\begin{document}

\title{Bayesian Dynamic Network Modelling: an application to metabolic associations in cardiovascular diseases}

\author[1]{Marco Molinari}
\author[2]{Andrea Cremaschi}
\author[1,2,3]{Maria De Iorio}
\author[4]{Nishi Chaturvedi}
\author[4]{Alun Hughes}
\author[4]{Therese Tillin}

\affil[1]{Department of Statistical Science, University College London, UK}
\affil[2]{Singapore Institute for Clinical Sciences, A*STAR, Singapore}
\affil[3]{Yong Loo Lin School of Medicine, National University of Singapore, Singapore}
\affil[4]{Department of Population Science and Experimental Medicine,
	University College London, UK}

\date{}

\maketitle

\textbf{Abstract}: 
We propose a novel approach to the estimation of multiple Graphical Models to analyse temporal patterns of association among a set of metabolites over different groups of patients. Our motivating application is the Southall And Brent REvisited (SABRE) study, a tri-ethnic cohort study conducted in the UK. We are interested in identifying potential ethnic differences in metabolite levels and associations as well as their evolution over time, with the aim of gaining a better understanding of different risk of cardio-metabolic disorders across ethnicities. Within a Bayesian framework, we employ a nodewise regression approach to infer the structure of the graphs, borrowing information across time as well as across ethnicities. The response variables of interest are metabolite levels measured at two time points and for two ethnic groups, Europeans and South-Asians. We use nodewise regression to estimate the high-dimensional precision matrices of the metabolites, imposing sparsity on the regression coefficients through the dynamic horseshoe prior, thus favouring sparser graphs. We provide the code to fit the proposed model using the software {\ttfamily Stan}, which performs posterior  inference  using Hamiltonian Monte Carlo sampling, as well as  a detailed description of a block Gibbs sampling scheme. 

\textbf{Keywords}: 	
Dynamic Shrinkage Priors; Gibbs Sampling; Graphical Models; Metabolomics; Nodewise Regression

\section{Introduction}
\label{sec:Intro}

The Southall And Brent REvisited (SABRE) \citep{tillin2012southall} population based cohort study was initiated in the late 1980s in north-west London with the aim of studying ethnic differences in cardiovascular diseases and diabetes. The study includes individuals of European, South-Asian and African-Caribbean descent, aged 40-69 years at baseline. Recently, metabolites measurements, obtained using nuclear magnetic resonance spectroscopy \citep{soininen2015quantitative}, have been collected on over 3000 stored blood samples at baseline and 20-years follow-up. 
All participants gave written informed consent. Approval for the baseline study was obtained from Ealing, Hounslow and Spelthorne, Parkside, and University College London research ethics committees, and at follow-up from St. Mary’s Hospital Research Ethics Committee (reference 07/H0712/109).
In this work, we investigate the complex metabolic pathways involved in cardiac metabolism, combing metabolite concentration data with more traditional clinical makers. Metabolites are small molecules that participate in metabolic reactions and are involved in biochemical pathways associated with metabolism in health and disease \cite{ellul2019metabolomics}. Indeed, there is increasing evidence that 
 most cardiovascular diseases, such as diabetes mellitus, involve disturbances in cardiac metabolism and that, on the other hand, heart diseases can affect the metabolism, hence initiating a vicious cycle \citep{mcgarrah2018cardiovascular}. In the last two decades, metabolomics (i.e.  the large-scale study of metabolites within cells, bio-fluids, tissues or organisms) has emerged as a powerful tool for defining changes in metabolism that occur across a range of cardiovascular disease states. Findings from metabolomic studies have contributed to a better understanding of the metabolic changes that occur in heart failure and ischaemic heart disease and have identified new cardiovascular disease biomarkers. 
 
Here, we focus on type 2 diabetes, a main clinical  outcome of interest in the SABRE study, as it  poses an enormous individual and societal burden, with high risk of major complications and diminished quality and length of life. Hence, it is imperative to understand causal mechanisms in order to identify subjects at highest risk and to tailor preventive and therapeutic measures for appropriate periods during the life course. The global epidemic of type 2 diabetes disproportionately affects non-European ethnic groups. South-Asians (from the Indian subcontinent) form the largest ethnic minority group in the UK with prevalence of diabetes estimated to be 2-4 times higher than that of the general population \citep{sproston2006health}.
Research to date suggests that insulin resistance and differences in body fat distribution explain some of the ethnic differences in diabetes risk, but the underlying mechanistic pathways are poorly understood, although they are likely to involve a complex interplay between environmental, behavioural, metabolic, genetic and epigenetic influences. Therefore, it is essential not to limit the analysis to differences in metabolite levels across ethnicities over time, but to also consider patterns of variations in metabolic associations to gain better insight in molecular mechanisms of disease pathogenesis and to formulate novel data-driven scientific hypotheses.

In the SABRE study measurements are available on over 200 metabolite concentrations, covering a wide range of chemical classes, at baseline and 20-years follow up. We represent association patterns among metabolites through a graph, which is the main object of statistical inference. Within a Bayesian framework, our model is based on nodewise regression, originally introduced by \cite{meinshausen2006high}. In the nodewise regression approach, estimating a graph is equivalent to estimating the precision matrix between variables, in our case metabolite levels. This is achieved by rewriting the problem in terms of $M$ independent linear regressions, where $M$ is the number of variables/metabolites, and each variable is regressed on all the others. It is a local method, because it infers the neighbourhood structure of each node (i.e. the connections involving the node) independently, as opposed to global methods that aim to infer  jointly the  association patterns across all the nodes. The element $j,l$ in the precision matrix is estimated to be non-zero (implying an edge in the graph between the two variables) if either the estimated regression coefficient of variable $j$ on $l$, or the estimated coefficient of variable $l$ on $j$, is non-zero. Alternative approaches to graphical models estimation are available in the literature. Graphical Lasso is a popular global method in both frequentist and Bayesian domain, based on a penalised maximum likelihood estimator \citep{friedman2008sparse} or on a double exponential prior \citep{wang2012bayesian}, respectively. Graphical Lasso has the advantage of ensuring a positive definite estimate of $\Omega$, but requires a greater computational effort and it is less flexible in estimating the individual scaling levels (i.e. the diagonal elements of $\Omega$)  compared to the nodewise approach \citep{jankova2018inference}. Other Bayesian approaches to graphical model estimation rely on the specification of a suitable prior distribution over the graph space and, conditional on the graph, a prior for precision matrix is selected. For example, \cite{lenkoski2011computational} propose an algorithm to perform posterior inference on graph space assuming a  G-Wishart prior, a generalisation of the Hyper-Wishart distribution, that allows to deal with non-decomposable graphs. However, the convergence of associated MCMC algorithms can be slow due to the single edge update and the intractable normalising constant in the marginal posterior that requires numerical approximations. A more efficient algorithm, based on a birth-death MCMC, has been proposed by \cite{mohammadi2015bayesian}.

In several applications, including the one presented in this work, information regarding the grouping of the subjects is available, e.g. different ethnic groups. Therefore, it is of interest to estimate a group-specific graph, by specifying a joint model for multiple graphs able to identify common structure as well as  group-specific connections. This is particularly relevant in our application as it allows to highlight biological mechanism which differ between ethnicity and potentially lead to different disease development. Examples of models for multiple graphs are found in the work of \cite{saegusa2016joint}, which specify a global penalisation and use optimization techniques, and of \cite{peterson2015bayesian}, which in a Bayesian framework propose a joint model involving a Markov random field prior to encourage sharing of information among edges. A Markov random field prior modelling both spatial and temporal dependence is also used by \cite{lin2017joint}. Additional examples are the approaches of \cite{tan2017bayesian}, where shared structures are modelled via a multiplicative prior, and of \cite{bilgrau2020targeted}, where multiple precision matrices are estimated via a penalisation approach which allows the inclusion of information a-priori.

The main methodological contribution of this work is to extend nodewise regression to 
(i)  achieve sparsity in the resulting precision matrix (ii) accommodate multiple groups (iii) account for different time points. This will allow not only to better understand group-specific and global dependence structure, but also how they evolve across time. To this end, we exploit recent contribution on shrinkage prior and dynamic stochastic volatility models. In a nodewise regression framework, inducing sparsity in the graph is equivalent to estimating some regression coefficients equal to zero. A wealth of proposals is available in the literature to impose sparsity on  regression coefficients (see, for instance, \cite{o2009review} for a review on shrinkage priors). We opt for the Horseshoe prior, which belongs to the class of continuous global-local shrinkage priors (see, for example, \citep{carvalho2010horseshoe,bhadra2017horseshoe+,armagan2013generalized}). The Horseshoe prior is characterised by an accentuated spike at zero to strongly shrink small or negligible coefficients, while leaving important coefficients unaffected thanks to its heavy tails. Moreover, this prior allows for efficient computations. Linking the hyper-parameters of the Horseshoe priors across groups and employing the dynamic extension of the Horseshoe prior proposed by \cite{kowal2017dynamic}, we are able to accurately estimate dynamic evolving complex precision matrices, across  multiple groups of observations of different sample sizes at different time points.
A similar approach is taken by \cite{lin2017joint}, where the neighbourhood selection of each node is based on the nodewise regression method of \cite{meinshausen2006high}, but it is different from our approach in two ways: (i) a  Spike and Slab prior is imposed on the regression coefficients to induce sparsity on edge selection, increasing computational cost; (ii) information is shared across multiple groups and time points through a Markov Random Field structure, which also involves neighbourhood selection. Our choice of a Horseshoe prior, which is  continuous and does not require hyper-parameter tuning (in its original formulation), significantly speeds up computations, allowing the use of  different inference approaches, such as Markov chain Monte Carlo and Hamiltonian Monte Carlo (HMC). Furthermore, information across multiple groups is shared by imposing a hierarchical structure on the hyper-parameters of the horseshoe prior, while time dependence is incorporated through an autoregressive model specified on the regression coefficients as  proposed by \cite{kowal2017dynamic}. Both extensions are computationally efficient, being also suitable for parallel computations on multicore machines. Moreover, the approach proposed by \cite{kowal2017dynamic} scales linearly in the number of time points.
A limitation of the neighbourhood selection based on nodewise regression  is the lack of a direct posterior estimate of the joint precision matrix. As detailed in \cite{lin2017joint}, the Bayesian version of  nodewise regression allows to accurately estimate the  posterior distribution of the edge selection, but does not directly provide an estimate of the precision matrix. Nonetheless, we can use the estimated posterior distribution of the regression coefficients to approximate the precision matrix and as a  measure of the strength and direction of each specific connection in the graph.

The paper is organised as follows. In Section 2 we introduce nodewise regression, its extensions and discuss prior specification. In Section 3 we show the performance of the proposed model in simulations. Section 4 illustrates the application to the SABRE study. Section 5 concludes the paper summarising main results and contributions.

\section{Methodology}
In this Section, we introduce nodewise regression and its use for graph estimation. We first  provide a detailed description of the basic model for a single graph. Next, we explain our strategy to extend nodewise regression to multiple graphs and to the more general dynamic multiple graphs framework. Hence, we explain how to perform posterior inference on such models. 

\subsection{Nodewise regression for graphical models}
\label{sec:model}
 We first introduce basic concepts from graph theory. For more details, we refer to   \citep{lauritzen1996graphical}.  Let $G=(V,E)$ be an undirected graph, with vertex set $V=(1, \dots, M)$ and edge set $E \subset \{ (j,l) \in V \times V : j < l \}$. The vertices of the graph are associated with a $M$-dimensional vector of variables $ \boldsymbol{y} = (y_1, \dots, y_M)$ assumed to follow a multivariate Normal distribution
\begin{equation}
	\boldsymbol{y} \sim \text{N}_M\left( \boldsymbol{\mu}, \Omega \right)
	\label{eq:JointNormal1}
\end{equation}
where $\bm \mu$ is the mean vector and $\Omega = \left[\omega_{jl}\right]_{j,l = 1}^M$ is the $M \times M$ precision matrix. In the context of  Gaussian Graphical Models (GGM) \citep{dempster1972covariance}, it is standard practice to specify a prior distribution on $G$ and, then, conditional on $G$, a prior for $\Omega$ as there is a direct correspondence between the elements of the precision matrix $\Omega$ and the edges in the graph $G$. An edge is present between nodes $V_j$ and $V_l$, that is $(j,l) \in E$, if and only if $\omega_{jl} \ne 0$ \citep{wermuth1976analogies, lauritzen1996graphical}. If $\omega_{jl} = 0$ (absence of an edge), then  $y_j$ and $y_l$ are conditionally independent given the remaining variables $\boldsymbol{y}_{-jl}$, where $\boldsymbol{y}_{-jl}$ denotes the random vector $\boldsymbol{y}$ excluding the elements $j$ and $l$. In this context the main object of inference is the graph $G$ and usually $\Omega$ is treated as a latent variable. Posterior inference is often complicated by the complexity of graph space.  In this paper we opt for a different approach as it is more scalable to higher dimensions. 
To estimate the  graph $G$ we use nodewise regression \cite{meinshausen2006high}, a technique that exploits the relationship between the partial correlation coefficients and the regression coefficients of a linear regression, without directly specifying a probability model for $G$. Recall that for the standard linear regression
\begin{equation}
	y_l = \sum_{j \neq l} \beta_{jl} y_j + \varepsilon_l, \qquad \varepsilon_l \sim \text{N}(0, \sigma^2_l )
	\label{eq:reg_on_j}
\end{equation}
the regression coefficients $\beta_{jl}$ can be expressed in terms of partial correlations 
 $\omega_{jl}$, i.e.  $ \beta_{jl} = \dfrac{-\omega_{jl} }{ \omega_{ll} }$ and analogously $ \beta_{lj} = \dfrac{-\omega_{lj}}{\omega_{jj} }$. Then
\begin{equation*}
	\omega_{jl} \neq 0 \iff  \beta_{lj} \neq 0  \iff  \beta_{jl} \neq 0
\end{equation*}
This result can be also derived from the moments of the conditional Normal distribution. Here, $\sigma^2_l$ denotes the error variance. 
In our context, 
consider the partition where the random variable $y_{l}$ is the $l$-th coordinate of $\boldsymbol{y}$ and $\boldsymbol{y}_{-l}$ corresponds to the remaining coordinates. The conditional distribution of $ y_l$ given $\boldsymbol{y}_{-l}$ is
\begin{equation}
	y_l \mid \boldsymbol{y}_{-l}, \boldsymbol{\mu}, \Omega \sim \text{N} \left( \mu_l - \sum_{j \ne l} \dfrac{ \omega_{jl}}{\omega_{ll}} ( y_{j} - \mu_{j} ), \dfrac{1}{\omega_{ll}} \right)
	\label{eq:conditional_mean}
\end{equation}
where $\omega_{jl} / \omega_{ll} = \beta_{jl}$, for $j \ne l$.  In many application, as in Section \ref{sec:Sabre},   $\mu_l$ captures covariate effects on the response $y_l$, i.e. $\mu_l= \bm Z_l \bm \eta_l$, where $\bm Z_l$ is the covariate vector and $\bm \eta_l$ is the vector of regression coefficients. In this case, the nodewise regression model is defined on the responses centred on the covariate term. For ease of explanation, in the following sections we assume, without loss of generality, that $\boldsymbol{\mu} = (\mu_1 ,\ldots,\mu_M)=\boldsymbol{0}$.

This framework allows us to express the problem of graphical model selection as $M$ independent linear regression problems, since a regression coefficient estimated equal to zero implies a zero in the corresponding element of the precision matrix and, implicitly, a lack of an edge in the underlying graph.
For this reason, we are interested  in identifying regression coefficients equal to zero. 
In the Bayesian framework, a wealth of sparse Bayesian regression techniques is available. When performing variable selection, a popular choice is to impose shrinkage priors on the regression coefficients. Typical examples include the class of two components discrete mixture priors, known as spike and slab \citep{GeoMc93}, and the class of continuous shrinkage priors, of which examples are the Horseshoe and the Horseshoe+ prior see \cite{bhadra2017horseshoe+} for a review. The spike and slab approach implies a positive probability for the regression coefficient to be zero, but it can be computationally demanding with a high number of parameters, due to the large state space. In contrast, continuous priors are easier to implement and are usually more computationally efficient, although the probability for the coefficient to be exactly zero is null and further thresholding is required. We opt for the Horseshoe prior \cite{carvalho2010horseshoe}, which is a scale mixture prior defined as
\begin{eqnarray*}
	\beta_j \mid \lambda_j, \tau & \sim & \text{N}\left( 0,  \lambda_j^2 \tau^2 \right)\\
	\lambda_j, \tau & \stackrel{ind}{\sim} & \text{C}^+(0,1)
\end{eqnarray*}
where $\text{C}^+(0,1)$ denotes the standard half-Cauchy distribution, with probability density function
\begin{equation*}
	p(\lambda_j) = \dfrac{2}{\pi( 1 + \lambda_j^2 ) }, \qquad \lambda_j > 0
\end{equation*}
The Horseshoe belongs to the family of global-local shrinkage priors, as the  global (i.e. common to all $\beta_j$) parameter $\tau$ shrinks all the coefficients towards zero, while the thick half-Cauchy tails for the	local scales $\lambda_j$ allow some regression coefficients  $\beta_j$ to escape the shrinkage 
and counter-balance the effect of $\tau$.
Moreover,
the Horseshoe prior is characterised by a singularity at zero to strongly shrink small or negligible coefficients, while leaving the important ones unaffected thanks to its heavy tails (given by the half-Cauchy distribution).

We now describe the proposed model for single graph estimation. Let $\bm Y$ be a $n \times M$ matrix of observations, where $n$ is the sample size and $M$ is the number of random variables/nodes. Each column $\boldsymbol{y}_l = (y_{1l}, y_{2l}, \dots, y_{nl})^T$, for $l=1,\dots,M$, contains the measurements of the $l$-th variable. The regression model for the $l$-th column can be written as
\begin{eqnarray}
	\boldsymbol{y}_l \mid \boldsymbol{\beta}_l, \sigma^2_l & \sim & \text{N}_n\left( \bm X \boldsymbol{\beta}_l, \sigma^2_l I_n \right) \nonumber \\
	\beta_{jl} \mid \lambda_{jl}, \tau_l & \sim & \text{N}\left(0, \lambda^2_{jl} \tau_l^2 \right) \nonumber \\
	\sigma^2_l \mid a_{\sigma} , b_{\sigma} & \sim & \text{Inverse-Gamma}\left( a_{\sigma} , b_{\sigma} \right)\\
	\lambda_{jl} &\sim & \text{C}^+ \left(0, 1\right) \nonumber \\
	\tau_l &\sim & \text{C}^+ \left(0, 1\right) \nonumber
	\label{eq:Reg_HS1}
\end{eqnarray}
where $\bm  X$ is the matrix of explanatory variables given by $\bm  Y_{-l}$ (i.e. $\bm  Y$ excluding the $l$-th column) and $\boldsymbol{\beta}_l = (\beta_{1l}, \beta_{2l}, \dots, \beta_{pl})$, with $p = M - 1$, is a vector of regression coefficients for the $l$-th regression. A pseudo inclusion probability parameter $\kappa_{jl}$ is defined by \cite{carvalho2010horseshoe} as
\begin{equation}
	\kappa_{jl} = \dfrac{1}{1 + \text{Var}(\beta_{jl} | \lambda_{jl}, \tau_l ) } = \dfrac{1}{1 + \lambda^2_{jl} \tau_l^2 }
	\label{eq:pseudo_prob}
\end{equation}
which is interpretable as the amount of shrinkage towards zero, with  $\kappa_{jl} \approx 1$ yielding maximal shrinkage and $\kappa_{jl} \approx 0$ corresponding to minimal shrinkage and leading to the inclusion of the variable/edge in the graph. As previously shown \cite{carvalho2010horseshoe}, variable selection based on \eqref{eq:pseudo_prob} (with a threshold level of $0.5$) performs similarly to explicit variable selection based on spike and slab prior. Throughout this work, we follow this criterion to decide which edges to include in the graph.

To improve computational efficiency, we use the representation of the standard half-Cauchy distribution employed by \cite{gelman2006prior, piironen2017sparsity}. The standard half-Cauchy distribution can be expressed as the product of a standard half-Normal distribution times the square root of an Inverse-Gamma distribution. Let $z \sim \text{N}^+(0,1)$ and $y \sim \text{Inverse-Gamma}(1/2, 1/2)$ and define $x = z \sqrt{y}$, then $x \sim \text{C}^+(0,1)$. $\text{N}^+(0,1)$ denotes the standard half-Normal distribution, which is defined as the absolute value of a Normal distribution \citep{leone1961folded}. This re-parametrisation can help to avoid divergent transitions in an HMC algorithm (a problem commonly encountered with funnel shaped distributions). As suggested by \cite{piironen2017sparsity}, we  also allow for a tunable global scale parameter $\tau_l$, which can help achieving the desired level of sparsity.  Thus the prior distribution on the regression parameters becomes
\begin{align}
	\begin{split}
		\beta_{jl} \mid \lambda_{jl}, \tau_l & \sim \text{N}(0, \lambda_{jl}^2 \tau_l^2 ) \\ 
		\lambda_{jl} &= \lambda_{jl}^{a} \sqrt{\lambda_{jl}^{b} }\\ 
		\tau_l  &= \tau_{l}^{a} \sqrt{\tau_{l}^{b}} \tau_0
	\end{split}
	\begin{split} &\\
		\lambda_{jl}^{a} &\sim \text{N}^+(0,1) \\
		\lambda_{jl}^{b}  &\sim \text{Inverse-Gamma}(1/2,1/2) \\
		\tau_{l}^{a}  &\sim \text{N}^+(0,1)\\
		\tau_{l}^{b}  &\sim \text{Inverse-Gamma}(1/2, 1/2)
	\end{split}
	\label{eq:Reg_HS2}
\end{align}
where $\tau_0 = \frac{p_0}{p - p_0}\frac{\sigma}{\sqrt{n}}$ and $p_0$ is a prior guess about the number of non-zero coefficients. The choice of $p_0$ is extensively discussed by \cite{piironen2017sparsity}. A schematic representation of model \eqref{eq:Reg_HS2} is shown in Figure 1(a).

\begin{figure}
	\centering
\subfloat[Static Horseshoe nodewise regression model]{
\tikzset{
	latentnode/.style  ={draw,minimum width=2.5em, shape=circle,thick, black,fill=white},
	visiblenode/.style ={draw, minimum width=2.5em, minimum height=2.5em, shape=rectangle,thick, black,fill=black!20},
	visiblenode2/.style ={draw,minimum width=2.5em, minimum height=2.5em, shape=rectangle,thick, black,fill=white}
}

\scalebox{.7}{
\begin{tikzpicture}[auto,thick,node distance=5em]
	\node[visiblenode] (yl) {$y_l$};
	\node[visiblenode, above of = yl, right of = yl] (yminusl) {$\bm y_{-l}$};
	
	\node[latentnode, above of = yl, left of = yl] (sig2l) {$\sigma^2_l$};
	\node[latentnode, below of = yl] (betajl) {$\beta_{jl}$};
	\plate [inner sep=0.2cm] {platebeta} {(betajl)} {$j=1:p$};
	\node[latentnode, below of = betajl, left of = betajl] (taul) {$\tau_l$};
	\node[latentnode, below of = betajl, right of = betajl] (lambdajl) {$\lambda_{jl}$};
	\plate [inner sep=0.2cm] {platelambdajl} {(lambdajl)} {$j=1:p$};
	\node[latentnode, below of = taul, left of = taul] (taual) {$\tau^a_l$};
	\node[latentnode, left of = taul] (taubl) {$\tau^b_l$};
	\node[visiblenode2, below of = taul] (tau0) {$\tau_0$};
	\node[latentnode, below of = lambdajl] (lambdajal) {$\lambda^a_{jl}$};
	\plate [inner sep=0.2cm] {platelambdajal} {(lambdajal)} {$j=1:p$};
	\node[latentnode, right of = lambdajal] (lambdajbl) {$\lambda^b_{jl}$};
	\plate [inner sep=0.2cm] {platelambdajbl} {(lambdajbl)} {$j=1:p$};

	\draw [->] (yminusl) -- (yl);
	\draw [->] (sig2l) -- (yl);
	\draw [->] (betajl) -- (yl);
	\draw [->] (taul) -- (betajl);
	\draw [->] (lambdajl) -- (betajl);
	\draw [->] (taual) -- (taul);
	\draw [->] (taubl) -- (taul);
	\draw [->] (tau0) -- (taul);
	\draw [->] (lambdajal) -- (lambdajl);    
	\draw [->] (lambdajbl) -- (lambdajl);

\end{tikzpicture}
}
}
\subfloat[Dynamic Horseshoe nodewise regression model]{
\tikzset{
	latentnode/.style  ={draw,minimum width=2.5em, shape=circle,thick, black,fill=white},
	visiblenode/.style ={draw, minimum width=2.5em, minimum height=2.5em, shape=rectangle,thick, black,fill=black!20}
}

\scalebox{.7}{	
	\begin{tikzpicture}[auto,thick,node distance=5em]
		\node[visiblenode] (ylrt) {$y_{lrt}$};
		\node[visiblenode, above of = ylrt, right of = ylrt] (yminuslrt) {$\bm y_{-lrt}$};
		
		\node[latentnode, above of = yl, left of = yl] (sig2lrt) {$\sigma^2_{lrt}$};
		\node[latentnode, below of = yl] (betaljrt) {$\beta_{ljrt}$};
		\node[latentnode, left of = betaljrt] (betaljrtminus1) {$\beta_{ljrt-1}$};
		\plate [inner sep=0.12cm, label={[label distance=-5mm]270:$j=1,\dots,p$}] {platebeta} {(betaljrt) (betaljrtminus1)} {};
		\node[latentnode, below of = betajlrt, left of = betaljrt] (gammaljrt) {$\gamma_{ljrt}$};
		\node[latentnode, below of = betajlrt, right of = betaljrt] (hljrt) {$h_{ljrt}$};		
		\node[latentnode, below of = hljrt] (tauljr) {$\tau_{ljr}$};
		\node[latentnode, left of = tauljr] (tau0) {$\tau_0$};
		\node[latentnode, below of = hljrt, right of = hljrt] (lambdaljrt) {$\lambda_{ljrt}$};	
		\node[latentnode, right of = lambdaljrt] (philjrt) {$\phi_{ljrt}$};		
		\plate [inner sep=0.12cm, label={[label distance=-5mm]270:$j=1,\dots,p$}] {plate1} {(gammaljrt) (hljrt)} {};
		\plate [inner sep=0.12cm, label={[label distance=-5mm]270:$j=1,\dots,p$}] {plate2} {(tauljr) (lambdaljrt) (philjrt)} {};
		
		\draw [->] (yminuslrt) -- (ylrt);
		\draw [->] (sig2lrt) -- (ylrt);
		\draw [->] (betaljrt) -- (ylrt);
		\draw [->] (betaljrtminus1) -- (betaljrt);
		\draw [->] (gammaljrt) -- (betaljrt);
		\draw [->] (hljrt) -- (betaljrt);
		\draw [->] (tau0) -- (hljrt);
		\draw [->] (tauljr) -- (hljrt);
		\draw [->] (lambdaljrt) -- (hljrt);
		\draw [->] (philjrt) -- (hljrt);    
		
	\end{tikzpicture}
}
}
\caption{Summary of the nodewise regression models, highlighting the relationship between the observations and the parameters of the model. Circles represent random variables, while squares correspond to observations or fixed hyper-parameters. (a) Static Horseshoe model for each observed variable $y_l$, with $l = 1, \dots, M$, when only a group of observations is under investigation at one time point. (\textbf{b}) Dynamic Horseshoe model for each observed variable $y_{lrt}$ at time $t$ and within group $r$, with $l = 1, \dots, M$.}
\label{fig:Graphical_Repres}
\end{figure}
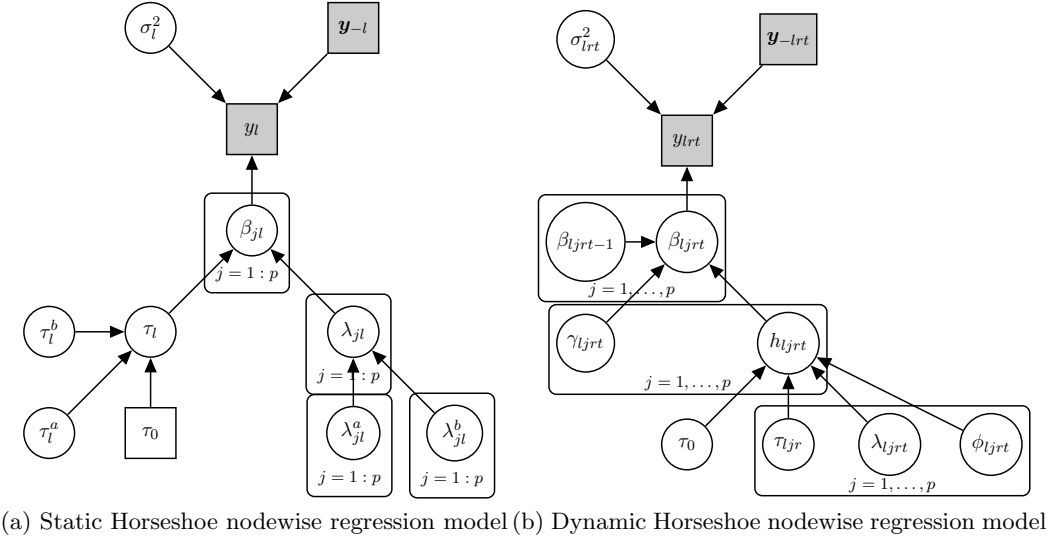

\subsubsection{Extension to multiple groups}
We now extend the Horseshoe prior to allow borrowing information across multiple groups of observations. The groups are usually defined by the problem under investigation, for example they might correspond to different biological conditions, disease status or spatial regions. Estimating a single graphical model would lead to an implicit assumption of homogeneity of the underlying graphs across  groups, with a consequent loss of information about their heterogeneity and a high risk of false positives. On the other hand, inferring each graph individually might lead to a loss of power given the reduction in sample size. Several approaches for joint inference of multiple GGMs have been recently developed (see, for example, \cite{guo2011joint,telesca2012modeling,danaher2014joint,mohan2014node,chun2015gene,peterson2015bayesian,tan2017bayesian,warnick2018bayesian}). We opt for nodewise regression for computational convenience and ease of extension to more complex setups. 

Let $R$ be the number of groups and let $\bm Y_r$ be a matrix of dimension $n_r \times M$ containing only the observations belonging to group $r$, with $r=1, \dots, R$. We introduce dependence across multiple graphs through the group specific global shrinkage parameter $\tau_{lr}$ of the Horseshoe prior, extending easily model \eqref{eq:Reg_HS2} to this setup. The model is now defined as follows
\begin{align}
	\begin{split}
		\boldsymbol{y}_{lr} \mid \boldsymbol{\beta}_{lr} , \sigma^2_{lr} & \sim \text{N}_p\left( \bm X_r \boldsymbol{\beta}_{lr} , \sigma^2_{lr} I_{n_r} \right)\\
		\beta_{jlr} \mid \lambda_{jlr}, \tau_{lr} & \sim \text{N}\left( 0, \lambda_{jlr}^2 \tau_{lr}^2 \right) \\
		\lambda_{jlr} & = \lambda_{jlr}^a \sqrt{ \lambda_{jlr}^b }\\
		\tau_{lr} & = \tau_{l}^a \sqrt{ \tau_{l}^b } \tau_{0r}
	\end{split}
	\begin{split} \sigma^2_{lr} \mid a_{\sigma} , b_{\sigma} & \ind \text{Inverse-Gamma}\left( a_{\sigma} , b_{\sigma} \right)\\ 
		\lambda_{jlr}^a & \sim \text{N}^+(0,1) \\
		\lambda_{jlr}^b & \sim \text{Inverse-Gamma}(1/2,1/2) \\
		\tau_{l}^a & \sim \text{N}^+(0,1)\\
		\tau_{l}^b & \sim \text{Inverse-Gamma}(1/2, 1/2)
	\end{split}
	\label{eq:Reg_HS_Multiple}
\end{align}
where $\bm X_r$ is a $n_r \times p$ matrix corresponding to $\bm Y_{-lr}$ (i.e. $\bm Y_r$ excluding the $l$-th column) and $\boldsymbol{\beta}_{lr} = (\beta_{1lr}, \beta_{2lr}, \dots, \beta_{plr} )$ is a vector of regression coefficients specific to equation $l$ and group $r$. We exploit the structure of the Horseshoe prior, retaining group-coefficient specific local shrinkage parameters $\lambda_{jlr}$, while we link the parameters $\tau_{lr}$. The rationale  behind this  strategy is the following. The global shrinkage parameter $\tau_{lr}$ pulls all the coefficients globally towards zero, while the thick half-Cauchy tails for the local variances $\lambda_{jlr}^2$ allow the important coefficients to escape the global shrinkage independently in each group \citep{carvalho2010horseshoe}. 
In practice, 
we expect the graphs to have group specific connection patterns, allowed by the group-coefficient specific local shrinkage parameters $\lambda_{jlr}$. We also expect groups to share some common structures and to be sparse. For this reason, we link  the global shrinkage parameters $\tau_{lr}$, allowing borrowing information across groups about the global level of sparsity of the graphs.

\subsubsection{Extension to multiple time points - Dynamic Horseshoe prior}
A natural extension of the model above is to introduce a temporal dimension, which allows joint inference of time dependent graphs from multiple groups. Some recent proposals for inference of dynamic Graphical Models can be found, for instance, in \cite{harrington2010spatio,abbruzzo2013dynamic,greenewald2017time,huang2017learning,lin2017joint,epskamp2018gaussian}.

In this work, we consider the evolution over time of the patterns of metabolic associations for two ethnic groups. Our goal is to estimate sparse multiple-graphs evolving over time. Here, we provide an extension of the nodewise regression which enables estimation of time dependent graphs for multiple groups, enabling borrowing of information across time and groups.  One of the main goals is to gain a  better understanding how the differences in associations among groups evolve over time. To this end, we extend the model in \eqref{eq:Reg_HS_Multiple} by imposing a time structure over the shrinkage scale parameters of the Horseshoe prior following the approach proposed by \cite{kowal2017dynamic}. Time dependence is introduced by specifying a stochastic volatility model on the log-variance of each regression coefficient, as well as an autoregressive term in the distribution of $\beta_{jrt}$. Let $t = 1, \dots, T$ be the time index and let $\bm Y_{rt}$ be a $n_{rt} \times M$ matrix containing  the observations belonging to group $r$ at time $t$. In what follows, we  omit the equation subscript $l$ for ease of  notation. Define
  $h_{jrt}$ as 
\begin{align}
	h_{jrt} = \log \left(\tau_{jr}^2 \tau_{0}^2 / \sqrt{pn_{rt}} \right) + \phi_{jr} \left( h_{jrt-1} - \log \left( \tau_{jr}^2 \tau_{0}^2 / \sqrt{pn_{rt}} \right) \right) + \log\left( \lambda_{jrt}^2 \right)
	\label{eq:log_var_dynModel}
\end{align}
where $p=M-1$,  $\phi_{jr}$ is an autoregressive coefficient specific to the $j$-th covariate and $r$-th group, $\tau_{0}$ is a global shrinkage parameter common to all regression coefficients and shared by all groups, $\tau_{jr}$ is a group-coefficient specific shrinkage parameter and $\lambda_{jrt}^2$ is a time-group-coefficient specific local shrinkage parameter. Note that $\tau_0$  allows sharing information across groups, while the time dependence is captured by the parameter $\phi_{jr}$. 
Finally, the time-dependent multiple-groups nodewise regression model becomes
\begin{align}\label{eq:Reg_HS_Time}
		\boldsymbol{y}_{rt} \mid \boldsymbol{\beta}_{rt}, \sigma^2_{rt} & \sim \text{N}_p\left( \bm X_{rt} \boldsymbol{\beta}_{rt}, \sigma^2_{rt} I_{n_rt} \right) \nonumber\\
		\sigma^2_{rt} \mid a_{\sigma}, b_{\sigma} & \sim \text{Inverse-Gamma}\left( a_{\sigma}, b_{\sigma} \right) \nonumber\\
		\beta_{jrt} & = \beta_{jrt-1} + \gamma_{jrt} \exp( h_{jrt}/2 ) \nonumber\\
		\gamma_{jrt} & \ind \text{N}(0,1)\\
	h_{jrt}&  = \log \left(\tau_{jr}^2 \tau_{0}^2 / \sqrt{pn_{rt}} \right) + \phi_{jr} \left( h_{jrt-1} - \log \left( \tau_{jr}^2 \tau_{0}^2 / \sqrt{pn_{rt}} \right) \right) + \log\left( \lambda_{jrt}^2 \right) \nonumber\\
		(\phi_{jr} + 1)/2 \mid \phi_a, \phi_b & \sim \text{Beta}(\phi_a, \phi_b) \nonumber\\
		\tau_{0} & \sim \text{C}^+ \left(0, 1 \right) \nonumber\\
		\tau_{jr} & \sim \text{C}^+(0,1) \nonumber\\
		\lambda_{jrt} & \sim \text{C}^+(0,1)  \nonumber
\end{align}
where $\boldsymbol{\beta}_{rt} = (\beta_{1rt},\dots, \beta_{prt})$ is a time and group specific vector of regression parameters and  $\bm X_{rt}$ is a $n_{rt} \times p$ matrix corresponding to $\bm Y_{-lrt}$ (i.e. $\bm Y_{rt}$ excluding the $l$-th column). A schematic representation of model \eqref{eq:Reg_HS_Time} is shown in Figure 1(b).

The distribution of the logarithm of the square of a half-Cauchy random variable is a Z-distribution \citep{kowal2017dynamic}. In particular, if $\zeta = \log(\lambda^2)$, where $\lambda \sim \text{C}^+(0, 1)$, then $\zeta$ has probability density function
\begin{equation}
	g(\zeta) = \pi^{-1} \exp\left(\zeta\right) \left[1 + \exp\left(\zeta \right) \right]^{-1}, \quad \zeta \in \mathbb{R}
\end{equation}

The Z-distribution can be represented as a mean-variance mixture of Gaussian distributions \citep{barndorff1982normal} and, thanks to the Polya-Gamma expansion proposed by \cite{kowal2017dynamic}, we can develop a multiple groups hierarchy similar to the one in model \eqref{eq:Reg_HS_Multiple}. To this end, we define $\zeta_{jrt} = \log\left( \lambda_{jrt}^2\right)$, $\mu_{0} = \log\left( \tau_{0}^2 \right)$ and $\mu_{jr} = \log\left( \tau_{jr}^2 \tau_{0}^2 \right)$ and we re-write the prior distributions for the parameters on a log scale as
\begin{align}
	\begin{split}
		\zeta_{jrt} \mid \xi_{\zeta_{jrt}} & \sim \text{N} \left( 0, \xi_{\zeta_{jrt}}^{-1} \right)\\
		\mu_{jr} \mid \mu_{0}, \xi_{\mu_{jr}} & \sim \text{N} \left( \mu_{0}, \xi_{\mu_{jr}}^{-1} \right)\\
		\mu_{0} \mid \xi_{\mu_0} & \sim \text{N} \left( 0, \xi_{\mu_0}^{-1} \right)\\
	\end{split}
	\begin{split}
		\xi_{\zeta_{jrt}} & \sim \text{Polya-Gamma}(1,0)\\
		\xi_{\mu_{jr}} & \sim \text{Polya-Gamma}(1,0)\\
		\xi_{\mu_0} & \sim \text{Polya-Gamma}(1,0)
	\end{split}
	\label{eq:polya_gamma}
\end{align}

This modelling strategy allows us  to propagate the shrinkage profile of each regression coefficient over time, allowing fast structural changes or slowly adjusting processes.  The theoretical properties of the dynamic Horseshoe prior are discussed by~\cite{kowal2017dynamic}, who also  show its good performance  when compared to alternative priors. Moreover, the Polya-Gamma expansion in \eqref{eq:polya_gamma} leads to efficient computations as it allows to design  a fast block-Gibbs sampler (although, for the application in this work, posterior inference is performed through Hamiltonian Monte Carlo methods as implemented in the software {\tt Stan}).

\subsection{Posterior Inference}
For datasets of moderate size, e.g. $M \le 30$ and sample size $n \le 1000$, posterior inference for the nodewise regression model with Horseshoe prior (static and dynamic) can be  performed efficiently
using Bayesian softwares like {\tt Stan} \citep{JSSv076i01} or JAGS  (\url{http://mcmc-jags.sourceforge.net/}). The main computational bottleneck for our model is sampling from the posterior distribution of the regression coefficients, which can be expensive for large values of $p$ with complexity equal to $O(p^3)$ in general settings \cite{rue2001fast, bhattacharya2016fast}. \cite{johndrow2020scalable} propose an approximation scheme for the horseshoe posterior, which exploits the structural sparsity of the posterior to reduce the cost per step and that can scale to hundreds of thousands of predictors. The per step computational cost of the approximate algorithm is of the order $(\kappa \vee p) N$, where $\kappa$ is the number of included variables at every step.

In Section 1 of Supplemental Material, we provide sample code to implement the proposed approach in {\tt Stan}, which implements Hamiltonian Monte Carlo (HMC, \cite{brooks2011handbook}, chapter 5). When $M$ is large ($>100$), implementation in a low level language is advisable. We also develop a block Gibbs sampling, extending the algorithm provided by \cite{kowal2017dynamic} to allow for  multiple groups of different sample sizes across time. Details of the MCMC algorithm are provided in Section 1 of Supplemental Material. 
The computational efficiency of the two algorithms is assessed through a simulation study. We consider $N = 100$ observations split equally between two groups and different values for the number of nodes $M$ and of time points $T$. We record the computational times (in seconds) per iteration for both the {\tt stan} implementation and the block Gibbs sampler MCMC algorithm (implemented in {\tt R}). The computer on which we have run the simulations is a Linux machine with Intel Core i7 1.8 GHZ. As it is shown in Table 1 of Supplemental Section 1, both algorithms are computationally efficient when $T=3$, with the Gibbs sampler outperforming {\tt Stan} for larger number of time points. This is due to the ability of the Gibbs sampler to perform joint updates without looping over the number of time points. Recall also that the computational burden of the dynamic Horseshoe approach scales linearly with the number of time points \cite{kowal2017dynamic}. The results can be further improved by implementing a parallel programming strategy, thanks to the conditional independence of the different regressions.

Finally, posterior inference on the implied graph can be performed by 
applying either the \textit{OR} rule or the \textit{AND} rule. In practice, at every iteration of the MCMC algorithm, we obtain a draw for the graph, by including an edge between nodes $j$ and $l$ if and only if the draws  ${\beta}_{jl} \neq 0 \ or \  {\beta}_{lj} \neq 0$, in the case of the \textit{OR} rule. When using the \textit{AND} rule, we include an edge between nodes $l$ and $j$ if and only if the draws  ${\beta}_{jl} \neq 0 \ and \  {\beta}_{lj} \neq 0$. Moreover, given ${\beta}_{jl}$ and ${\sigma}^2_l$, we obtain a sampled value for ${\Omega}$ by setting the diagonal elements equal to $1/{{\sigma}}^2_l$ and off-diagonal elements equal to $-{{\beta}}_{jl} / {{\sigma}}^2_l$. 

\section{Simulation Study}
\label{sec:sim}
We investigate the performance of the proposed models on synthetic datasets. We evaluate the ability of each model to estimate the precision matrix $\Omega$ through the Mean Absolute Error (MAE), calculated between the true matrix and its posterior mean estimate. We also investigate the ability of the model to recover the true graph structure $G$ using the Area Under the Curve (AUC), which is a normalised measure of the area under the Receiver Operating Characteristic (ROC) curve. The ROC curve is obtained by plotting the true positive rate against the false positive rate evaluated at different thresholds for the edge inclusion probability. We also include results relative to the false positive rates (FPR) and false negative rates (FNR) when comparing different network estimates. For all the simulation scenarios presented in this Section, we show the results obtained by applying the \textit{AND} rule. Using the \textit{OR} rule yield comparable results.

In the first simulation, we compare the estimate of the multiple groups model in \eqref{eq:Reg_HS_Multiple} with that of the R package \texttt{BDgraph}, which implements a birth-death MCMC algorithm for Bayesian structure learning in graphical models, and with the Graphical Group Lasso \citep{danaher2014joint}. We construct three precision matrices $\Omega_1$, $\Omega_2$ and $\Omega_3$, corresponding to graphs $G_1$, $G_2$ and $G_3$, of $M=20$ nodes. Following \citep{peterson2015bayesian}, we first define the precision matrix $\Omega_1$ and then derive the others as a perturbation of the first. We set the main diagonal elements of $\Omega_1$ equal to 1, first off-diagonal elements $\omega_{i,i+1} = \omega_{i+1,i} = 0.5$, for $i=1,\dots,19$ and second off-diagonal elements $\omega_{i,i+2} = \omega_{i+2,i} = 0.5$, for $i=1,\dots, 18$. Then we set all $\omega_{i,j} = 0.9$, for $i<j<6$, while the rest of the elements are set to zero. $\Omega_2$ is derived from $\Omega_1$, setting the second off-diagonal elements $\omega_{i,i+2} = \omega_{i+2,i} = 0$, for $i=1,\dots, 18$, all the remaining elements being equal. $\Omega_3$ is derived from $\Omega_1$, setting the first off-diagonal elements $\omega_{i,i+1} = \omega_{i+1,i} = 0$, for $i=1,\dots,19$, all the remaining elements being equal. The newly created matrices are not positive definite and, therefore, we compute the nearest positive-definite approximation through the R function \texttt{nearPD} (\cite{higham2002computing}, from the R package \texttt{Matrix}. The precision matrices $\Omega_2$ and $\Omega_3$ constructed with this procedure are a perturbation of $\Omega_1$: as a result they exhibit some common edges and some group specific connections. The number of observations is fixed to $60, 40, 30$ for group 1, 2 and 3, respectively. Each graph is characterised by a dense group of edges on nodes 1 to 6, representing a set of high partial correlations (absolute value of $0.9$).

In the second simulation scenario we compare the multiple groups dynamic nodewise model in \eqref{eq:Reg_HS_Time} with (i) the static multiple groups nodewise model in \eqref{eq:Reg_HS_Multiple} (where we assume the three time periods to be independent); (ii) the {\tt BDgraph} package and (iii) the Graphical Group LASSO. We consider $M=20$ nodes, two groups and we construct two matrices $\Omega_{1t_1}$ and $\Omega_{2t_1}$, one for each group at time 1. First we set the main diagonal elements of $\Omega_{1t_1}$ equal to 1 and we add $12$ non-zero off-diagonal elements, chosen randomly from the $K$ possible edges and setting them equal to 0.5. Then $\Omega_{2t_1}$ is constructed removing two edges from $\Omega_{1t_1}$ at random  and adding three new edges chosen randomly as before. These three new edges are set equal to 0.5. Finally, we simulate the evolution over time of the two precision matrices, removing two edges and adding a new edge randomly chosen for each time point (setting the corresponding elements in the precision matrix equal to 0.5) for a total of $T=3$ time points. In Figure 3 in Section 2 of Supplemental Material, we show the networks generated with such procedure from which we simulate the datasets. The number of observations is fixed to $50, 40, 30$ respectively for $t_1, t_2$ and $t_3$ (where each group has half of the total sample size at each time point).

In the third simulation we construct the dynamic precision matrices following the same procedure as the second scenario, changing the number of time points to $T=10$. The number of observations is fixed to $40$ per time point (equally split between two groups). The generated graphs are characterised by a slowly changing pattern, where only one edge is added or removed at each time point.

\subsection{Simulation results}
In Figure 1 in Supplemental Section 2 we report the results of the first simulation scenario by displaying the boxplots of the MAE and AUC, calculated over twenty replicates obtained with the multiple groups nodewise model \eqref{eq:Reg_HS_Multiple} and with the \texttt{BDgraph} package. The nodewise model works better in terms of MAE, for which a value closer to 0 denotes an estimate of $\Omega$ close to the truth, and in terms of AUC, for which a value close to 1 denotes a better recovery of the true graph. The values of the FPR and FNR for the two models, shown in Figure 2 in Section 2 of Supplemental Material, are comparable in all groups, and in particular equal to zero for groups $G_2$ and $G_3$.

In Figures~4 and 5 in Supplemental Section 2 we show the results for the second simulation scenario. We compare the results obtained using the dynamic and static multiple groups nodewise models in \eqref{eq:Reg_HS_Multiple} and \eqref{eq:Reg_HS_Time}, the \texttt{BDgraph} package and the Graphical Group LASSO method. We notice a competitive recovery offered by the proposed model of the true precision matrices (Figure~4, MAE, top two panels). The graphical structure (Figure~4, AUC, bottom two panel) is recovered well by the proposed method in two of the time points for the first group, and in the second time point for the second group. In this simulation setting, no method consistently outperforms the others. This is not surprising given the sample sizes. 
In Figure~5 we display False Positive and False Negative Rates for the four models. As expected, given the use of a shrinkage prior,the proposed models perform better in terms of FPR and, in general, similarly to BDgraph. We note that Graphical Group LASSO induces less sparsity in graph reconstruction.

The results of the comparison between the dynamic and static nodewise models obtained from the third simulation scenario are shown in Figures~2 and 3, where we report the boxplots of the MAE, AUC, FPR and FNR calculated over twenty replicates. The dynamic model has the lowest MAE values in both groups and at all time points, while {\tt BDgraph} shows the worst performance in terms of MAE. In terms of graph recovery, as measured by AUC, the methods are comparable.In terms of FPR, the dynamic Horseshoe model performs most shrinkage, although it suffers the highest FNR.

\begin{figure}
	\includegraphics[width=1\linewidth]{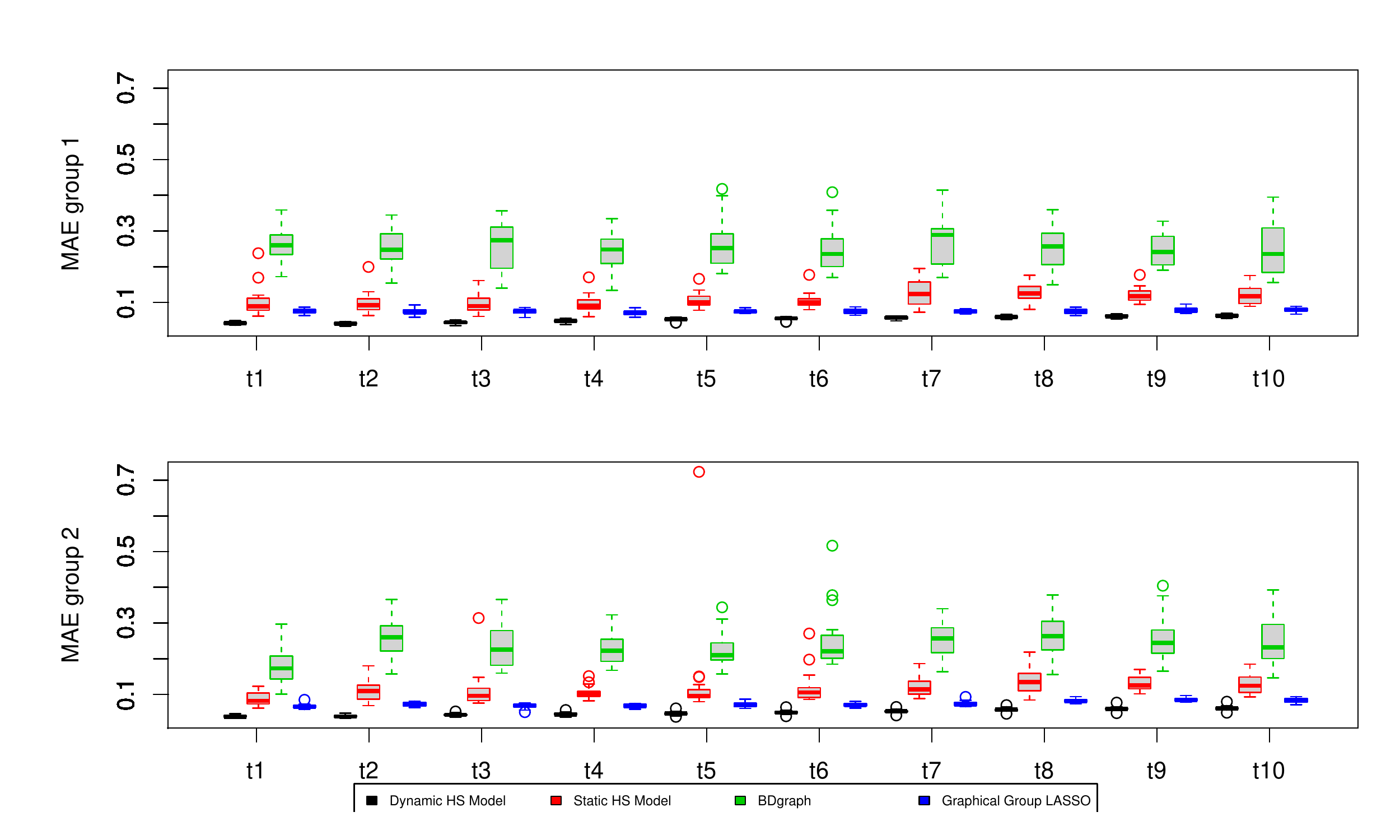}	\includegraphics[width=1\linewidth]{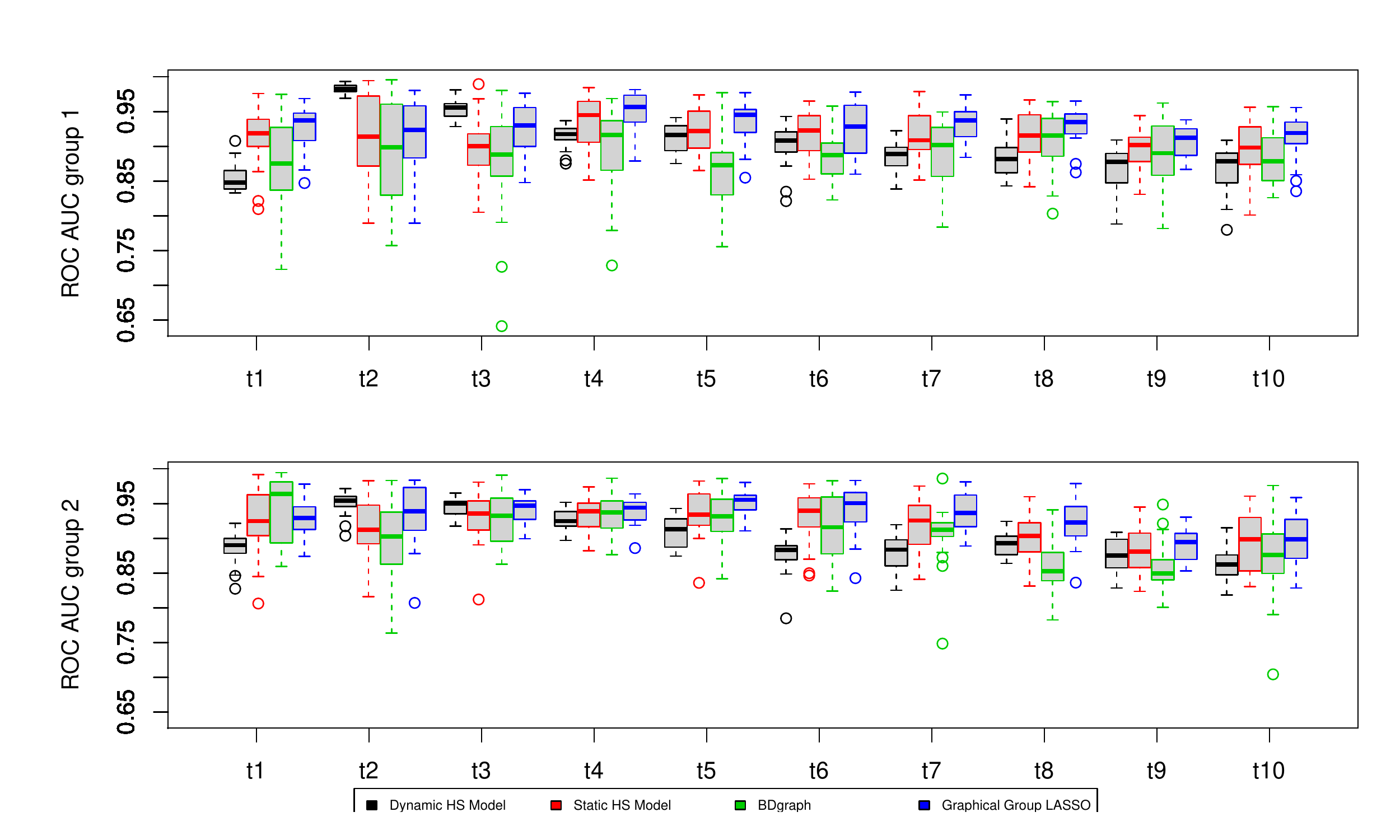}
	\caption{Third simulation scenario. MAE and AUC comparison between the results obtained using the dynamic model in \eqref{eq:Reg_HS_Time} (black), the static model in \eqref{eq:Reg_HS_Multiple} (red),  {\tt BDgraph} (green) and  Graphical Group LASSO (blue). Each row refers to one of the groups and shows the boxplots for each of the ten time points. The results are obtained over twenty replicates and by adopting the \textit{AND} rule. In terms of MAE, the dynamic model yields the lowest values, indicating overall a good performance in estimating the precision matrix. On the other hand, {\tt BDgraph} shows the worst performance in terms of MAE. The methods are comparable in terms of AUC.}
	\label{fig:AUC_NT10_m20}
\end{figure}

\begin{figure}
	\includegraphics[width=1\linewidth]{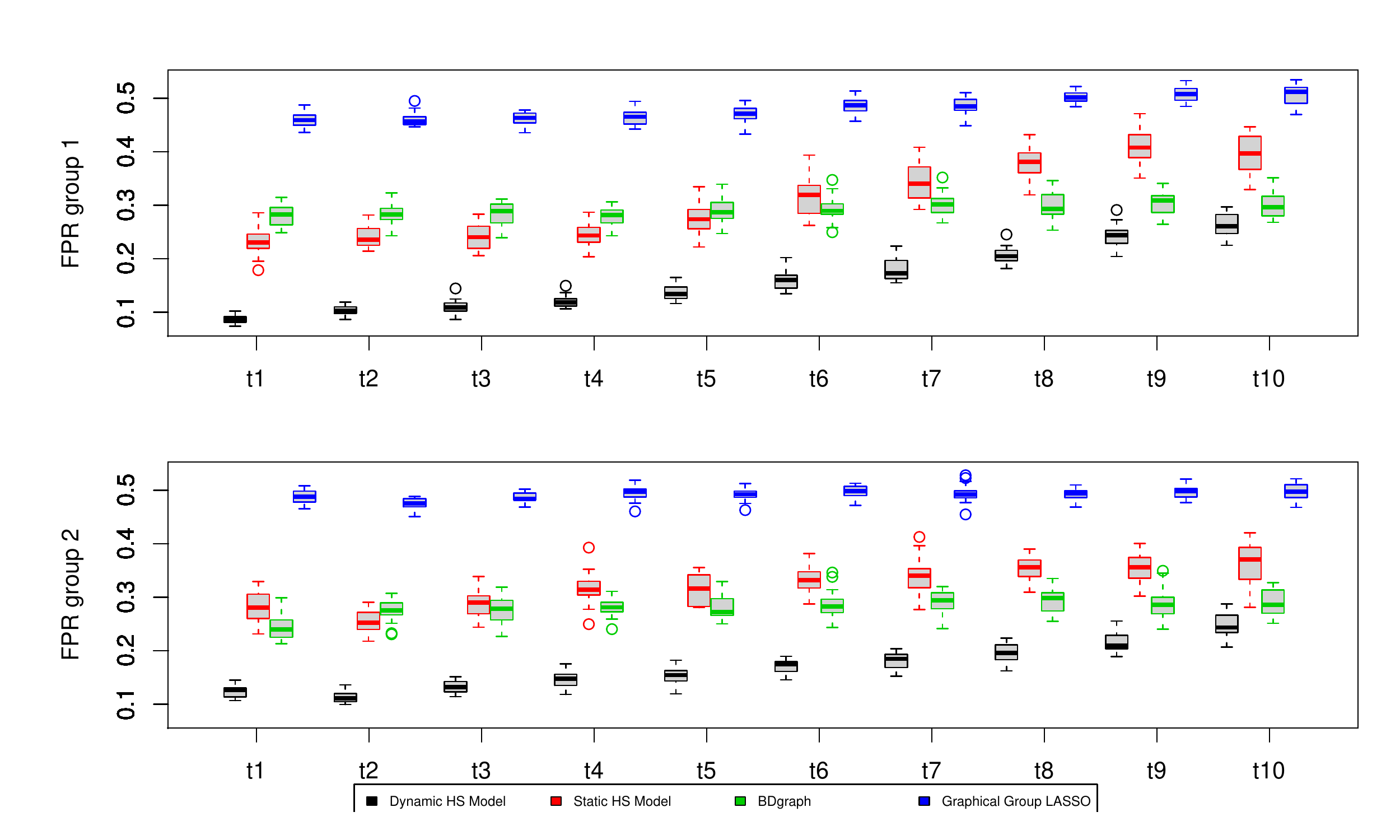}	\includegraphics[width=1\linewidth]{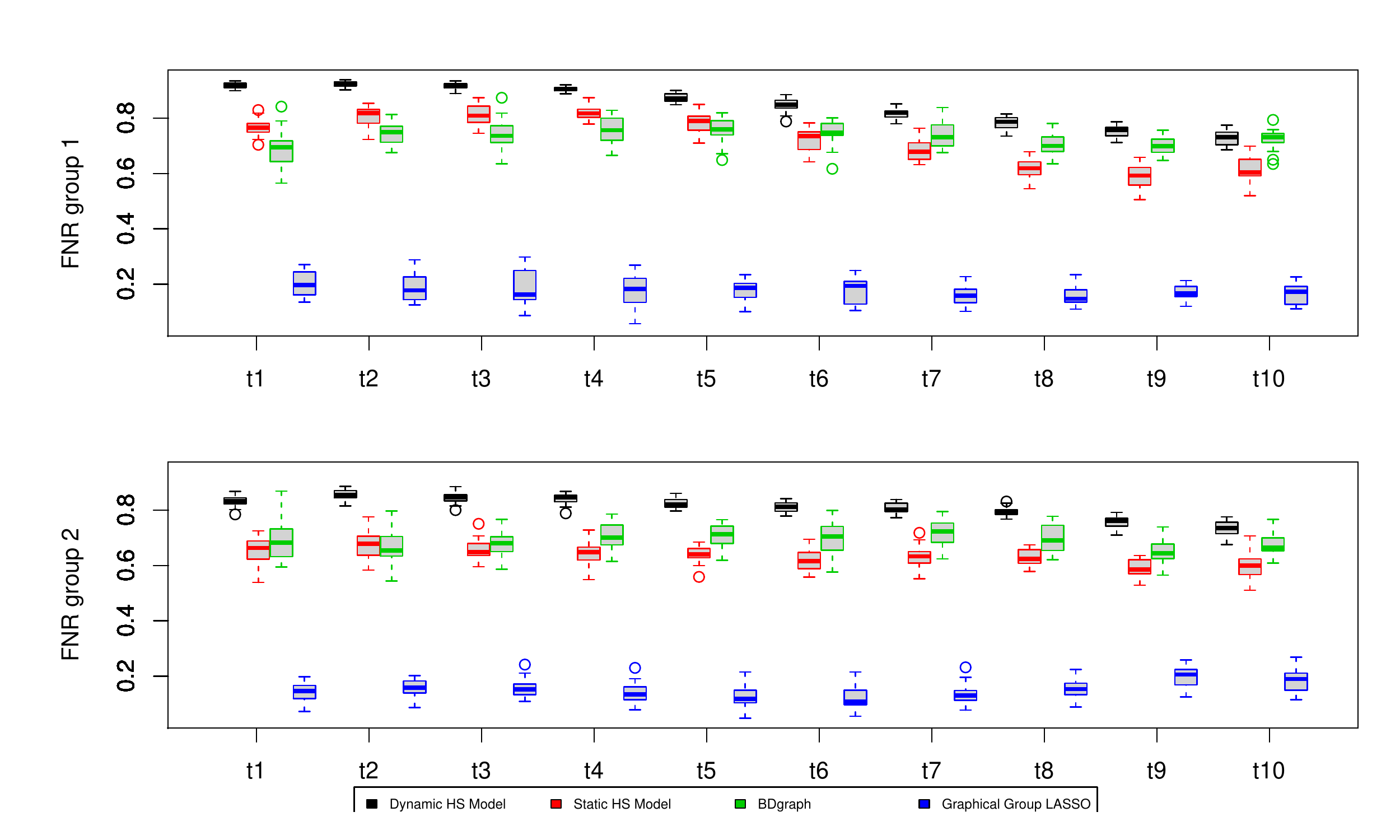}
	\caption{Third simulation scenario. False positive rates (FPR) and false negative rates (FNR) for the dynamic model in \eqref{eq:Reg_HS_Time} (black), the static model in \eqref{eq:Reg_HS_Multiple} (red),  {\tt BDgraph} (green) and Graphical Group LASSO (blue). Each row refers to one of the groups and shows the boxplots for each of the ten time points. The results are obtained over twenty replicates and by adopting the \textit{AND} rule. In both groups, the dynamic and Graphical Group LASSO models differ the most in terms of FPR and FNR. On the other hand, the results for the static model and  BDgraph are comparable at the first time points, and diverge slightly at later time points.
	}
	\label{fig:FPNR_NT10_m20}
\end{figure}

\section{Application}\label{sec:Sabre}
In this Section, we apply the nodewise regression model for dynamic multiple graphs to the SABRE metabolic data. The dataset has a total of 1246 observations at baseline ($T_1$) and 875 at follow-up time ($T_2$). In this application, groups are defined by ethnicity. Individuals are stratified into two ethnicities at each time point, 690 Europeans and 556 South-Asians at $T_1$ and 503 Europeans and 372 South-Asians at $T_2$. Measurements of different classes of metabolites are available. Lipoproteins, which constitute the majority of the metabolomic  dataset, are classified according to their density (very-low-density lipoprotein (VLDL), low-density lipoprotein (LDL), intermediate-density lipoprotein (IDL) and high-density lipoprotein (HDL)). Each lipoprotein subclass can be further characterised by its lipid composition (i.e. triglycerides, phospholipids, free cholesterol and cholesteryl esters) and its particle size. The full list of metabolites included in the analysis is reported in Table 2 in Section 2 of Supplemental Material. We include in the analysis clinical markers, such as the homeostasis model assessment (Homa IR) as an index of insulin resistance
\citep{Matthews1985} - an important risk factor for the development of type 2 diabetes -, waist to hip ratio (WHR) as a measure of body fat distribution, control variables such as gender, age, smoking habits, physical activity and alcohol consumption. We control for previous disease status, in particular we include indicators of coronary heart disease (CHD), stroke and diabetes mellitus. We also include indicators of drug treatments for blood pressure, diabetes and blood lipids. The full list of covariates included is given in Table 3 in Section 2 of Supplemental Material.

The nodes of the graph correspond to the different metabolites and there are a total of $M = 88$ nodes (i.e. the number of equations in the nodewise regression). When analysing these data, it is important  to control for clinical events of interest (e.g. development of diabetes) that occur before $T_1$ and between $T_1$ and $T_2$. To this end, we model the mean $\mu_l$ of the multivariate Normal distribution in \eqref{eq:conditional_mean} through a linear predictor and assume that $\mu_{lrt} = \bm Z \boldsymbol{\eta}_{lrt}$, where $\bm Z$ is a matrix of predictors common to all equations and $\boldsymbol{\eta}_{lrt}$ is a $p_z$-dimensional vector of regression coefficients for the mean level. The model is completed by specifying a time dependent structure and a prior distribution on $\boldsymbol{\eta}_{lrt}$ as follows
\begin{align*}
	\eta_{klrt} & = \eta_{klrt-1} + \gamma_{klrt}^z\\
	\gamma_{klrt}^z & \sim \text{N}(0, s_0)
\end{align*}
for $k=1,\dots,p_z$, where $s_0$ is the prior variance, here specified to induce a flat Normal distribution. Posterior inference is performed by updating jointly the regression coefficients $\boldsymbol{\beta}_{lrt}$ and $\boldsymbol{\eta}_{lrt}$.  We also include an intercept term so that the total number of covariates is $p_z = 20$. We run the MCMC for 10000 iterations, including a burn-in period of 2000 iterations and thinning every 4 iterations. 
We perform posterior analysis applying both the \textit{AND} and \textit{OR} rule.
In Figures 6 to 9 in Supplemental Section 2, we report the posterior expectation of $\kappa$ for each group and time point. Furthermore, in Figures~10 to 17 we display the inferred individual networks for each ethnic group at both baseline and follow-up times. 
These networks are characterised by a high number of edges, in particular, we can notice a very highly connected group of lipoproteins in all graphs.
In addition to the individual networks we also estimate the differential networks \citep{de2010differential, valcarcel2011differential,tan2017bayesian} derived from the pairwise comparison between the graphs corresponding to the two ethnicities for each time point and the pairwise comparison between $T_1$ and $T_2$ for each ethnicity. Following \cite{tan2017bayesian}, a differential network includes all the edges that are present only in one of the two groups/times (i.e. present in one group/time and not the other and vice-versa), thus shedding light on the main differences between ethnicities and the evolution of the metabolic associations over time. Differential networks obtained with the \textit{AND} are shown in Figures~4 to 7, while those obtained with \textit{OR} are reported in Figures~18 to 21. In a differential network,  an edge between two nodes is added if the posterior probability of an edge being in one graph but not in the other is higher than $0.5$. In this Section, we only discuss the results obtained with the \textit{AND} rule since it implies a more stringent selection criteria.

In Figures 4 and 5 we show the differential networks between $T_1$  and $T_2$ for Europeans and South-Asians, respectively.
\begin{figure}
	\centering
	\includegraphics[width=0.9\linewidth]{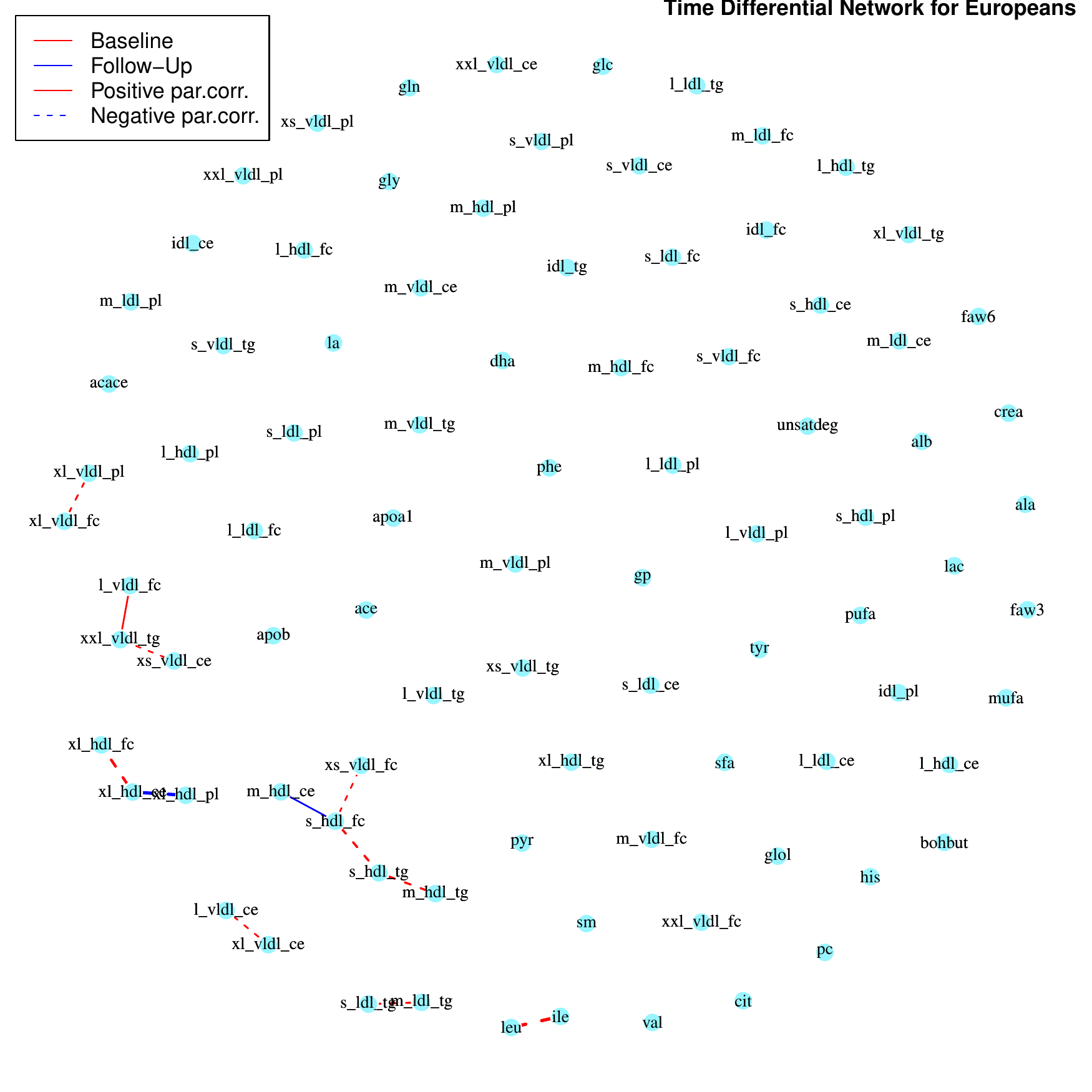}
	\caption{European Differential network between baseline and follow-up obtained with the \textit{AND} rule. Red lines correspond to edges only present in the baseline network, while blue lines to those only present in the follow-up network. Continuous lines represent differential positive partial correlations, while dashed lines indicate negative ones. Note that the majority of the differential edges comes from those only present in the baseline network.}	\label{fig:time_diff_net_EU_AND}
\end{figure}
\begin{figure}
	\centering
	\includegraphics[width=0.9\linewidth]{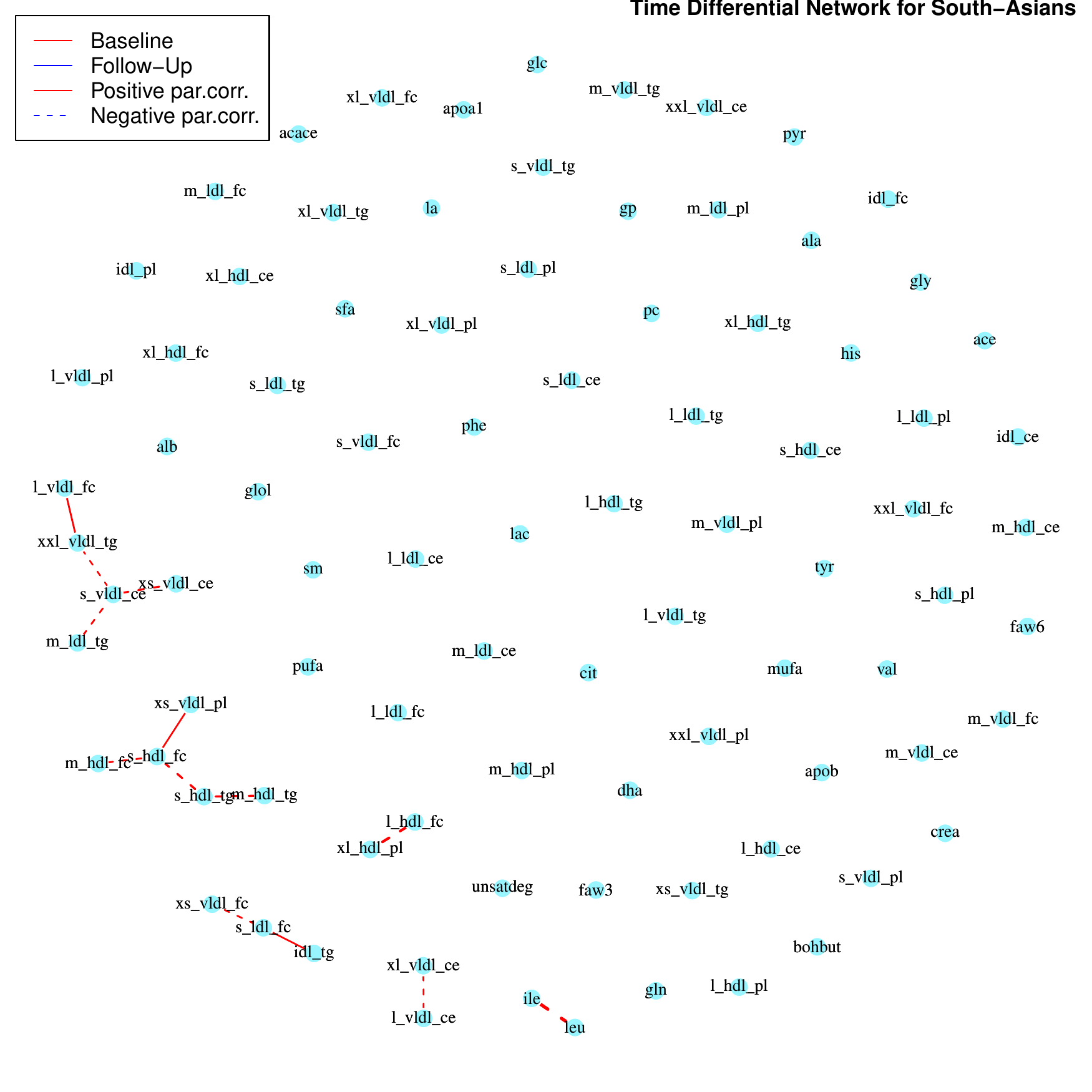}
	\caption{South-Asians Differential network between baseline and the follow-up obtained with the \textit{AND} rule. Red lines correspond to edges only present in the baseline network, while blue lines to those only present in the follow-up network. Continuous lines represent differential positive partial correlations, while dashed lines indicate negative ones. Note that the only difference between the two networks comes from edges only present in the baseline network.}
	\label{fig:time_diff_net_AS_AND}
\end{figure}
It is worth noticing that there are no edges among the majority of the metabolites in both  differential networks, which implies that the presence or absence of those connections is shared by the respective ethnicity both at $T_1$ and $T_2$. Moreover, for both ethnic groups, the majority of the edges in the differential networks derives from edges present at baseline, but not at follow-up. The connected metabolites in the differential networks for both ethnicities (Figures~4 and 5) belong predominantly to the groups of very low density lipoproteins and high density lipoproteins, with the addition of an edge between leucine and isoleucine. These are, together with valine, essential amino acids which account for 35–40\% of the dietary indispensable amino acids in body protein and 14\% of the total amino acids in skeletal muscle \citep{manoli2016disorders}. Their association with a number of disorders has often been reported in the literature, including insulin resistance, type-2 diabetes \citep{guasch2016metabolomics} and cardiovascular diseases \citep{tobias2018circulating}.

In Figures 6 and 7 we report the differential networks between Europeans and South-Asians, respectively at time $T_1$ and $T_2$.
\begin{figure}
	\centering
	\includegraphics[width=0.9\linewidth]{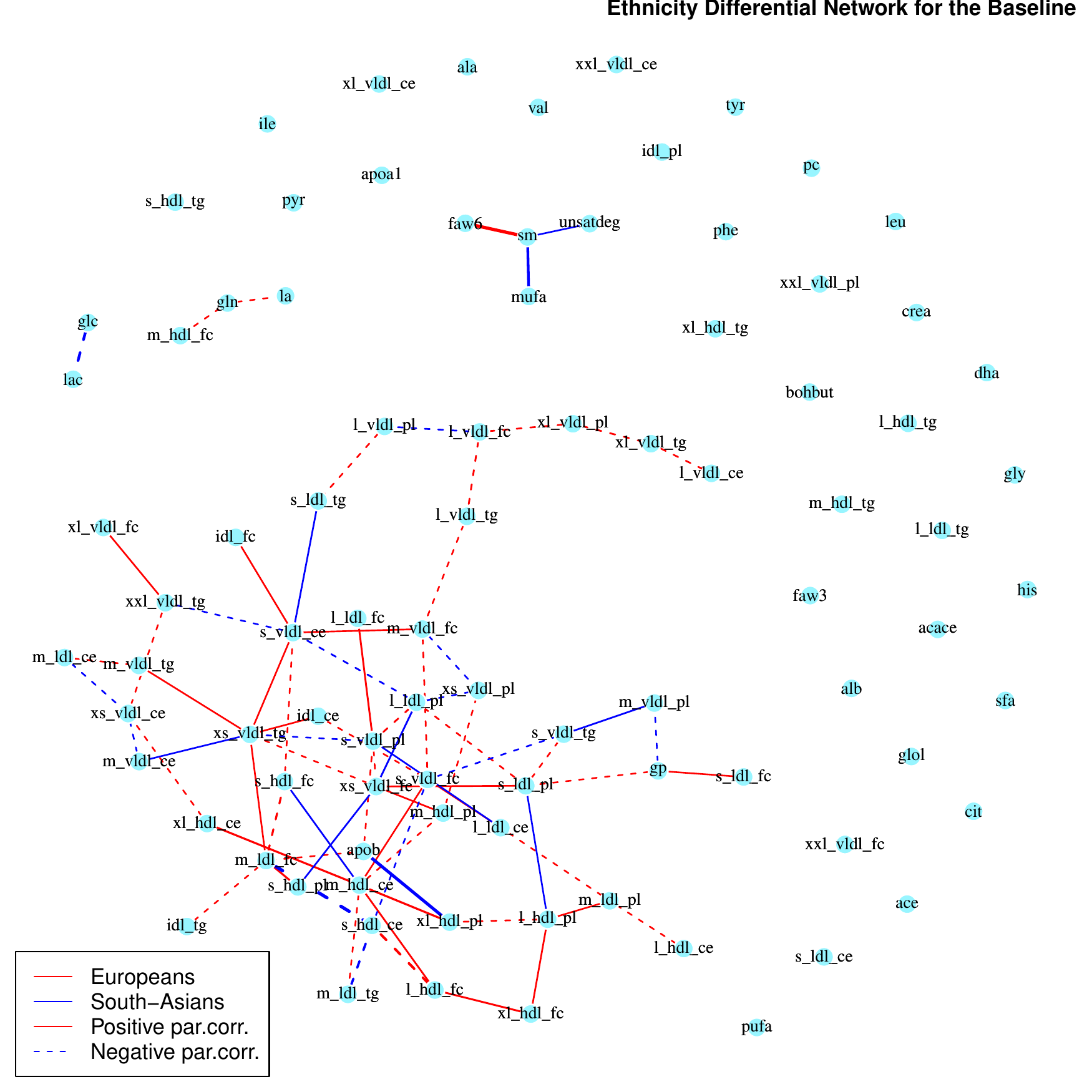}
	\caption{Differential network between Europeans  and South-Asians at  baseline obtained with the \textit{AND} rule. Red lines correspond to edges only present in the European network, while blue lines to those only present in the South-Asian network. Continuous lines represent differential positive partial correlations, while dashed lines indicate negative ones.}
	\label{fig:diff_net_EU_AS_t1_AND}
\end{figure}
\begin{figure}
	\centering
	\includegraphics[width=0.9\linewidth]{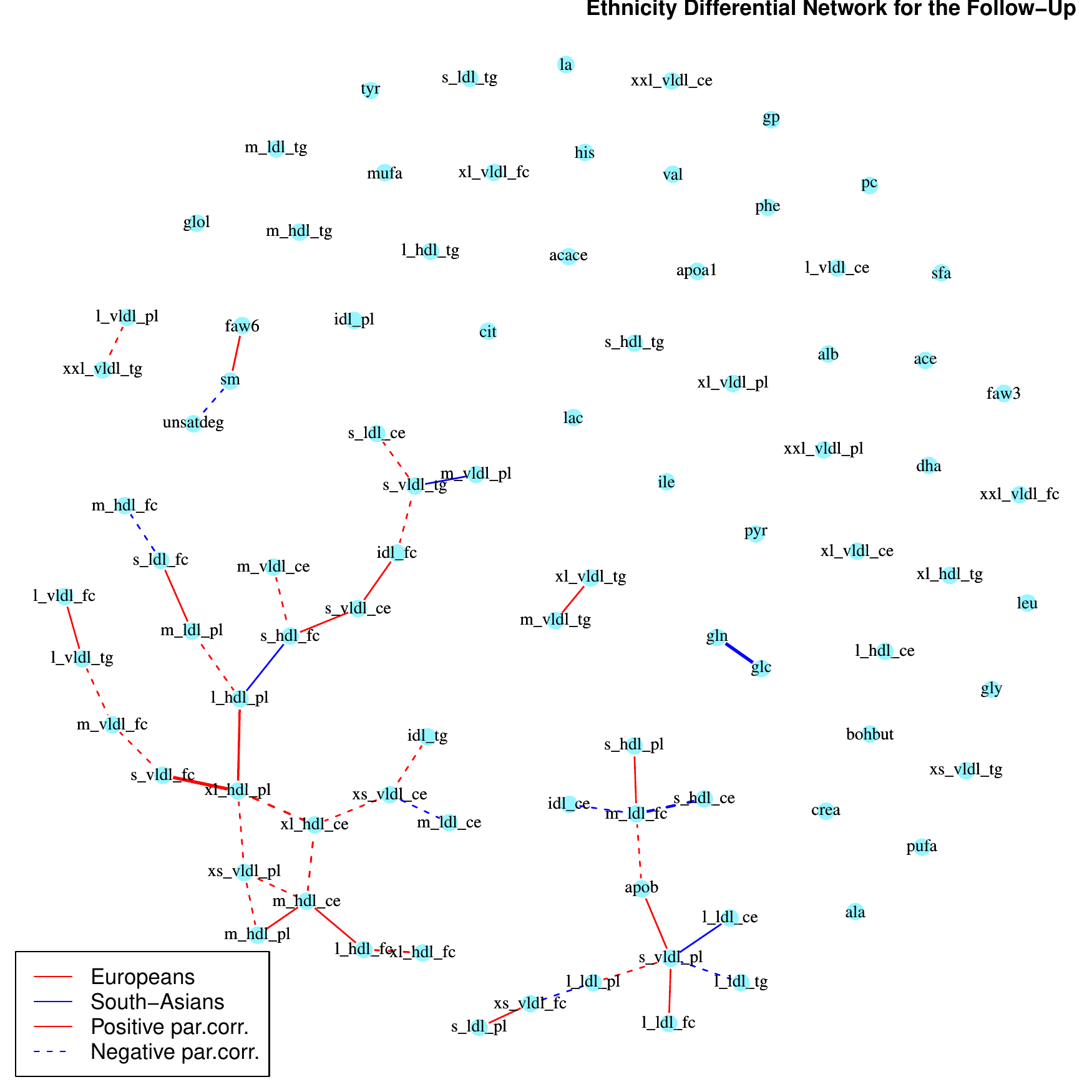}
	\caption{Differential network between Europeans and South-Asians at follow-up obtained with the \textit{AND} rule. Red lines correspond to edges only present in the European network, while blue lines to those only present in the South-Asian network. Continuous lines represent differential positive partial correlations, while dashed lines indicate negative ones.}
	\label{fig:diff_net_EU_AS_t2_AND}
\end{figure}
These networks present edges connecting amino-acids and lipoproteins sub-fractions, highlighting potential differences in underlying metabolic processes. In the last column of Table 2 in Supplemental Section 2 we report the metabolites differentially connected at both time points. To gain a better understanding of the estimated connections and to relate the estimated graph to known metabolic pathways, we conduct a pathway over-representation analysis (ORA) using the online software \texttt{IMPaLA} \citep[Integrated Molecular Pathway Level Analysis,][]{kamburov2011integrated}. We include in the analysis all metabolites that have a connection in the differential network. ORA evaluates statistically the fraction of metabolites in a particular pathway found among the user-specified set of metabolites, in our case the metabolites with connections in the differential network. For each pathway, input metabolites that are part of the pathway are counted. Next, every pathway is tested for over- or under-representation in the list of input metabolites using the hypergeometric test. The most represented pathways are the ones with smaller p-value and higher number of over-represented metabolites. Here we discuss the first four top-ranked pathways at time $T_2$: \textit{Cholesterol metabolism}, \textit{Fat digestion and absorption metabolism}, \textit{Lipid and atherosclerosis metabolism} and \textit{Vitamin digestion and absorption}. Note that the differential network at time $T_1$ contains only one extra metabolites as compared to $T_2$, leading to substantially the same pathways.

Cholesterol metabolism in humans is a complex process which involves multiple metabolic pathways subject to many points of regulation under genetic and metabolic control \citep{goldstein2015century}. In brief, cholesterol is either supplied from  diet or synthesized by many cells of the body, with the liver being one of the major sites of cholesterol synthesis. A diet characterised by high intakes of cholesterol itself, or of saturated fats and excessive calories, may increase the level of cholesterol in the blood. The response to changes in dietary cholesterol is heterogeneous in humans and many studies show that cholesterol consumption should be restricted by diabetics and others at risk for cardiovascular disease \citep{elkin2017cholesterol}. Moreover,  levels of cholesterol in blood are also influenced by environmental factors such as dietary fatty acids and metabolic perturbations such as diabetes, obesity and genetic factors \citep{ARNOLD20031226}. Lipids and lipoproteins crucially contribute to atherosclerosis, the pathological basis of cardiovascular diseases and influence inflammatory processes as well as function of leukocytes, vascular and cardiac cells, thereby impacting on vessels and heart. Clinical studies have found a clear causal relationship between hypercholesterolaemia and atherosclerotic disease \citep[see, for instance,][]{rader2003monogenic}. Furthermore, beyond LDL cholesterol, there is evidence that also other lipid mediators contribute to cardiovascular risk \citep{soppert2020lipoproteins}. Consistently with our analysis, ethnic differences in dyslipidemia patterns, lipid metabolism, total and HDL cholesterol have been reported in the literature \citep{ujcic2010ethnic, frank2014racial, enkhmaa2016lipoprotein}.

Lipids have physical, chemical, and physiological characteristics that make them important factors in human nutrition. They form a group of compounds of varied chemical nature, with the common property of being soluble in organic solvents but insoluble in water. This property affects their digestion, absorption and transport in the blood and metabolism at cellular level \citep{ramirez2001absorption}. Dietary fats consist of a wide range of polar and nonpolar lipids \citep{carey1983small}. Triacylglycerol is the dominant fat in the diet, contributing 90–95\% of the total energy derived from dietary fat. Dietary fats also include phospholipids, sterols (e.g. cholesterol), and many other lipids (e.g. fat-soluble vitamins). Lipid digestion, absorption, and metabolism have been linked to lipid-associated disorders, including dyslipidemias and cardiovascular diseases. Efficient absorption of dietary fats is important as fat can be used as a source of energy to support various cellular functions or stored until it is needed to support intracellular processes \citep{iqbal2009intestinal}. Most circulating and stored fatty acids are derived from the diet \citep{visioli2006lipid} and in recent years the  relationship between fatty acids and cardiovascular disease has received increasing interest \citep{visioli2020fatty}. For instance, a major risk factor for obesity and obesity-related metabolic disorders is the regular consumption of fat-rich meals. Hence, a tremendous effort has been devoted towards limiting the amount of dietary lipids absorbed through the gastrointestinal tract.
In general, most focus has been placed on limiting triglyceride digestion in the intestinal lumen and the transport and absorption of fatty acids and cholesterol through the intestinal mucosa. Other fats also contribute to obesity-related disorders, such as phospholipids, which are major constituents in the intestinal lumen after meal consumption, and products of phospholipid metabolism in the intestine, which directly contribute to cardio-metabolic diseases through multiple mechanisms \citep{hui2016intestinal}.

Regarding the lipid and atherosclerosis pathway, atherosclerosis is a chronic inflammatory disease marked by a narrowing of the arteries from lipid-rich plaques present within the walls of arterial blood vessels \citep{kanehisa2000kegg}. Elevated levels of low density lipoprotein (LDL) cholesterol constitute a major risk factor for genesis of atherosclerosis as LDL can accumulate within the blood vessel wall and undergo modification by oxidation. It is generally accepted that atherosclerotic lesions are initiated via an enhancement of LDL uptake by monocytes and macrophages \citep{choy2004lipids, linton2019role}. Ethnic differences have been reported in the risk of atherosclerosis. For example, Lipoprotein(a) (LP(a)) is a well-known risk factor for atherosclerosis and Lp(a)-associated risk may vary by ethnicity \citep[e.g.,][]{forbang2016sex}.  Lp(a) has been related to greater risk of carotid plaque and its progression in whites and Lp(a)-associated risk of plaque outcomes vary significantly in white and black individuals \citep{steffen2019race}. Moreover, there are distinct patterns of lipid profiles associated with ethnicity regardless of the glucose levels \citep{zhang2010distinct} and in the biological relationships
underlying the serum lipids-disease association \citep{bentley2017interethnic}.

The fourth identified pathway is \textit{Vitamin digestion and absorption}. Vitamins are organic substances which are essential for health and well-being as they catalyse numerous biochemical reactions. Vitamins must be obtained from exogenous sources via intestinal absorption since humans and other mammals cannot synthesize these compounds \citep{kanehisa2000kegg}. The importance of vitamins in health has been widely reported \citep{taylor1979importance, maqbool2017biological}.

These findings suggest plausible metabolic pathways which may be disordered to a greater extent in South-Asians and contribute to their excess risk of diabetes and cardiovascular disease.

 In Figures 22, 23, 24 and 25 we plot the posterior mean of the regression coefficients $\eta_{jlrt}$, for each equation $l$ and covariate $j$, grouped by ethnicity and time. The measure of body-fat distribution WHR has a negative effect on many metabolites for both ethnicities, particularity at $T_1$, while a few metabolites are affected at $T_2$. Blood lipids (triglycerides, cholesterols) and HDL are important for both groups and time periods. The presence of diabetes, or diabetes treatment, also affects the mean level of some metabolites, in particular in Europeans. Homa IR has an effect on an elevated number of metabolites at $T_1$ and $T_2$ in both Europeans and South-Asians. Overall, Homa IR, blood lipids and serum HDL are the control variables that have stronger effect (see the $95\%$ credible region) on the metabolites. High blood triglycerides and low HDL are among the risk factors that determine the metabolic syndrome \citep{roberts2013metabolic}, which can lead to the development of type 2 diabetes. In summary, these findings highlight the presence of complex interplays between metabolic processes, anthropometric factors and clinical markers, which can have different impacts on the risk of diabetes and other cardiovascular diseases  across ethnicities and across time.

\section{Discussion and Conclusion}
\label{sec:Conclusions}

This paper extends nodewise regression technique to infer dynamic evolving multiple graphs. The model allows us to analyse multiple groups of different sample sizes observed at multiple time points, borrowing information across time and groups. We impose regularisation on the regression coefficients and the model allows for the inclusion of prior information about specific connections between pairs of nodes, when prior knowledge is available. The structure of nodewise regression ensures good scalability of the MCMC algorithm thanks to the possibility to infer each regression independently. The Horseshoe prior effectively shrinks small and negligible coefficients to zero (inducing sparsity in the graph), while leaving important coefficients unaffected due to its heavy tails, hence performing (group and time specific) variable selection. 

We illustrate the performance of the proposed model in a simulation study and compare it with an alternative Bayesian model for graph estimation as well as a frequentist method. The results highlight the ability of the model to recover the true underlying structure of the graphs and to accurately estimate the corresponding precision matrices. 
In most scenarios, the proposed methods outperforms the competitors, still maintaining a computational advantage. Finally, we employ the proposed dynamic model to analyse metabolic data from the SABRE cohort study, an information rich dataset on cardiovascular and metabolic diseases. Our clinical interest focuses on different patterns of metabolite associations which characterise  the European and South-Asian ethnicities and their evolution over time, from the baseline visit to  follow-up. Our approach enables us to detect an interpretable set of unique association patterns which can aid mechanistic understanding of between-group and between-times differences in the development of insulin resistance, diabetes and cardiovascular diseases and has the potential to help generating new scientific hypotheses. In doing this, we correct for potential confounders and clinical events that could affect metabolites levels. 

\section*{Supplemental Material}

The Supplemental Material file, {\texttt{supplementarydata.pdf}}, is available on the Journal website.

\section*{Acknowledgement(s)}
 We are grateful to all the people who took part in the study, and past and present members of the SABRE team who helped to collect the data.

\section*{Funding}
Alun Hughes and Nishi Chaturvedi receive support from the National Institute for Health Research University College London Hospitals Biomedical Research Centre, and work in a unit that receives support from the UK Medical Research Council (Programme Code $MC\_UU\_12019/1$). The SABRE study was funded at baseline by the Medical Research Council, Diabetes UK, and the British Heart Foundation. At follow-up the study was funded by the Wellcome Trust $(067100, 37055891)$ $\&$  $(086676/7/08/Z)$, the British Heart Foundation ($PG/06/145$, $PG/08/103/26133$, $PG/12/29/29497$ and $CS/13/1/30327$) and Diabetes UK $(13/0004774)$. The SABRE study team also acknowledges the support of the National Institute of Health Research Clinical Research Network (NIHR CRN). This work was supported by the Singapore Ministry of Education Academic Research Fund Tier~2 under Grant MOE2019-T2-2-100.

\clearpage
\bibliographystyle{plainnat}
\bibliography{Reference_Marco}

\end{document}


\title{Supplemental Material to 'Bayesian Dynamic Network Modelling: an application to metabolic associations in cardiovascular diseases'}

\author[1]{Marco Molinari}
\author[2]{Andrea Cremaschi}
\author[1,2,3]{Maria De Iorio}
\author[4]{Nishi Chaturvedi}
\author[4]{Alun Hughes}
\author[4]{Therese Tillin}

\affil[1]{Department of Statistical Science, University College London, UK}
\affil[2]{Singapore Institute for Clinical Sciences, A*STAR, Singapore}
\affil[3]{Yong Loo Lin School of Medicine, National University of Singapore, Singapore}
\affil[4]{Department of Population Science and Experimental Medicine,
	University College London, UK}

\maketitle

\textbf{Abstract}: 
This document contains additional information relative to the manuscript 'Bayesian Dynamic Network Modelling: an application to metabolic associations in cardiovascular diseases' and it is organised as follows. Section \ref{SMsec:mcmc-algorithm} contains the details of the MCMC algorithm needed to perform posterior inference. We provide both the {\tt Stan} code and the Gibbs sampling scheme. Section \ref{SMsec:additional-figures-and-tables} shows additional figures and tables referenced in the main text.

\textbf{Keywords}: 
Dynamic Shrinkage Priors; Gibbs Sampling; Graphical Models; Metabolomics; Nodewise Regression

\clearpage

\section{MCMC Algorithm}\label{SMsec:mcmc-algorithm}
Here we provide the {\tt Stan} code and the details of the Gibbs sampling algorithm to perform posterior inference of the multiple groups dynamic nodewise regression model.\\

\subsection*{{\tt Stan} code}
{\footnotesize
	\begin{verbatim}
		data {
			int < lower = 1 > NT;           	// Time points
			int < lower = 1 > n_t[ NT ];		// Number of observations for each time point
			int < lower = 1 > Ngr;			// Number of groups 
			int < lower = 1 > n_groups[ Ngr*NT ];	// Dimension of the groups for each time
			int < lower = 1 > m;  	              	// Number of regressors
			int < lower = 1, upper = Ngr > G_t[ sum(n_groups) ];    // Vector with groups membership
			vector [ sum(n_t) ] y_t; 	          	// Response variable
			row_vector[m] X_t[ sum(n_t) ]; 	      	// Input matrix (= Y_-j)
			// ---------------
			real < lower = 0 > scale_global_tau_0[ Ngr ];	// prior scale for the global shrinkage parameter Tau
			real < lower = 1 > df_global_0; 				// degrees of freedom for Tau
			real < lower = 1 > df_global_j;
			real < lower = 1 > df_local_t;					// degrees of freedom for Lambdas
			matrix <lower = 0> [m, Ngr] devs_X_t[ NT ];		// Deviances of X (diag of XtX )
			// ------ Params for the regularisation of the Horseshoe
			real <lower = 0> slab_scale_c_t; 				// Slab of student-t
			real <lower = 0> slab_df_c_t; 					// df of student-t
			// Parameters Beta on Phi_j
			real <lower = 0> phi_ab[ 2 ];
			real <lower = 0> sigma_prior;
		}
		transformed data{
			int n_cumsum_gr[ Ngr * NT ]; // Vector of cumulative sum of observations in groups per time
			int pos;
			pos = 1;
			for(tt in 1:NT){
				for(gr in 1:Ngr){
					if( pos == 1) n_cumsum_gr[ pos ] = 1;
					if( pos > 1) n_cumsum_gr[ pos ] = n_cumsum_gr[ pos - 1 ] + n_groups[ pos - 1 ];
					pos += 1;
				}
			}
		}
		parameters {
			vector < lower = 0 > [ Ngr * NT ] sigma_t;	// Time varying noise std
			// Parameters T
			vector[ m * Ngr ] z_t[ NT ];	//Innovations for Beta_t
			// Tau_j
			vector < lower = 0 > [ m * Ngr ] aux1_global_j;
			vector < lower = 0 > [ m * Ngr ] aux2_global_j;
			// Tau_0
			real < lower = 0 > aux1_global_0;
			real < lower = 0 > aux2_global_0;
			// Local Lambda_jt
			vector < lower = 0 > [ m * Ngr ] aux1_local_t[ NT ];
			vector < lower = 0 > [ m * Ngr ] aux2_local_t[ NT ];
			vector < lower = 0 > [ Ngr ] aux_c_t[ NT ];
			// Autoregressive Parameter phi_j for the dynamic HS
			vector < lower = 0 , upper = 1 > [ m * Ngr ] phi_j_pos;
		}
		transformed parameters {
			vector < lower = -1 , upper = 1 > [ m * Ngr ] phi_j;
			// Tau_j
			vector < lower = 0 > [ m ] tau_j[ Ngr ];
			// Tau_0
			real < lower = 0 > tau_0[ Ngr ];
			// Local Lambda_j,t
			vector < lower = 0 > [ m ] lambda_t[ NT * Ngr ];
			vector < lower = 0 > [ m ] lambda_tilde_t[ NT * Ngr ];	// truncated local shrinkage parameter
			vector < lower = 0 > [ Ngr ] c_t[ NT ];		// c for reg HS
			// log_scale variance h_j,t
			vector [ m ] log_h[ NT * Ngr ];
			vector < lower = 0 > [ m ] exp_h[ NT * Ngr ];
			// Beta_t
			vector[ m ] beta_t[ NT * Ngr ];
			vector [ sum( n_t ) ] f_t;
			vector < lower = 0 > [ sum(n_t) ] sigma_long;  // To store the sigma for each obs
			// <<<<<<<<<<<<<<<<<<<<<>>>>>>>>>>>>>>>>>>>>>>>>
			// <<<<<<<<<<<<<<< Now operations >>>>>>>>>>>>>>
			// <<<<<<<<<<<<<<<<<<<<<>>>>>>>>>>>>>>>>>>>>>>>>
			// Adjust phi_j
			phi_j = (phi_j_pos * 2) - 1;
			// In Time == 1
			c_t[ 1 ] = slab_scale_c_t * sqrt( aux_c_t[ 1 ] );
			for(gr in 1:Ngr){
				// Global Shrinkage parameters
				tau_j[ gr ] = aux1_global_j[((gr-1)*m + 1):((gr-1)*m + m)] .* 
				sqrt( aux2_global_j[((gr-1)*m + 1):((gr-1)*m + m)] );
				tau_0[ gr ] = aux1_global_0 * sqrt( aux2_global_0 ) * scale_global_tau_0[ gr ] * sigma_t[ gr ];
				// tau_j[ gr ] = aux1_global_j .* sqrt( aux2_global_j[ ((gr-1)*m + 1):((gr-1)*m + m) ] );
				lambda_t[ gr ] = aux1_local_t[ 1 ][((gr-1)*m + 1):((gr-1)*m + m)].* 
				sqrt( aux2_local_t[ 1 ][((gr-1)*m + 1):((gr-1)*m + m)] );
				lambda_tilde_t[ gr ] = sqrt( ( c_t[ 1 ][ gr ]^2 * 
				square( lambda_t[ gr ] ) ) ./ ( c_t[ 1 ][ gr ]^2 + 
				square( tau_j[ gr ] ) .* square( lambda_t[ gr ] )));
				// Begin the creation og log-volatility h
				log_h[ gr ] = 2*log( lambda_tilde_t[ gr ] ) + 2*( log( tau_j[ gr ] ) + log(tau_0[ gr ]) );
				exp_h[ gr ] = exp( 0.5 * log_h[ gr ] );
			}
			// For time > 1
			for(tt in 2:NT){
				c_t[ tt ] = slab_scale_c_t * sqrt( aux_c_t[ tt ] );
				for(gr in 1:Ngr ){
					lambda_t[ (tt-1)*Ngr + gr ] = aux1_local_t[ tt ][((gr-1)*m + 1):((gr-1)*m + m)] .*
					sqrt( aux2_local_t[ tt ][((gr-1)*m + 1):((gr-1)*m + m)] );
					lambda_tilde_t[ (tt-1)*Ngr + gr ] = sqrt( ( c_t[ tt ][ gr ]^2 * 
					square( lambda_t[ (tt-1)*Ngr + gr ] ) ) ./
					( c_t[ tt ][ gr ]^2 + square( tau_j[ gr ] ) .* square(lambda_t[ (tt-1)*Ngr + gr ])));
					for(jj in 1:m){
						log_h[ (tt-1)*Ngr + gr ][jj] = 2*log( lambda_tilde_t[ (tt-1)*Ngr + gr ][ jj ] ) + 
						2*( log( tau_j[ gr ][jj]) + log( tau_0[ gr ] ) ) + 
						phi_j[ (gr-1)*m + jj ] * ( log_h[ (tt-2)*Ngr + gr ][jj] - 
						2*( log( tau_j[ gr ][jj]) + log( tau_0[ gr ] )));
						exp_h[ (tt-1)*Ngr + gr ][jj] = exp( 0.5 * log_h[ (tt-1)*Ngr + gr ][jj] );
					}
				}// gr
			}// tt
			// Beta_t
			for(gr in 1:Ngr){
				beta_t[ gr ] = z_t[ 1 ][((gr-1)*m + 1):((gr-1)*m + m)] .* exp_h[ gr ];
			}
			for(tt in 2:NT){
				for(gr in 1:Ngr){
					beta_t[ (tt-1)*Ngr + gr ] = beta_t[ (tt-2)*Ngr + gr ] + z_t[ tt ][((gr-1)*m + 1):((gr-1)*m + m)] .* 
					exp_h[ (tt-1)*Ngr + gr ];
				}
			}
			// Mean function
			{
				int ii_ind;
				ii_ind = 1;
				for(tt in 1:NT ){
					for(gr in 1:Ngr){
						for(ii in 1:n_groups[ (tt-1)*Ngr + gr ] ){
							sigma_long[ ii_ind ] = sigma_t[ (tt-1)*Ngr + gr ];
							f_t[ ii_ind ] = X_t[ ii_ind ] * beta_t[ (tt-1)*Ngr + gr ];
							ii_ind += 1;
						}
					}
				}
			}
		}
		model {
			// Here we use auxiliary variables for tau and lambda
			// Locals shared across groups
			sigma_t ~ inv_gamma( 0.5, 0.5 );
			phi_j_pos ~ beta( phi_ab[1], phi_ab[2] );
			// One global tau for each group
			aux1_global_0 ~ std_normal();
			aux2_global_0 ~ inv_gamma( 0.5 * df_global_0, 0.5 * df_global_0 );
			aux1_global_j ~ std_normal();
			aux2_global_j ~ inv_gamma( 0.5 * df_global_j, 0.5 * df_global_j );
			for(tt in 1:NT){
				aux2_local_t[ tt ] ~ inv_gamma( 0.5 * df_local_t, 0.5 * df_local_t );
				aux1_local_t[ tt ] ~ std_normal();
				aux_c_t[ tt ] ~ inv_gamma( 0.5 * slab_df_c_t, 0.5 * slab_df_c_t );
				z_t[ tt ] ~ std_normal();
			}
			// Likelihood
			y_t ~ normal(f_t , sigma_long );
		}
		generated quantities {
			// Vector of k_j, the pseudo inclusion probability
			vector [ m ] k_j_t[ NT * Ngr ];
			// Vector to hold precision Omega elements for the current equation
			vector [ m ] omega_vec_t[ NT * Ngr ];
			// Elements of Omega
			for(tt in 1:NT){
				for(gr in 1:Ngr ){
					for(jj in 1:m){
						k_j_t[ (tt-1)*Ngr + gr ][jj] = 1/( 1 + 1/square( sigma_t[ gr ] ) * (exp_h[ (tt-1)*Ngr + gr ][jj] * 
						exp_h[ (tt-1)*Ngr + gr ][jj]) * devs_X_t[ tt ][ jj, gr ] );
						omega_vec_t[ (tt-1)*Ngr + gr ][ jj ] = (- beta_t[ (tt-1)*Ngr + gr ][ jj ] ) / square( sigma_t[ gr ] );
					}
				}
			}
		}
	\end{verbatim}
}

\subsection*{Gibbs sampling algorithm}
Here we provide details of the Gibbs sampling. Our starting point is  the algorithm described in \cite{kowal2017dynamic}  and we extend  it to allow for multiple groups of different sample sizes. The sampler consists of two main components: a stochastic volatility sampling algorithm \citep{kastner2014ancillarity} augmented with a Polya-Gamma sampler \citep{polson2013bayesian}, and a Cholesky Factor Algorithm \citep{rue2001fast} to sample the regression coefficients in the dynamic linear model. We provide details of the update steps that differ from the original paper, for the others refer to \cite{kowal2017dynamic}.
\begin{itemize}
	\item \textit{Resampling the log-variances $h_{jrt}$ given the rest}. \cite{omori2007stochastic} propose a method to sample directly from the full-conditional distribution of $\boldsymbol{h_{jr}} = (h_{jr1}, \dots, h_{jrT})$,  on the log-scale, where the ensuing log-chi-square distribution of the likelihood $\log(\gamma_{jrt}^2)$, from
	$$
	\log\left( (\beta_{jrt} - \beta_{jrt-1})^2 \right) = \log(\gamma_{jrt}^2)  + h_{jrt}
	$$
	is approximated via a known normal mixture approximation of 10 components. To sample the new log-variances we use the all-without-a-loop (AWOL) sampler of \cite{kastner2014ancillarity}, which allows to sample $\boldsymbol{h_{jr}}$ jointly without the need of a sequential algorithm. Conditional on all the other parameters the log-variances are independent across groups. Resampling proceeds as in \cite{kowal2017dynamic}.
	\item \textit{Resampling the mixture components indicator $s_{jrt}$ given the rest}. The discrete mixture probabilities are straightforward to update. The prior mixture probabilities are the pre-specified mixing proportions given by \cite{omori2007stochastic} and the likelihood is $ \log(\gamma_{jrt}^2 + c) \sim \text{N}( h_{jrt} + m_{s_{jrt}}, v_{s_{jrt}} )$, where $c$ is a small offset to avoid numerical underflows and $m$ and $v$ are the pre-specified mean and variance, respectively, of the 10 mixture components. Resampling proceeds as in \cite{kowal2017dynamic}, independently for each group.
	\item \textit{Resampling the mean parameters $\mu_{jr}$ given the rest}. The update of $\mu_{jr}$ is done independently for each group. We can re-write the log-variance equation in (8) as an ordinary linear regression, where the parameters $\mu_{jr}$ play the role of regression coefficients, as follows:
	\begin{align*}
		h_{jrt} & = \mu_{jr} + \phi_{jr}(h_{jrt-1} - \mu_{jr} ) + \eta_{jrt}\\
		\tilde{ h }_{jrt} & = \mu_{jr} (1 - \phi_{jr}) + \eta_{jrt}
	\end{align*}
	where $ \tilde{ h }_{jrt} = h_{jrt} - \phi_{jr}h_{jrt-1}$ and $ \tilde{ h }_{jrt} \sim \text{N}\left(\mu_{jr} (1 - \phi_{jr}), \xi_{\eta_{jrt}}^{-1} \right) $. Finally, we can write 
	$$ \tilde{ \tilde{ h } }_{jrt} \sim \text{N}\left( \mu_{jr} z_{jrt}, 1 \right) $$
	where $z_{jrt} = (1 - \phi_{jr})\sqrt{\xi_{\eta_{jrt}}}$ are the regression covariates, $\mu_{jr}$ are the regression coefficient and $\tilde{ \tilde{ h } }_{jrt} = \tilde{ h }_{jrt}\sqrt{\xi_{\eta_{jrt}}}$, for $t=2,\cdots, T$, are the response variables. The posterior follows from the standard update of a Normal prior with Normal likelihood. For $j=1,\dots,p$ and $r=1,\dots,R$ we sample from
	\begin{align*}
		\mu_{jr} \mid rest & \sim \text{N} \left(l_{\mu}/q_{\mu}, 1/q_{\mu} \right)\\
		q_{\mu} &= \xi_{\mu_{jr}} + \xi_{\eta_{jr1}} + \sum_{t=2}^{T} z_{jrt}^2\\
		l_{\mu} &= \xi_{\mu_{jr}}\mu_{0} + h_{jr1} \xi_{\eta_{jr1}} + \sum_{t=2}^{T} \tilde{ \tilde{ h } }_{jrt} z_{jrt}
	\end{align*}
	\item \textit{Resampling the mean parameter $\mu_{0}$ given the rest}. The posterior distribution of $\mu_{0}$ follows from the standard update of a Normal prior and Normal likelihood represented by $\mu_{jr}$. We sample from
	\begin{align*}
		\mu_{0} \mid rest & \sim \text{N}(l_0/q_0, 1/q_0)\\
		q_0 & = \xi_{\mu_{0}} + \sum_{r=1}^{R} \sum_{j=1}^{p} \xi_{\mu_{jr}} \\
		l_0 & = \sum_{r=1}^{R} \sum_{j=1}^{p} \xi_{\mu_{jr}} \mu_{jr}
	\end{align*}
	\item \textit{Resampling the autoregressive coefficients $\phi_{jr}$ given the rest}. The new value of $\phi_{jr}$ is drawn via slice sampler, independently for each group. The parametrisation $(\phi_{jr} +1)/2 \sim \text{Beta}(a_{\phi}, b_{\phi})$ implies that $|\phi_{jr}| < 1$, which ensures a stationary stochastic  process for $\boldsymbol{h_{jr}}$. Resampling is performed as in \cite{kowal2017dynamic}.
	\item \textit{Resampling the Polya-Gamma mixing parameters $\xi_{\eta_{jrt}}$, $\xi_{\mu_{jr}}$ and $\xi_{\mu_{0}}$ given the rest}. This step is a conjugate update of a Polya-Gamma prior given a Gaussian likelihood \citep{polson2013bayesian}.
	\begin{align*}
		\xi_{\eta_{jrt}} \mid rest & \sim \text{Polya-Gamma}(1, \eta_{jrt} ), \qquad \forall j,r,t\\
		\xi_{\mu_{jr}} \mid rest & \sim \text{Polya-Gamma}(1, \mu_{jr} - \mu_{0r} ), \qquad \forall j,r\\
		\xi_{\mu_{0}} \mid rest & \sim \text{Polya-Gamma}\left( 1, \mu_{0} \right)
	\end{align*}
	\item \textit{Resampling the regression coefficients $\boldsymbol{\beta}_{rt}$ given the rest}. Conditional on the rest, the regression coefficients of each group are independent. We perform a joint update of $\boldsymbol{\beta} = (\boldsymbol{\beta}_{1}^T,\dots, \boldsymbol{\beta}_{t}^T, \dots, \boldsymbol{\beta}_{T}^T )$, where $\boldsymbol{\beta}_{t} = (\beta_{1t}, \dots, \beta_{pt} )$, exploiting the block-diagonal structure of the posterior precision matrix of $\boldsymbol{\beta}$. The posterior distribution is (omitting the group subscript for ease of notation)
	\begin{align*}
		\boldsymbol{\beta} \mid rest &\sim \text{N}\left( Q_{\beta}^{-1} \boldsymbol{m}_{\beta}, Q_{\beta}^{-1} \right)\\
		Q_{\beta} &= A_{\sigma} + A_{h}
	\end{align*}
	where $A_{\sigma}$ is a $Tp \times Tp$ block-diagonal matrix defined as follows:
	$$
	A_{\sigma} = 
	\begin{bmatrix}%
		X_{1}^T X_{1} / \sigma^2_1 & 0 & \cdots & \cdots & 0 \\
		\vdots & \ddots & \vdots & \vdots & \vdots \\
		0 & \cdots & X_{t}^T X_{t} / \sigma^2_t & \cdots & 0 \\
		\vdots & \vdots & \vdots & \ddots & \vdots \\
		0 & 0 & \cdots & \cdots & X_{T}^T X_{T} / \sigma^2_T
	\end{bmatrix}
	$$
	and $A_{h}$ is a $Tp \times Tp$ matrix defined as follows:
	$$
	A_{h} = \left(D^{T} \otimes I_p\right) \Sigma_h^{-1} \left(D \otimes I_p\right)
	$$
	where $\otimes$ denotes the Kronecker product, $D$ is a $T \times T$ tri-diagonal matrix with diagonal entries equal to 1 and first off-diagonal elements equal to $-1$, $I_p$ is the $p \times p$ identity matrix and $\Sigma_h^{-1}$ is a $Tp \times Tp$ diagonal matrix with diagonal entries equal to 
	$$(\exp(h_{11}/2), \dots, \exp(h_{p1}/2), \dots, \exp(h_{1T}/2), \dots, \exp(h_{pT}/2) )
	$$
	The posterior mean $\boldsymbol{m}_{\beta}$ is a $Tp$-dimensional vector defined as:
	$$
	\boldsymbol{m}_{\beta} = \left( X_1^T \boldsymbol{y}_1 / \sigma^2_1, \dots, X_t^T \boldsymbol{y}_t / \sigma^2_t, \dots, X_T^T \boldsymbol{y}_T / \sigma^2_T \right)
	$$
	\item \textit{Resampling the observation error variances $\sigma^2_{rt}$ given the rest}. This step is a conjugate update of an Inverse-Gamma prior given a Gaussian likelihood. For each group $r$ and time $t$ we sample:
	\begin{align*}
		\sigma^2_{rt} \mid rest \sim \text{Inverse-Gamma} \left( a_{\sigma} + \dfrac{n_{rt}}{2}, b_{\sigma} + \dfrac{\sum_{i=1}^{n_{rt}} (y_{irt} - \boldsymbol{x}_{irt} \boldsymbol{\beta}_{rt} )^2}{2}  \right)
	\end{align*}
\end{itemize}

\begin{table}
	\centering
	\begin{tabular}{cc|cc}
		\multicolumn{2}{c}{Setting} & HMC & Gibbs \\
		\hline
		\multirow{3}{*}{$T = 3$} & $M = 20$ & 0.26 & 0.28 \\
		& $M = 50$ & 1.18 & 1.11 \\
		& $M = 100$ & 2.36 & 4.00 \\\hline
		\multirow{3}{*}{$T = 10$} & $M = 20$ & 1.68 & 0.34 \\
		& $M = 50$ & 9.75 & 1.60 \\
		& $M = 100$ & 75.10 & 7.27 \\\hline	 
		&&& 
	\end{tabular}
	\caption{Computational times (seconds per iteration) for the HMC algorithm as implemented in {\tt Stan} and the blocked Gibbs sampling MCMC scheme as implemented in {\tt R}. Simulations are run for  different values of the number $M$ of nodes and number $T$ of time points, with $N = 100$ and $R = 2$. Computational times are recorded on a Linux machine	with Intel Core i7 1.8 GHZ.}
	\label{tab:ComputTimes}
\end{table}

\clearpage

\section{Additional Figures and Tables}\label{SMsec:additional-figures-and-tables}

\begin{figure}[!p]
	\centering
	\begin{minipage}{.5\textwidth}
		\includegraphics[width=1\linewidth]{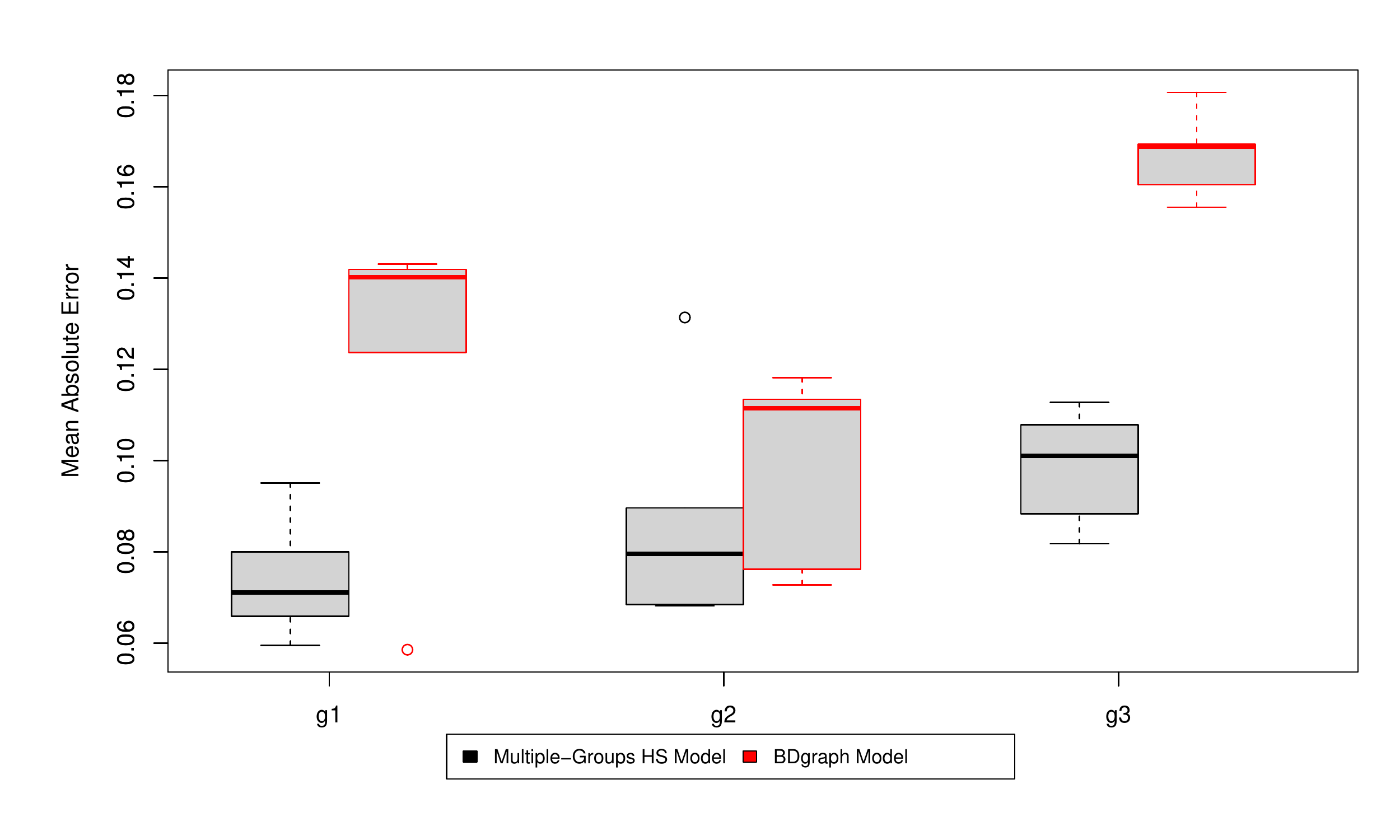}
	\end{minipage}%
	\begin{minipage}{.5\textwidth}
		\includegraphics[width=1\linewidth]{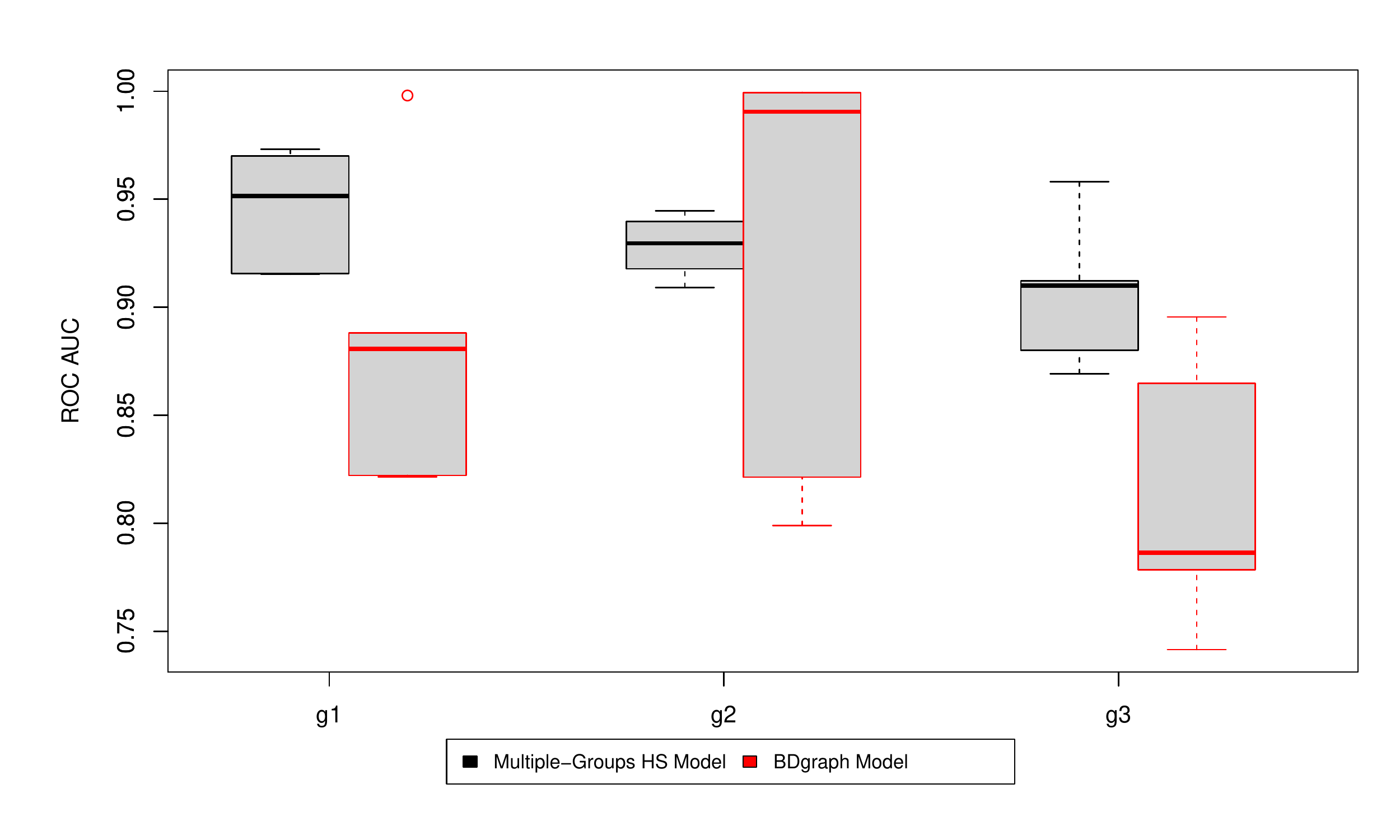}
	\end{minipage}
	\caption{First simulation scenario. Mean Absolute Error (left panel) and AUC (right panel) for the comparison between the multiple groups nodewise regression model in (7) and the package {\tt BDgraph}. The boxplots of the MAE and AUC are obtained over twenty replicates for each model and by adopting the \textit{AND} rule. The MAE and the AUC are compared within each of the three groups $G_1$, $G_2$ and $G_3$. The nodewise model works better both in terms of MAE, indicating a more precise estimation of $\Omega$, and in terms of AUC, denoting a better recovery of the graph structure within each group.}
	\label{fig:MAE_m20_HS_BD}
\end{figure}

\begin{figure}[!p]
	\centering
	\begin{minipage}{.5\textwidth}
		\includegraphics[width=1\linewidth]{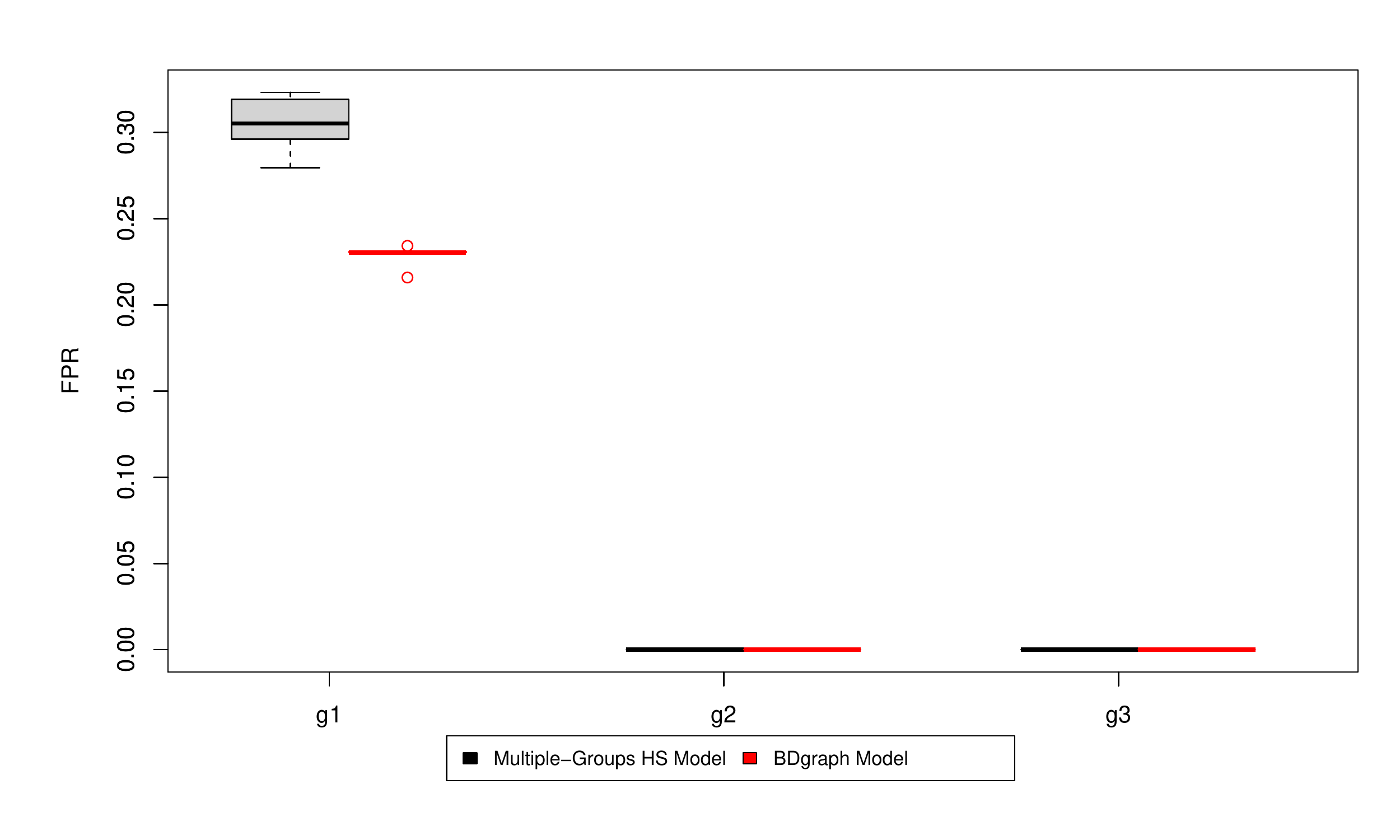}
	\end{minipage}%
	\begin{minipage}{.5\textwidth}
		\includegraphics[width=1\linewidth]{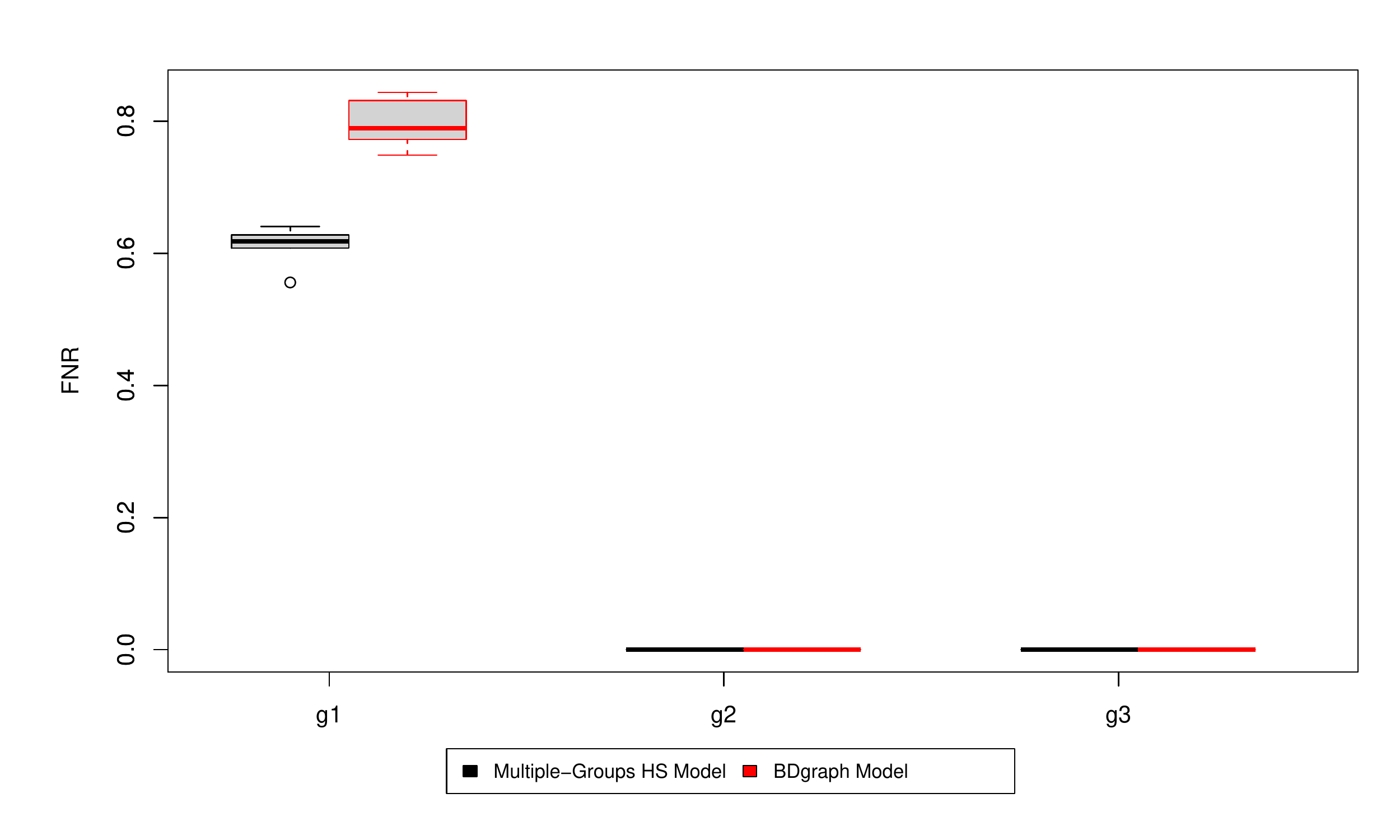}
	\end{minipage}
	\caption{First simulation scenario. False positive rates (FPR, left panel) and false negative rates (FNR, right panel) for the comparison between the multiple groups nodewise regression model in (7) and the package {\tt BDgraph}. The boxplots of the FPR and FNR are obtained over twenty replicates for each model and by adopting the \textit{AND} rule. The FPR and the FNR are compared within each of the three groups $G_1$, $G_2$ and $G_3$. The FPR and the FNR are comparable in both models, and in particular equal to zero for groups $G_2$ and $G_3$.}
	\label{fig:FPNR_m20_HS_BD}
\end{figure}

\begin{figure}[!p]
	\centering
	\includegraphics[width=1\linewidth]{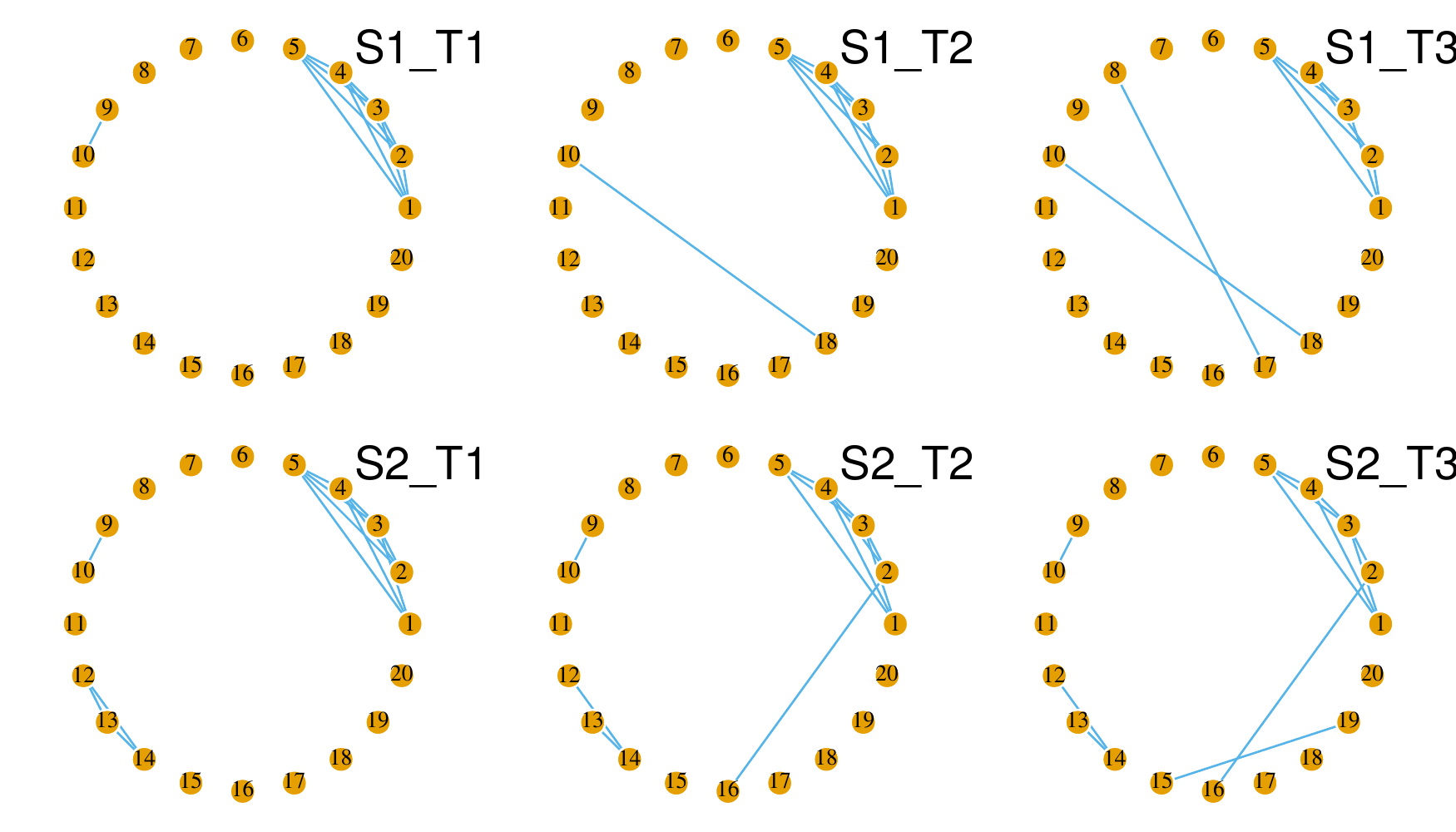}
	\caption{Networks generated in the second simulation scenario and used to simulate the dataset within each group and evolving over time.}
	\label{fig:graphs_sim_NT3}
\end{figure}

\begin{figure}[!p]
	\centering
	\includegraphics[width=0.9\linewidth]{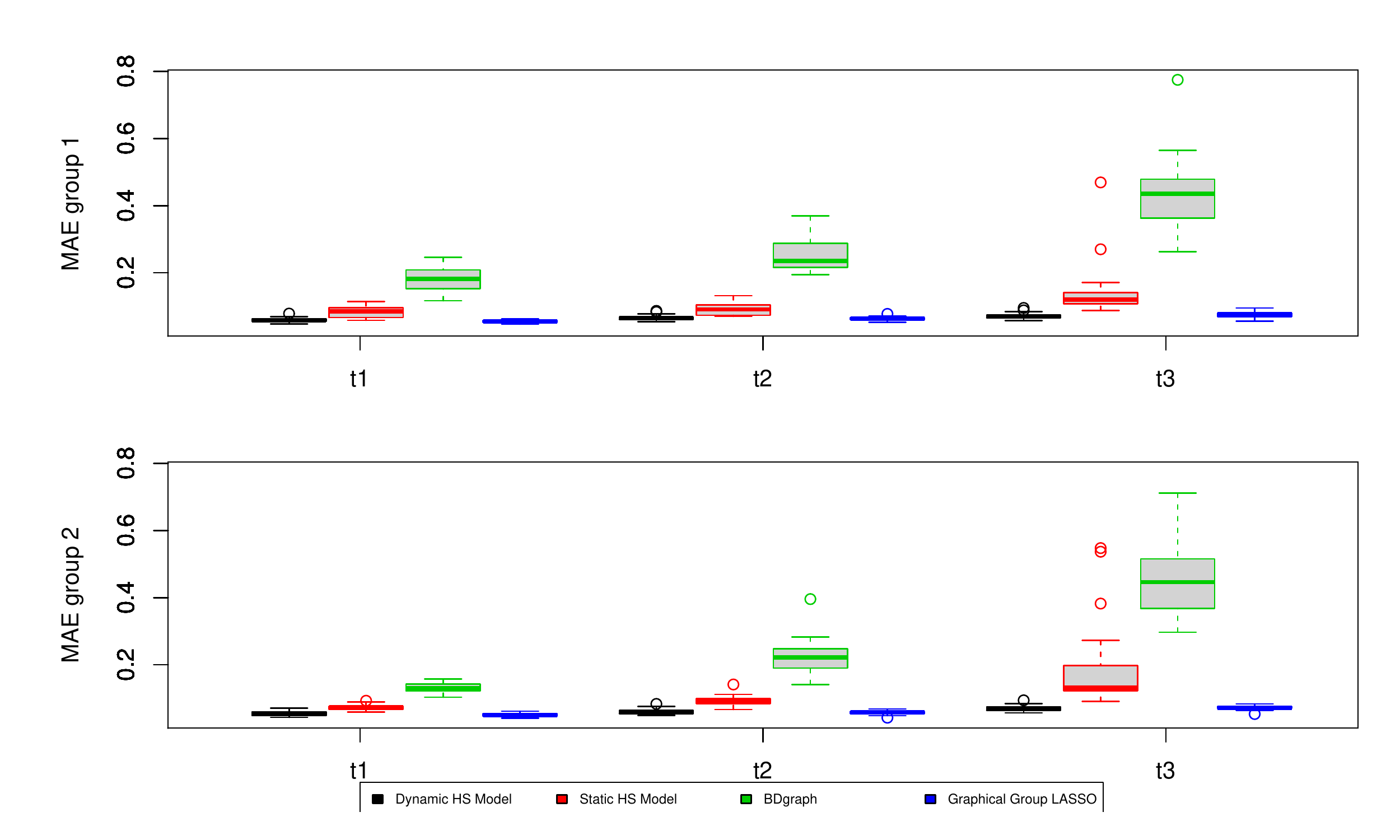}
	\includegraphics[width=0.9\linewidth]{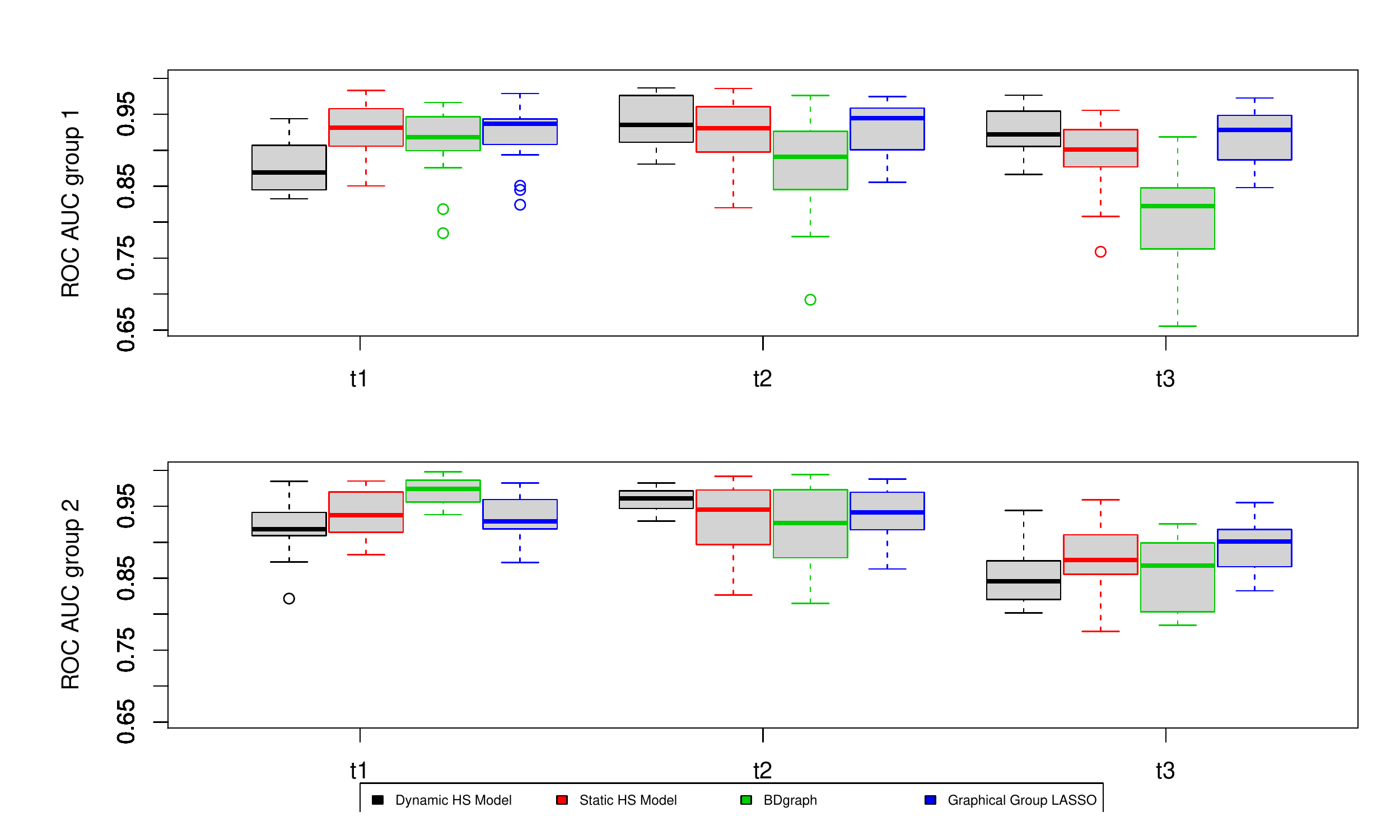}
	\caption{Second simulation scenario. Mean Absolute Error (top panel) and AUC (bottom panel) comparison between the dynamic model in (9) (black), the static model in (7) (red), {\tt BDgraph} (green) and  Graphical Group LASSO (blue). The boxplots of the MAE and AUC are obtained over twenty replicates for each model and by adopting the \textit{AND} rule. Each row in the Figure refers to a group, and the MAE and the AUC are compared at each time point. The MAE indicates that the proposed model performs well in the estimation of the precision matrix. In terms of AUC and graph recovery, the dynamic HS model yields better estimates in two of the time points for the first group, and in the second time point for the second group. In this simulation settings, no method consistently outperforms the others.}
	\label{fig:MAE_m20_NT3}
\end{figure}

\begin{figure}[!p]
	\centering
	\includegraphics[width=0.9\linewidth]{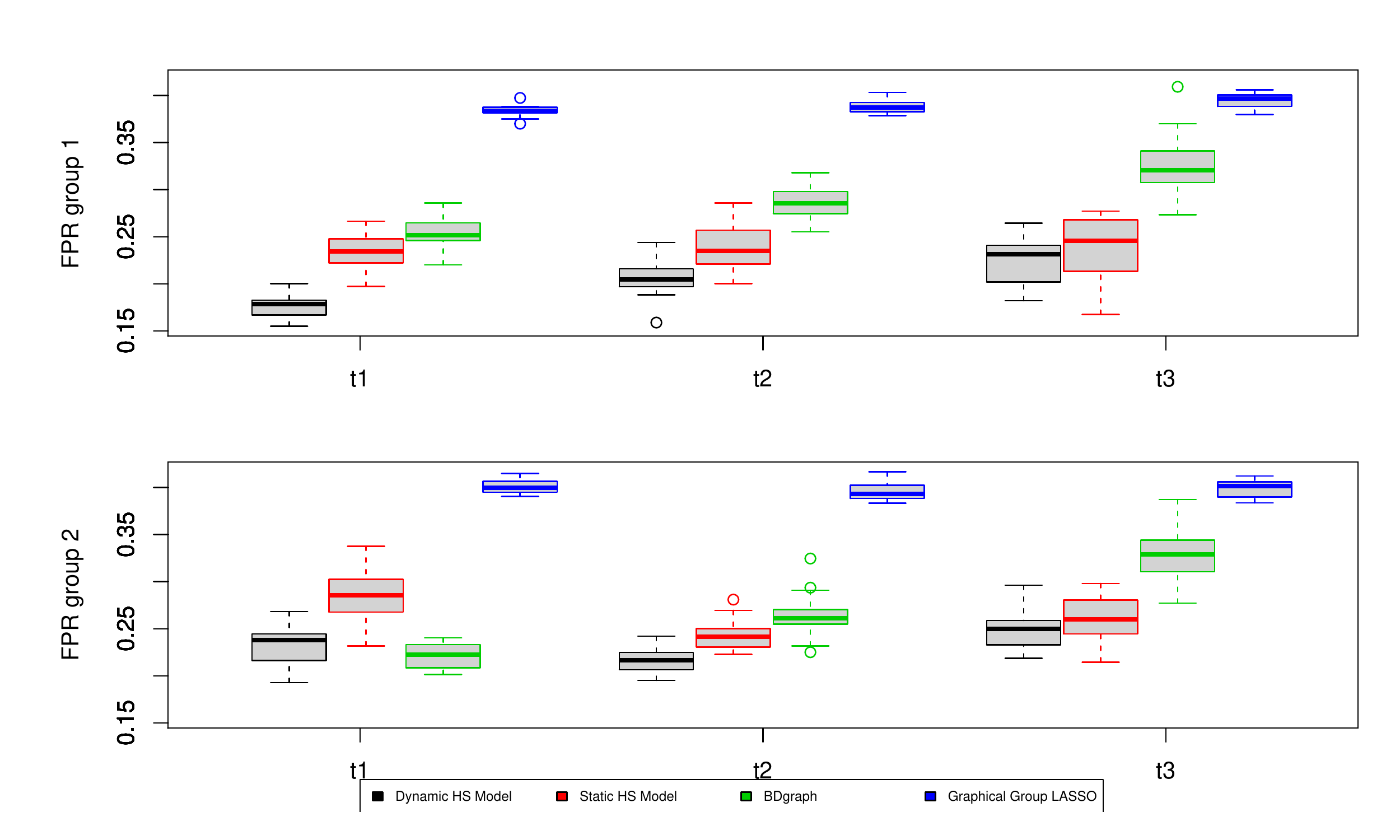}
	\includegraphics[width=0.9\linewidth]{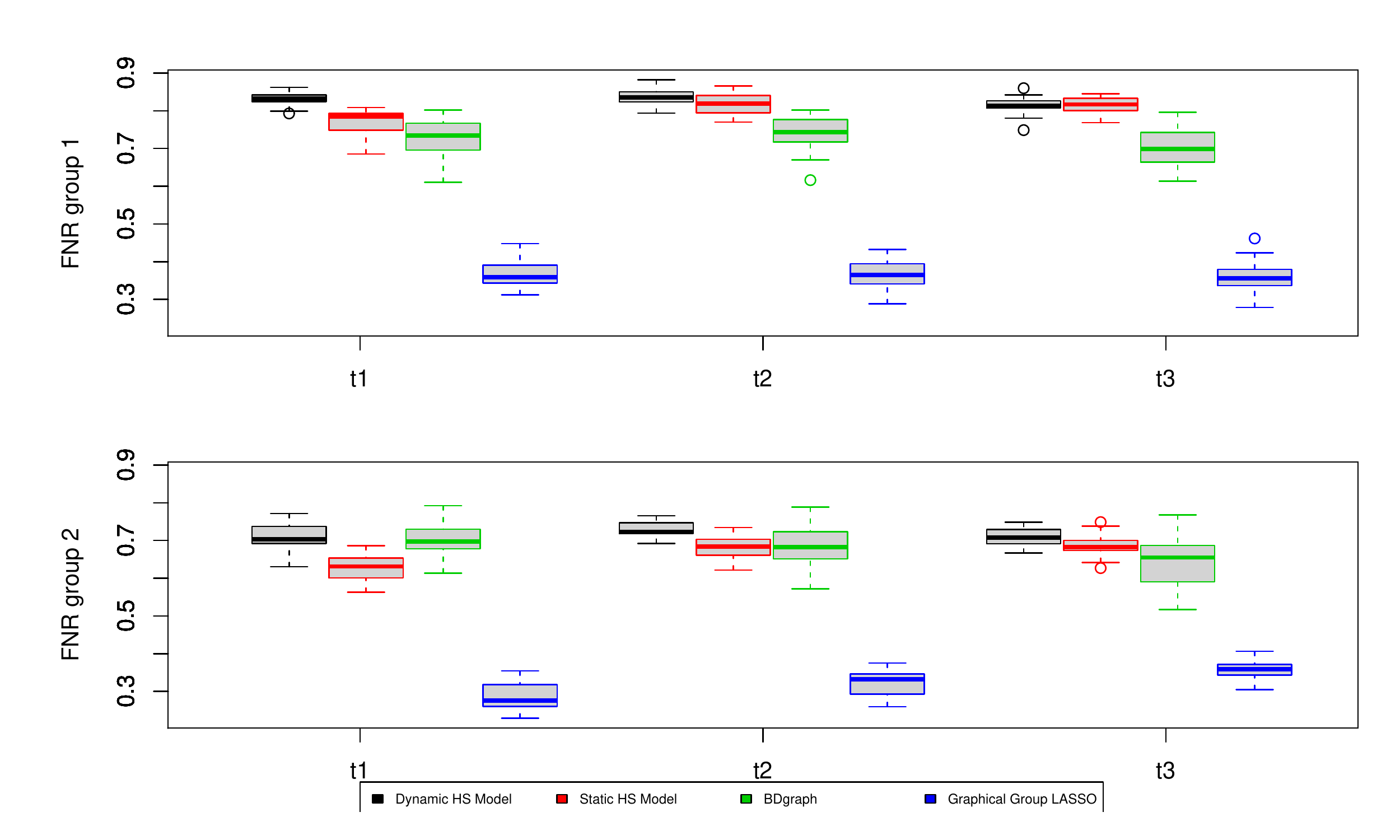}
	\caption{Second simulation scenario. False positive rates (FPR) and false negative rates (FNR) for the dynamic model in (9) (black), the static model in (7) (red), {\tt BDgraph} (green) and  Graphical Group LASSO (blue). The boxplots are obtained over twenty replicates for each model and by adopting the \textit{AND} rule. Each row in the Figure refers to a group, and the FPR and FNR are compared at each time point. The results obtained with the proposed dynamic and static models are comparable at most of time points and in both groups. Similar results are observed for the BDgraph model.  Graphical Group LASSO presents differs the most from the other models.}
	\label{fig:FPNR_m20_NT3}
\end{figure}

\begin{figure}[h!]
	\centering
	\includegraphics[width=1\linewidth]{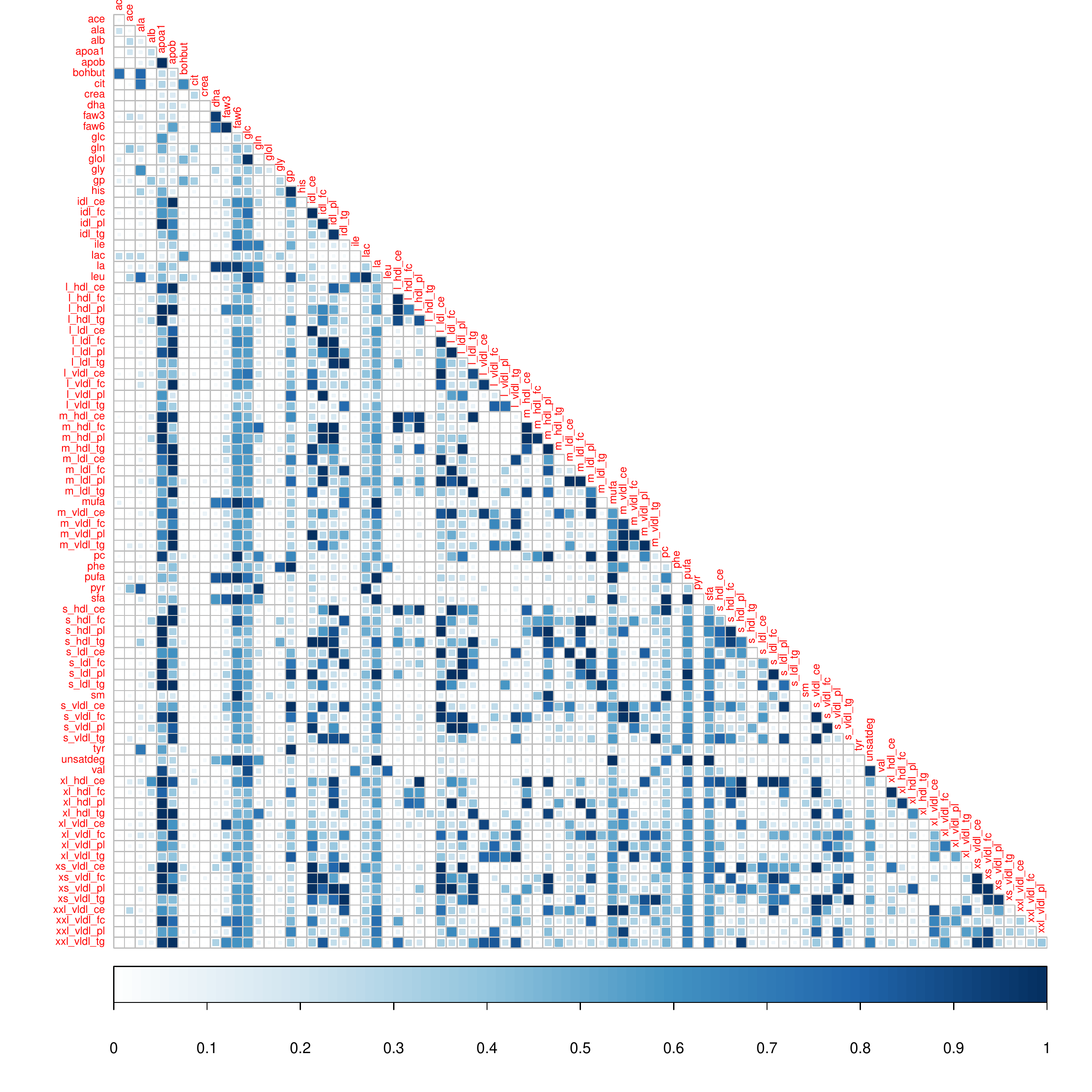}
	\caption{Posterior mean of the pseudo inclusion probability parameters $\kappa_{jl}$ (see Eq.~(5)) for the European ethnicity at  baseline. The edges in the individual networks are selected when the posterior mean of $(1 - \kappa_{jl})$ is greater than 0.5.}
	\label{fig:kappa_EU_t1}
\end{figure}
\begin{figure}[h!]
	\centering
	\includegraphics[width=1\linewidth]{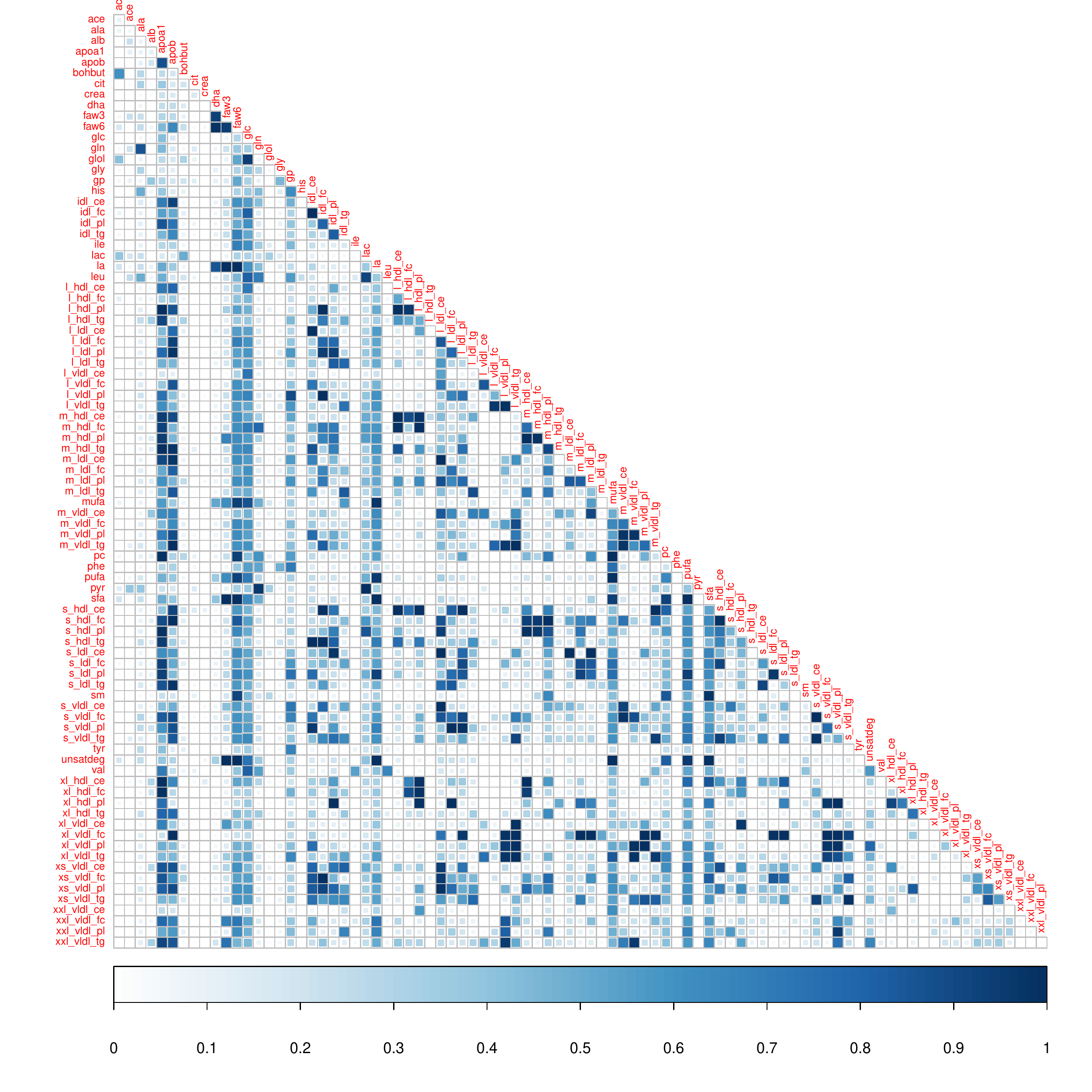}
	\caption{Posterior mean of the pseudo inclusion probability parameters $\kappa_{jl}$ (see Eq.~(5)) for the European ethnicity at  follow-up. The edges in the individual networks are selected when the posterior mean of $(1 - \kappa_{jl})$ is greater than 0.5.}
	\label{fig:kappa_EU_t2}
\end{figure}
\begin{figure}[h!]
	\centering
	\includegraphics[width=1\linewidth]{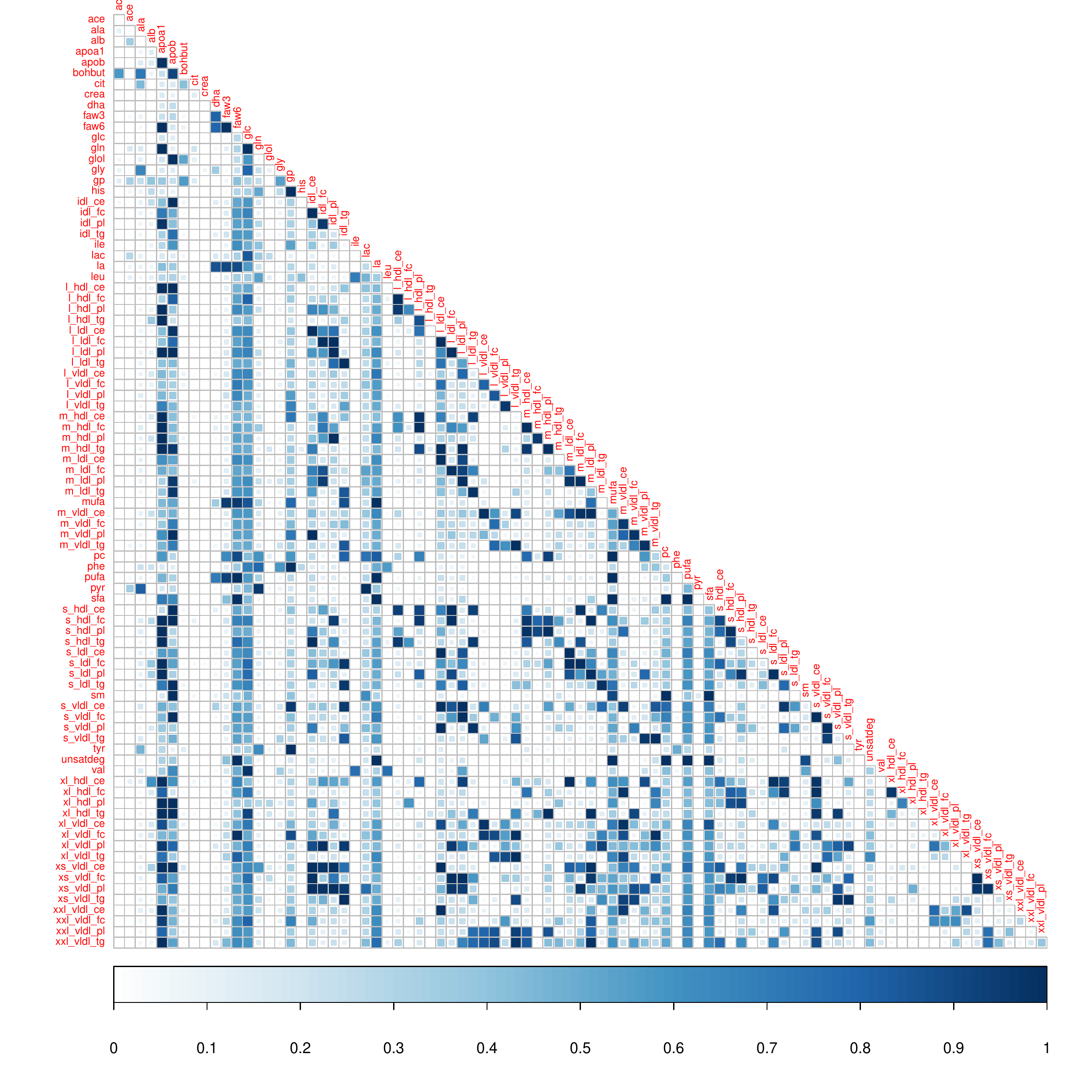}
	\caption{Posterior mean of the pseudo inclusion probability parameters $\kappa_{jl}$ (see Eq.~(5)) for the South-Asian ethnicity at  baseline. The edges in the individual networks are selected when the posterior mean of $(1 - \kappa_{jl})$ is greater than 0.5.}
	\label{fig:kappa_AS_t1}
\end{figure}
\begin{figure}[h!]
	\centering
	\includegraphics[width=1\linewidth]{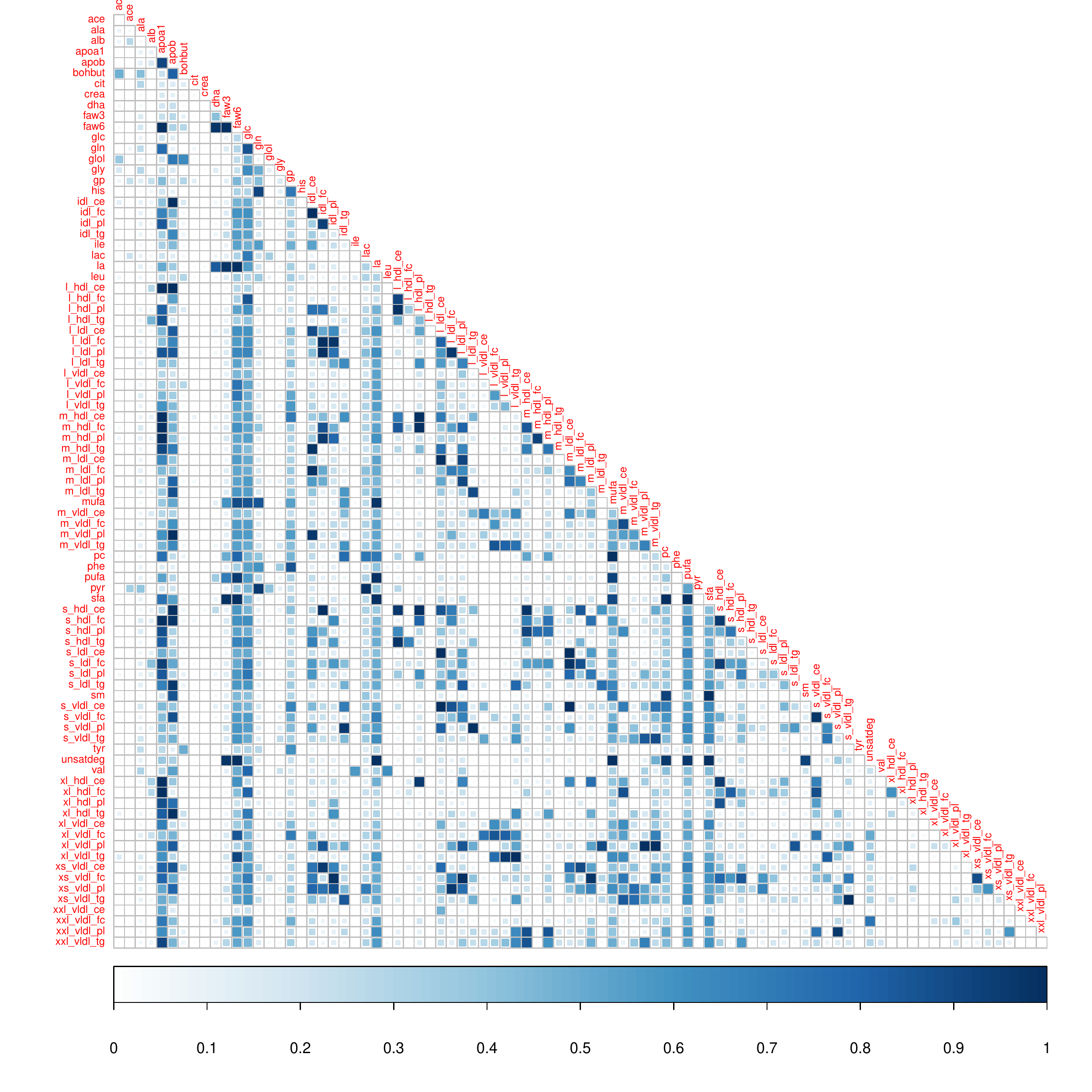}
	\caption{Posterior mean of the pseudo inclusion probability parameters $\kappa_{jl}$ (see Eq.~(5)) for the South-Asian ethnicity at  follow-up. The edges in the individual networks are selected when the posterior mean of $(1 - \kappa_{jl})$ is greater than 0.5.}
	\label{fig:kappa_AS_t2}
\end{figure}

\begin{figure}[h!]
	\centering
	\includegraphics[width=1\linewidth]{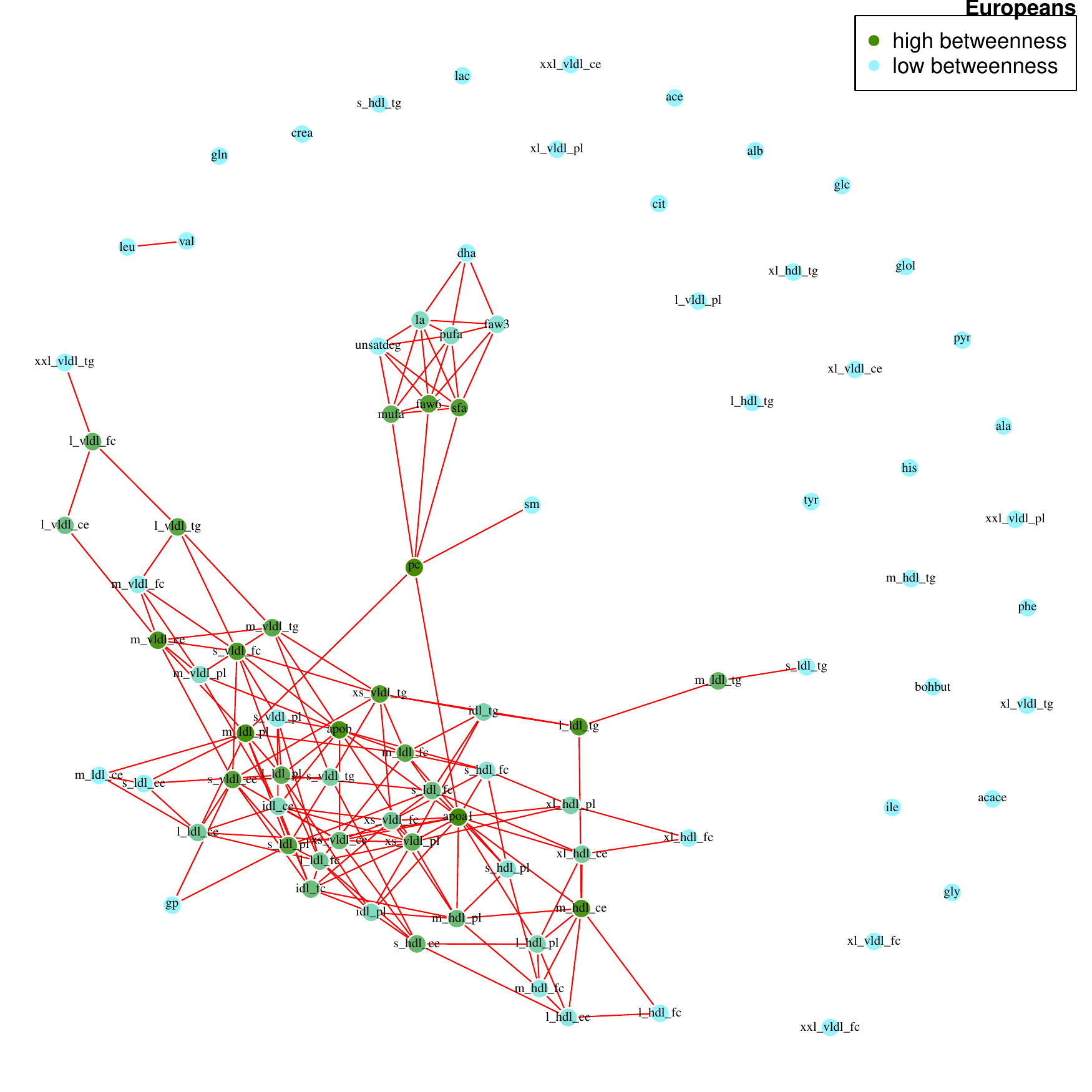}
	\caption{Individual network for the European ethnicity at  baseline obtained applying the \textit{AND} rule. An edge between two nodes is included in the graph if its posterior probability of inclusion is higher than 0.8. The colours of the nodes represent high/low levels of betweenness, indicating the degree of connectivity of the nodes within the graph.}
	\label{fig:network_EU_T1_AND}
\end{figure}
\begin{figure}[h!]
	\centering
	\includegraphics[width=1\linewidth]{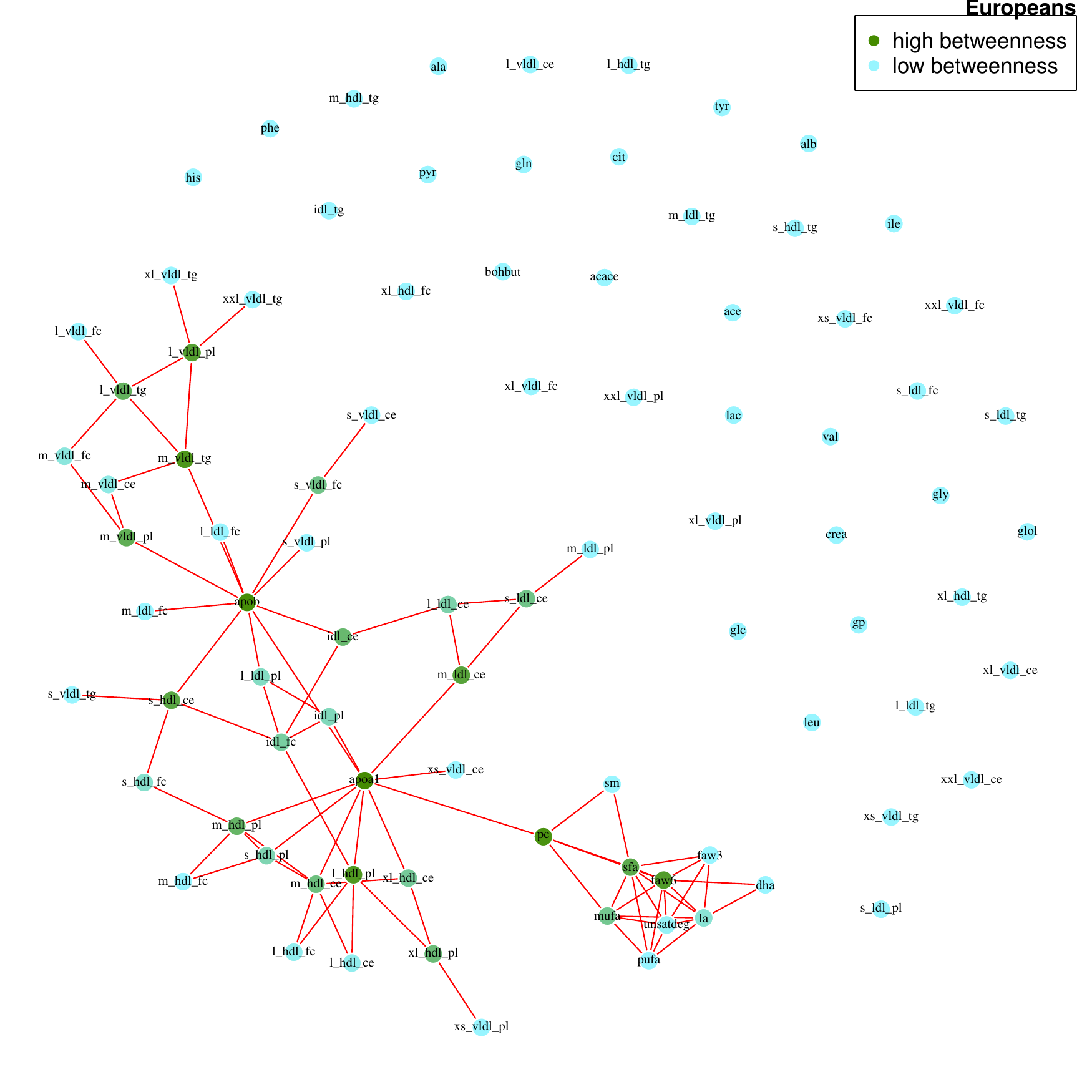}
	\caption{Individual network for the European ethnicity at follow-up obtained applying the \textit{AND} rule. An edge between two nodes is included in the graph if its posterior probability of inclusion is higher than 0.8. The colours of the nodes represent high/low levels of betweenness, indicating the degree of connectivity of the nodes within the graph.}
	\label{fig:network_EU_T2_AND}
\end{figure}

\begin{figure}[h!]
	\centering
	\includegraphics[width=1\linewidth]{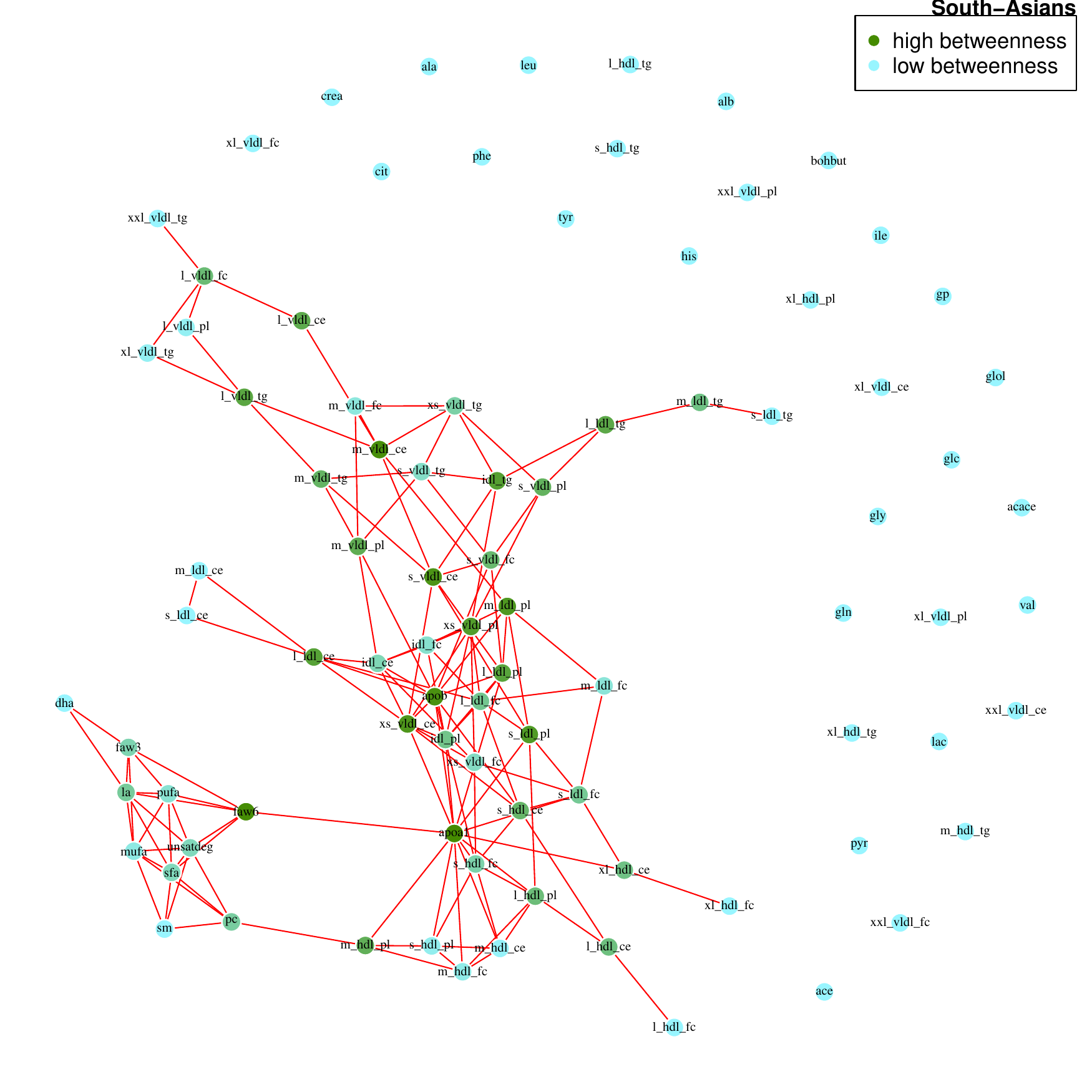}
	\caption{Individual network for the South-Asian ethnicity at  baseline obtained applying the \textit{AND} rule. An edge between two nodes is included in the graph if its posterior probability of inclusion is higher than 0.8. The colours of the nodes represent high/low levels of betweenness, indicating the degree of connectivity of the nodes within the graph.}
	\label{fig:network_AS_T1_AND}
\end{figure}
\begin{figure}[h!]
	\centering
	\includegraphics[width=1\linewidth]{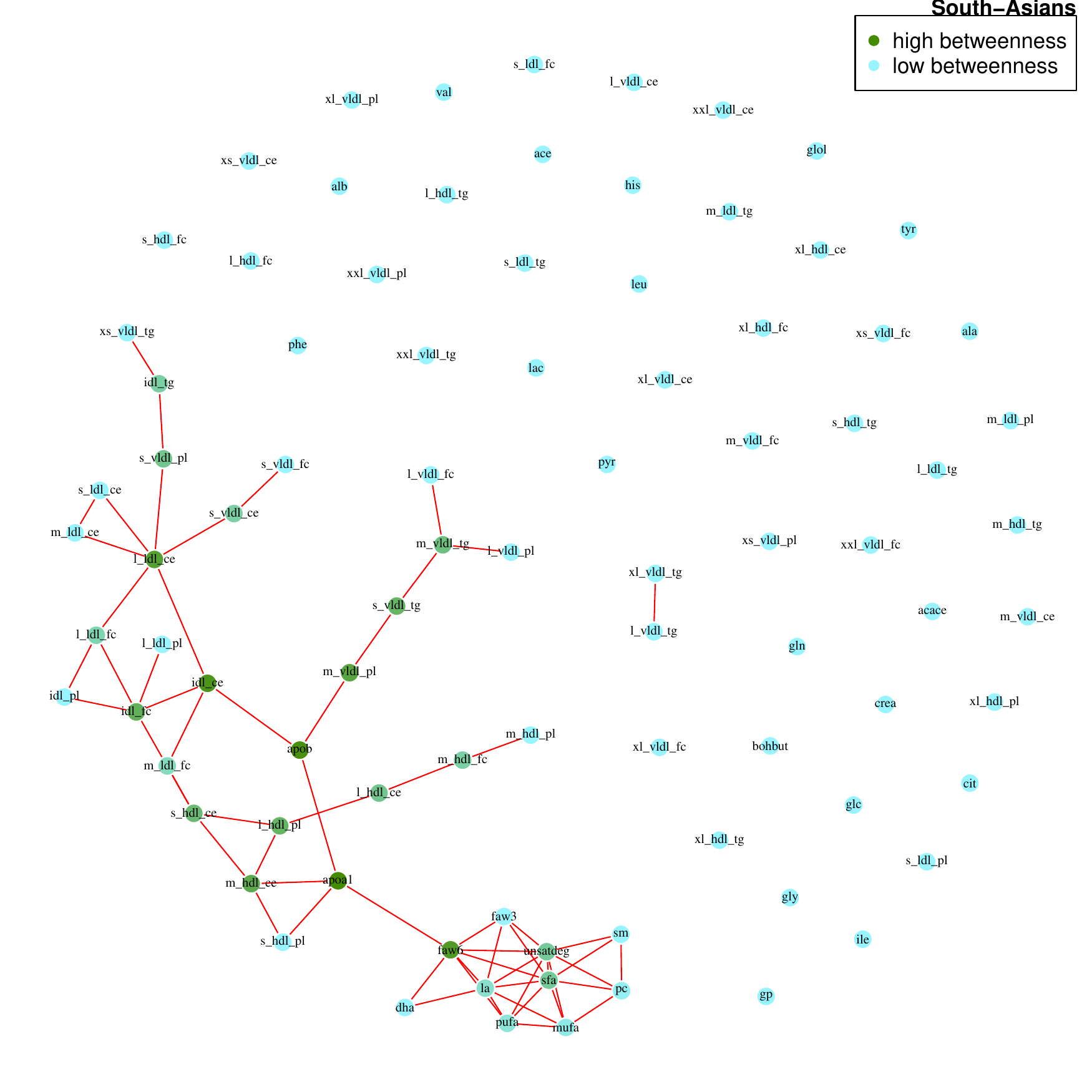}
	\caption{Individual network for the South-Asian ethnicity at follow-up obtained applying the \textit{AND} rule. An edge between two nodes is included in the graph if its posterior probability of inclusion is higher than 0.8. The colours of the nodes represent high/low levels of betweenness, indicating the degree of connectivity of the nodes within the graph.}
	\label{fig:network_AS_T2_AND}
\end{figure}

\begin{figure}[h!]
	\centering
	\includegraphics[width=1\linewidth]{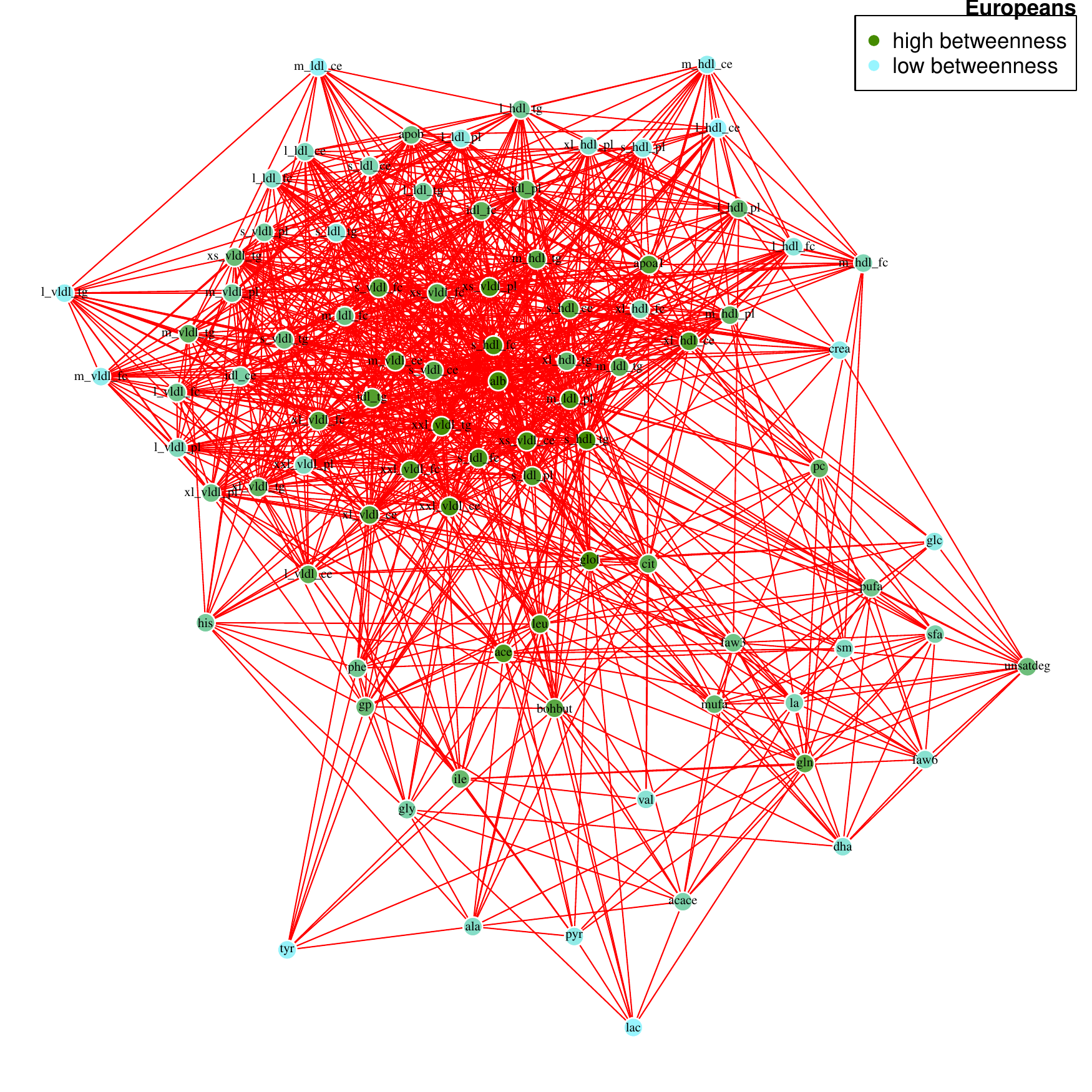}
	\caption{Individual network for the European ethnicity at  baseline obtained applying the \textit{OR} rule. An edge between two nodes is included in the graph if its posterior probability of inclusion is higher than 0.8. The colours of the nodes represent high/low levels of betweenness, indicating the degree of connectivity of the nodes within the graph.}
	\label{fig:network_EU_T1_OR}
\end{figure}
\begin{figure}[h!]
	\centering
	\includegraphics[width=1\linewidth]{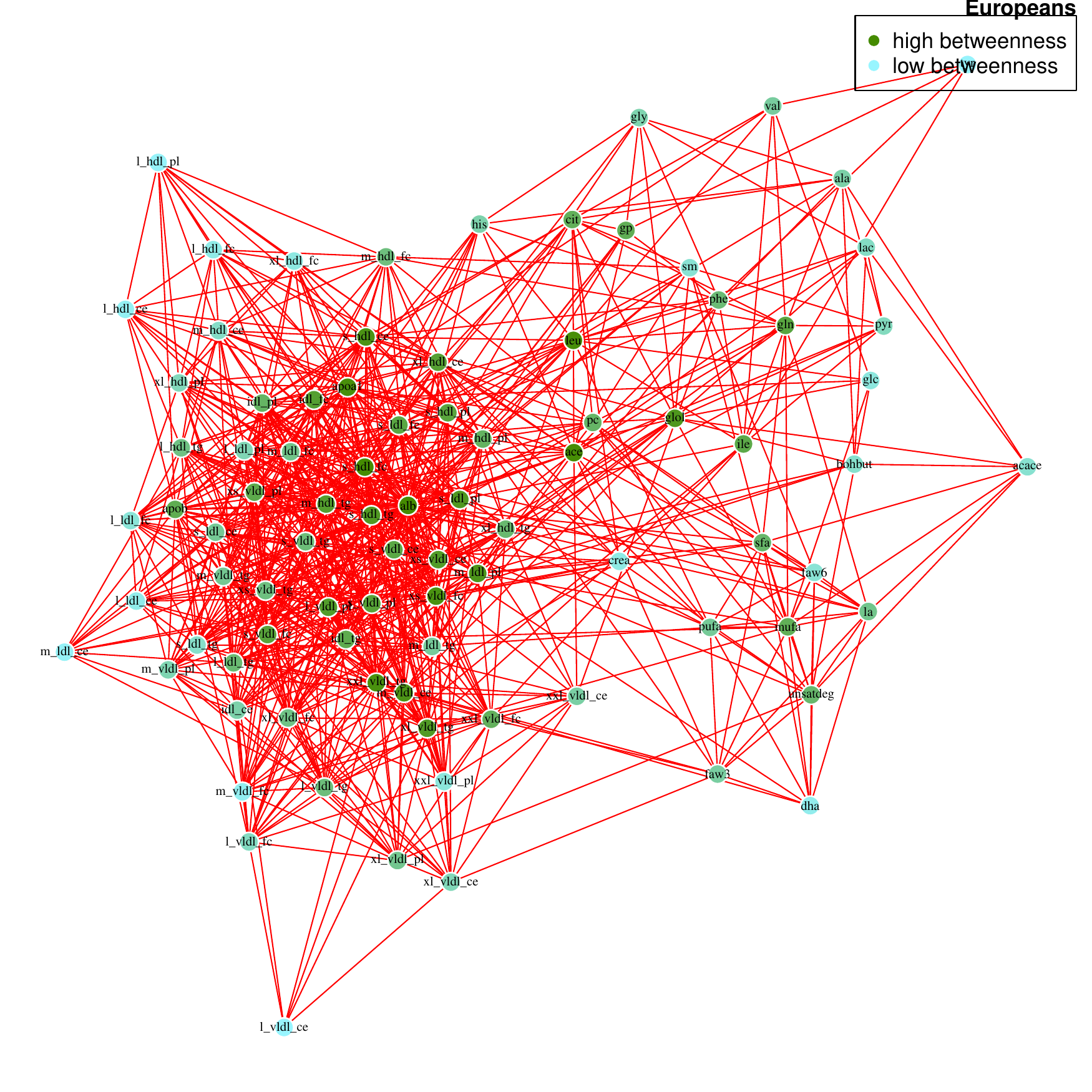}
	\caption{Individual network for the European ethnicity at follow-up  obtained applying the \textit{OR} rule. An edge between two nodes is included in the graph if its posterior probability of inclusion is higher than 0.8. The colours of the nodes represent high/low levels of betweenness, indicating the degree of connectivity of the nodes within the graph.}
	\label{fig:network_EU_T2_OR}
\end{figure}

\begin{figure}[h!]
	\centering
	\includegraphics[width=1\linewidth]{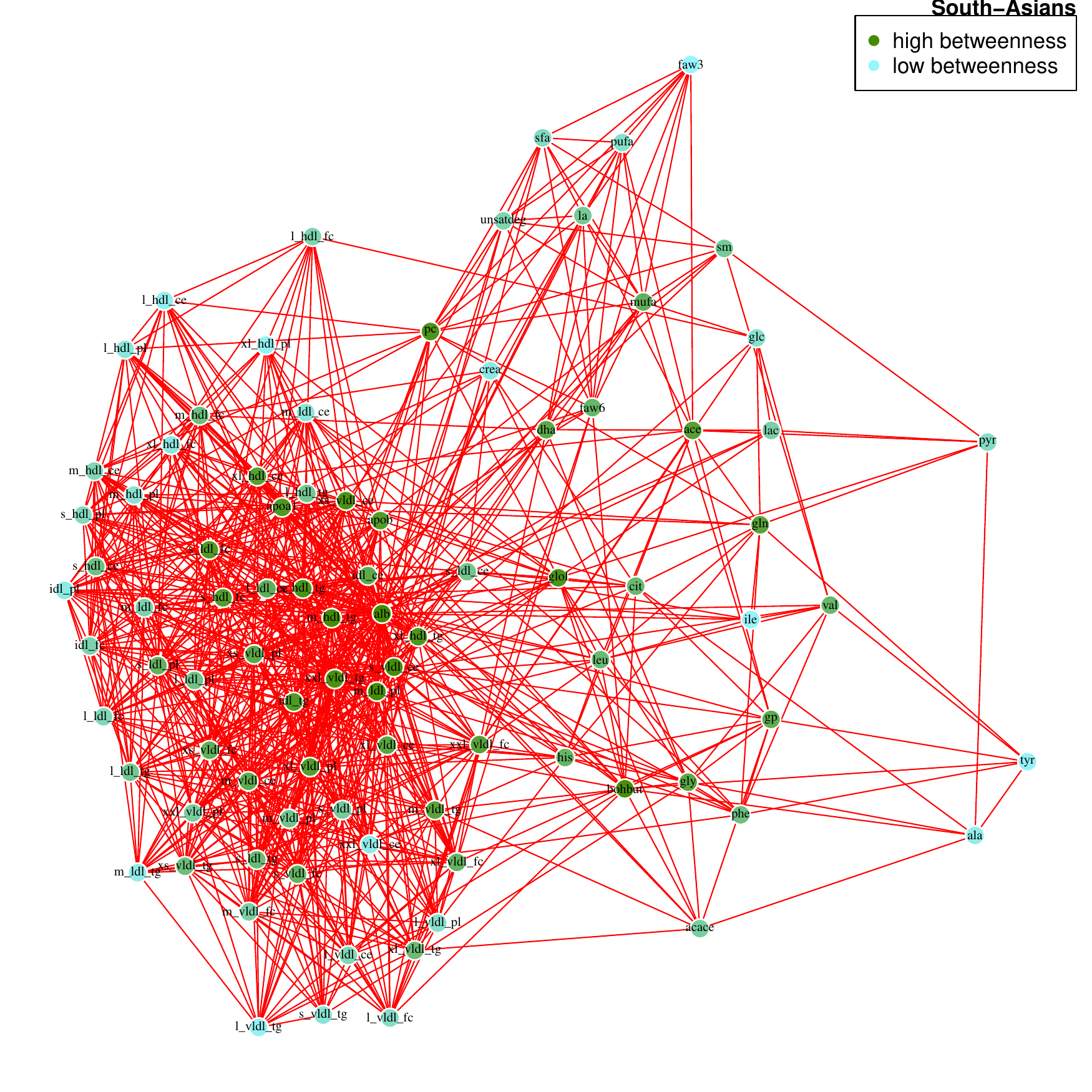}
	\caption{Individual network for the South-Asian ethnicity at  baseline  obtained applying the \textit{OR} rule. An edge between two nodes is included in the graph if its posterior probability of inclusion is higher than 0.8. The colours of the nodes represent high/low levels of betweenness, indicating the degree of connectivity of the nodes within the graph.}
	\label{fig:network_AS_T1_OR}
\end{figure}
\begin{figure}[h!]
	\centering
	\includegraphics[width=1\linewidth]{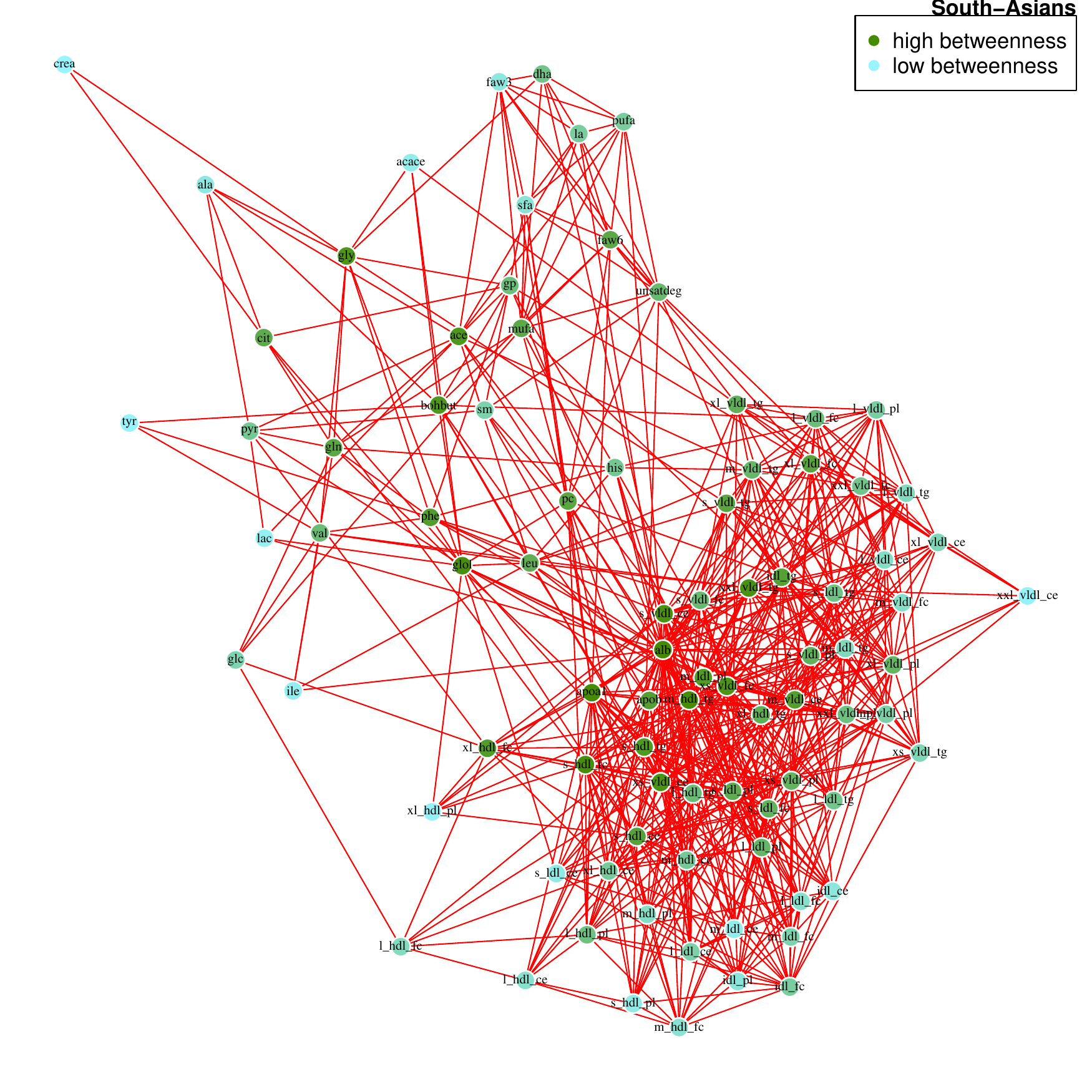}
	\caption{Individual network for the South-Asian ethnicity at follow-up  obtained applying the \textit{OR} rule. An edge between two nodes is included in the graph if its posterior probability of inclusion is higher than 0.8. The colours of the nodes represent high/low levels of betweenness, indicating the degree of connectivity of the nodes within the graph.}
	\label{fig:network_AS_T2_OR}
\end{figure}

\begin{figure}
	\centering
	\includegraphics[width=0.9\linewidth]{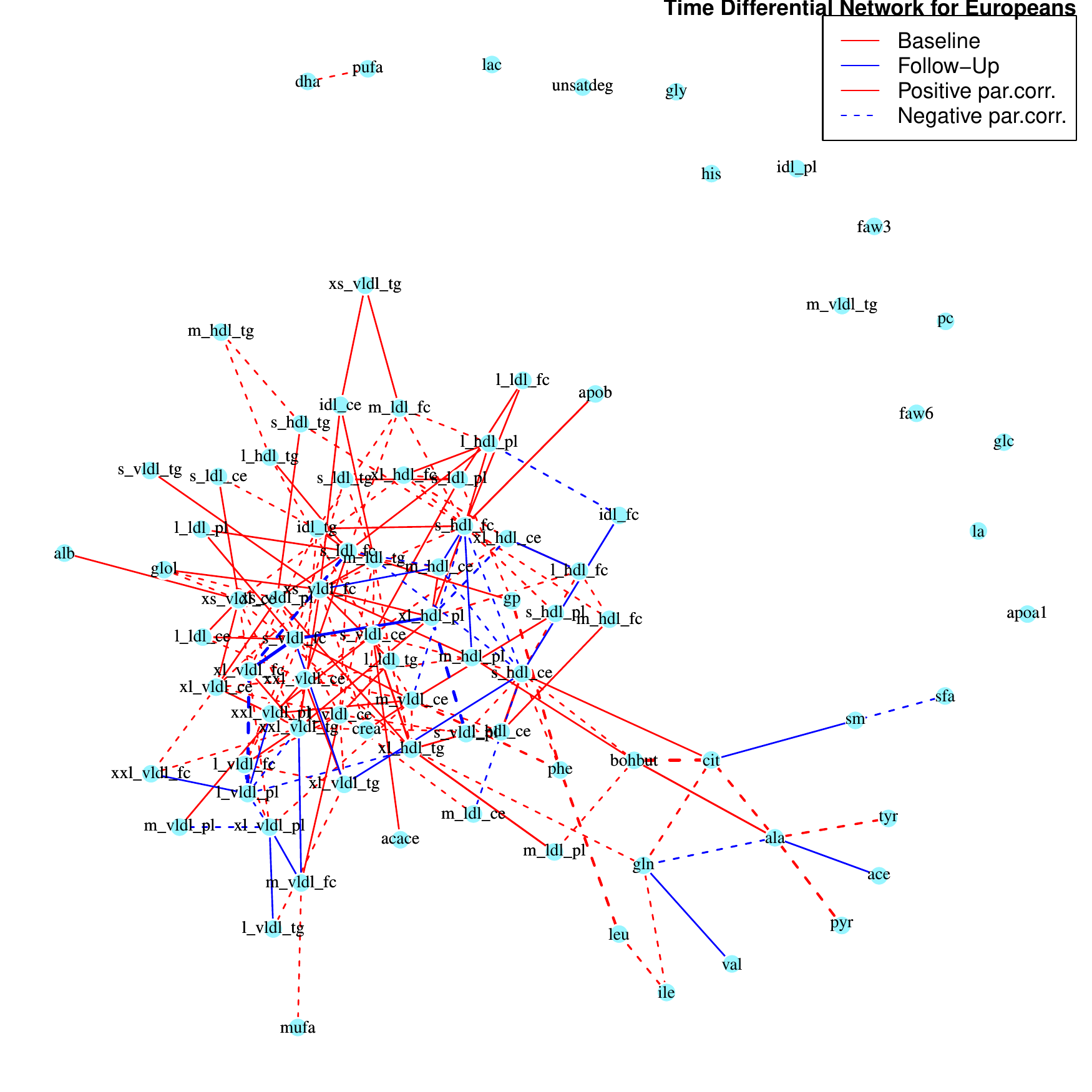}
	\caption{European Differential network between baseline and follow-up obtained with the \textit{OR} rule. Red lines correspond to edges only present in the baseline network, while blue lines to those only present in the follow-up network. Continuous lines represent differential positive partial correlations, while dashed lines indicate negative ones.}
	\label{fig:time_diff_net_EU_OR}
\end{figure}
\begin{figure}
	\centering
	\includegraphics[width=0.9\linewidth]{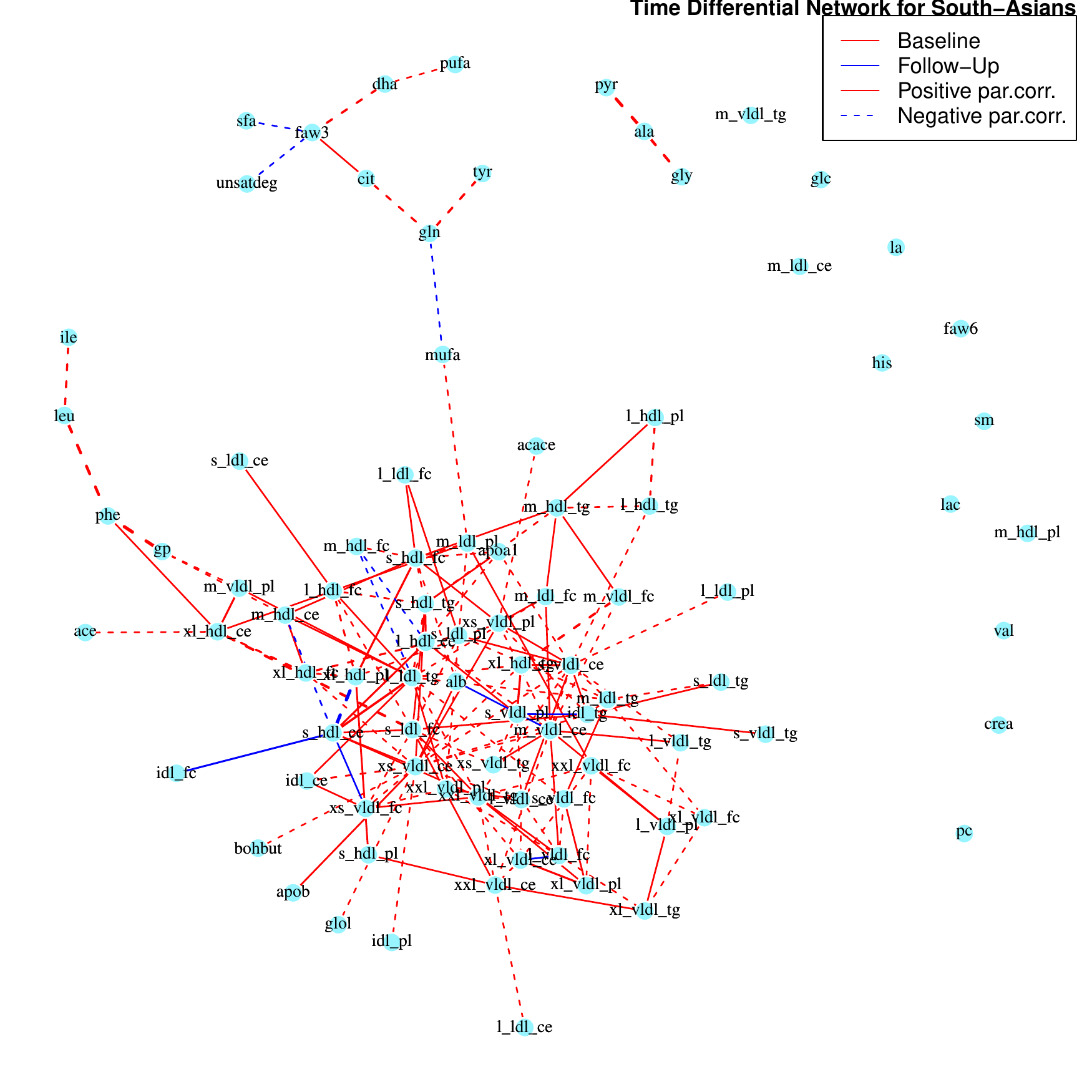}
	\caption{South-Asians Differential network between baseline and the follow-up obtained with the \textit{OR} rule. Red lines correspond to edges only present in the baseline network, while blue lines to those only present in the follow-up network. Continuous lines represent differential positive partial correlations, while dashed lines indicate negative ones.}
	\label{fig:time_diff_net_AS_OR}
\end{figure}
\begin{figure}
	\centering
	\includegraphics[width=0.9\linewidth]{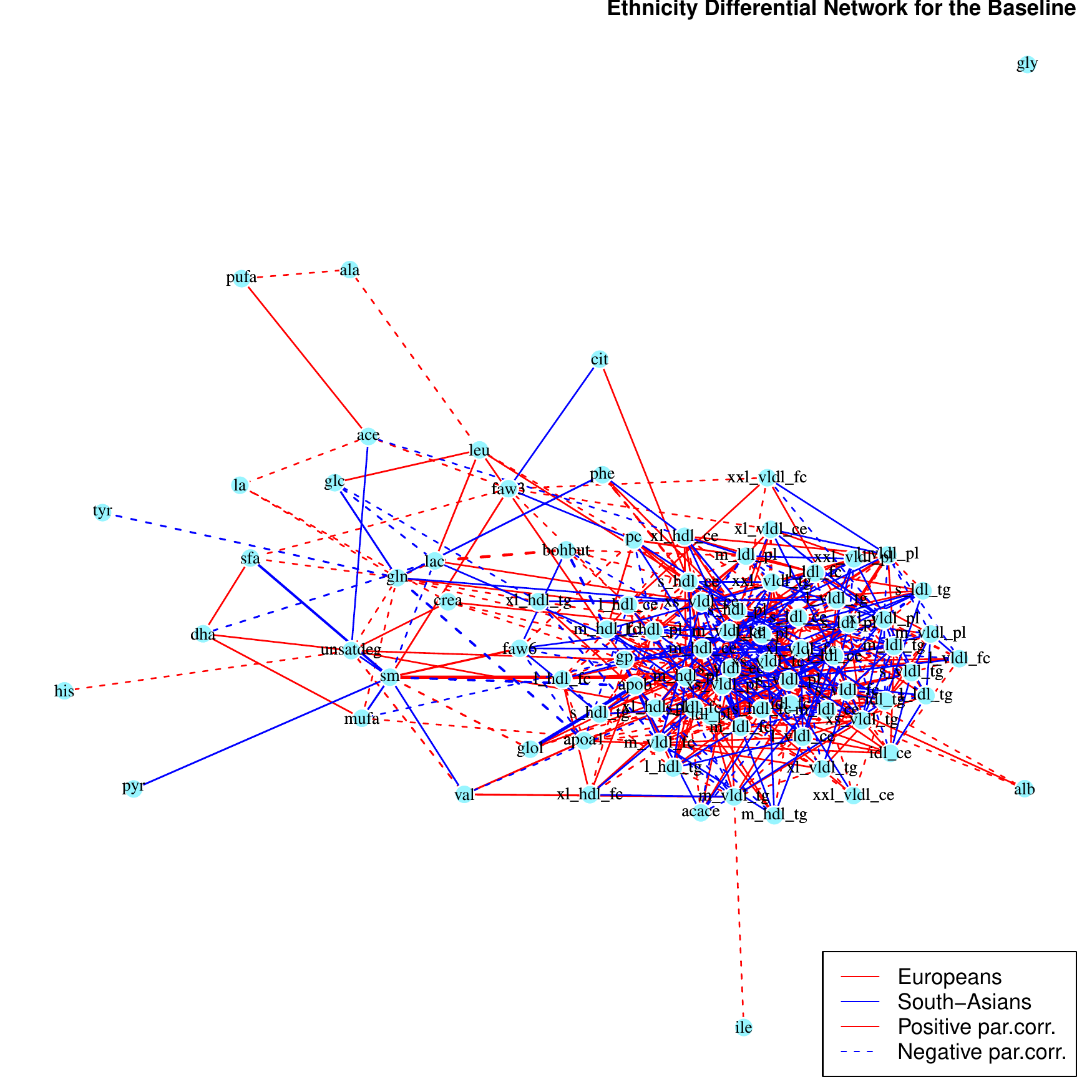}
	\caption{Differential network between Europeans and South-Asians at  baseline obtained with the \textit{OR} rule. Red lines correspond to edges only present in the European network, while blue lines to those only present in the South-Asian network. Continuous lines represent differential positive partial correlations, while dashed lines indicate negative ones.}
	\label{fig:diff_net_EU_AS_t1_OR}
\end{figure}
\begin{figure}
	\centering
	\includegraphics[width=0.9\linewidth]{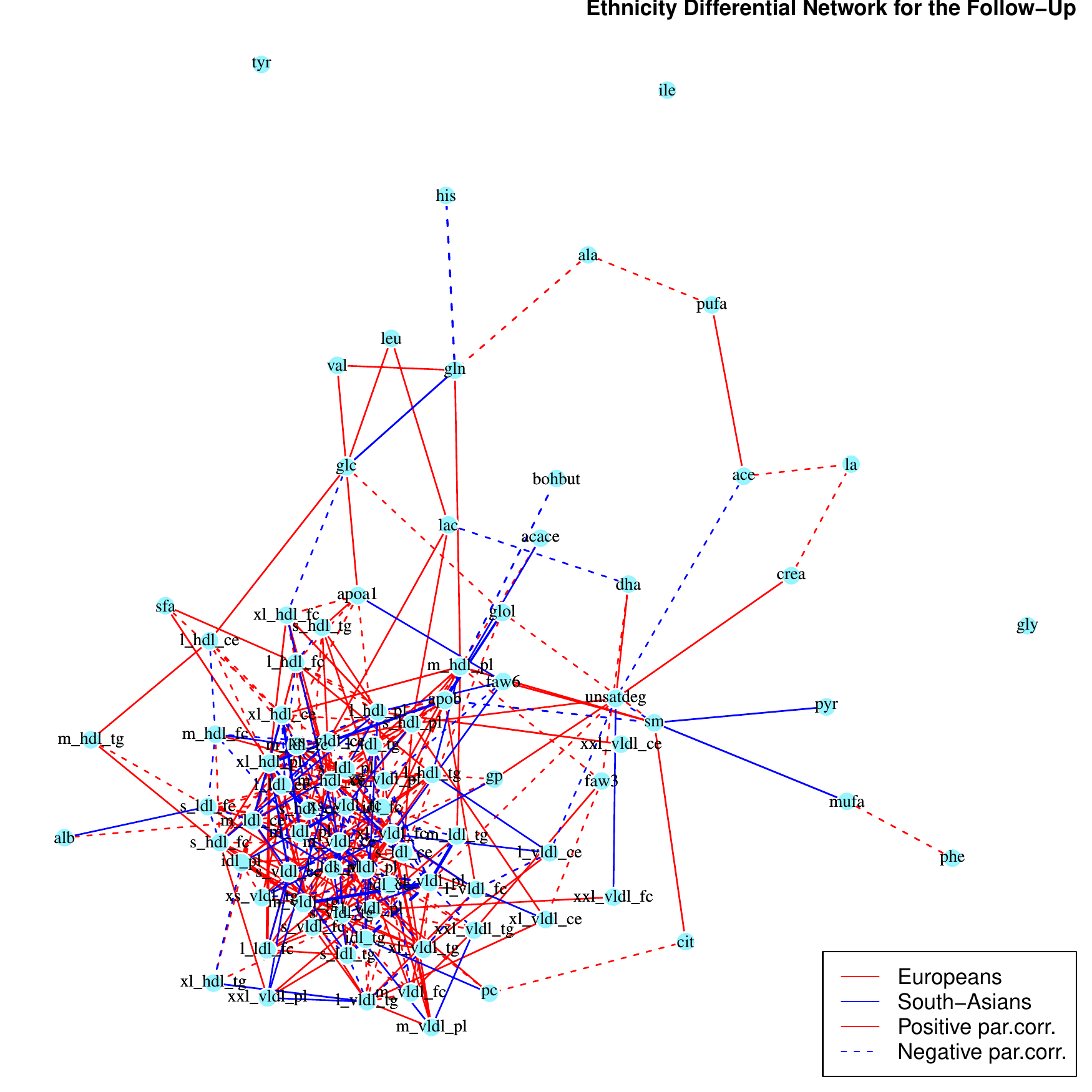}
	\caption{Differential network between Europeans and South-Asians at follow-up obtained with the \textit{OR} rule. Red lines correspond to edges only present in the European network, while blue lines to those only present in the South-Asian network. Continuous lines represent differential positive partial correlations, while dashed lines indicate negative ones.}
	\label{fig:diff_net_EU_AS_t2_OR}
\end{figure}

\begin{figure}[h!]
	\centering
	\includegraphics[width=1\linewidth]{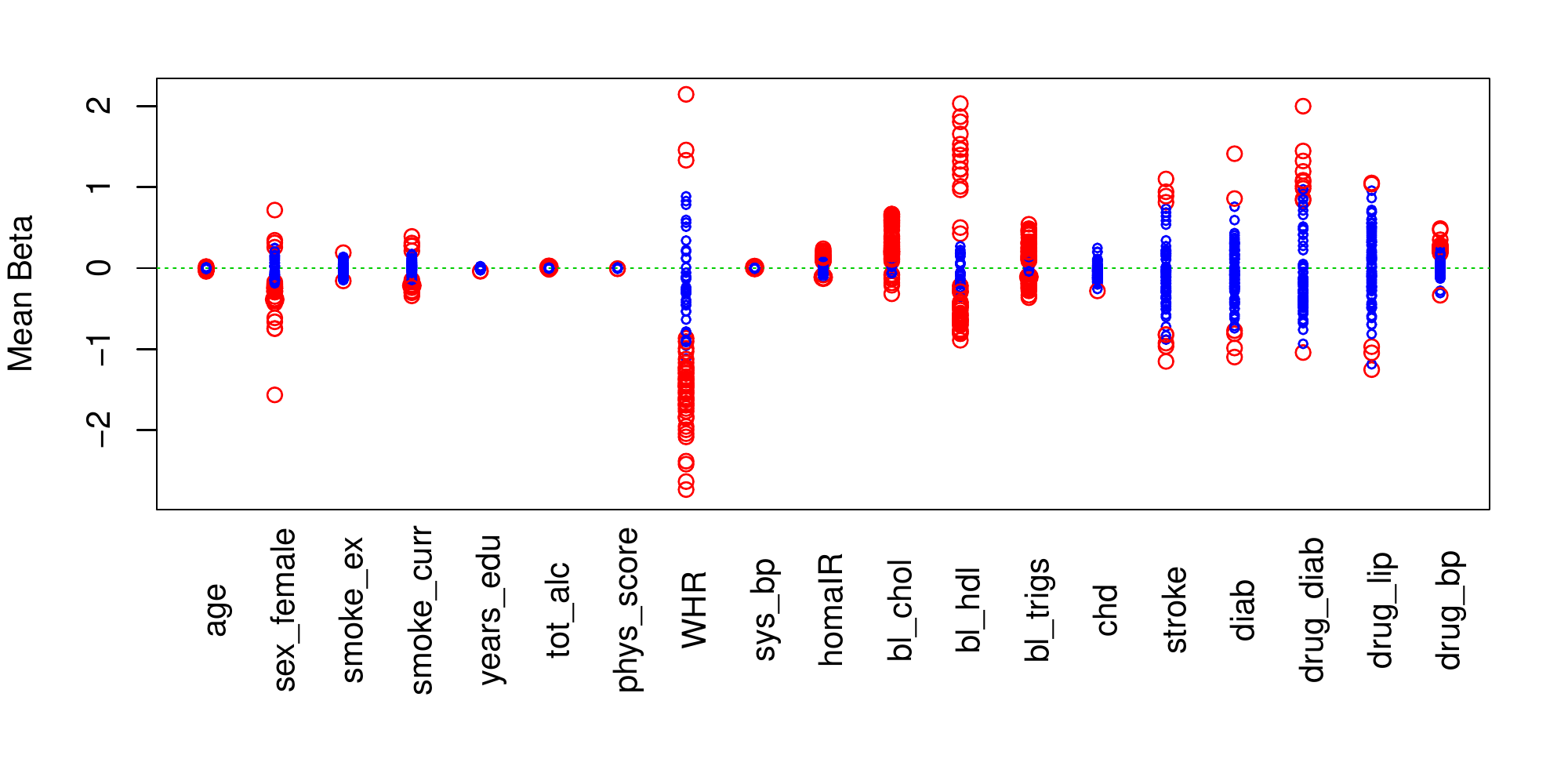}
	\caption{Posterior means of $\eta_{lj}$ for the Europeans at baseline. Each dot represents the mean of the posterior distribution of a coefficient $\eta_{lj}$, $l=1\ldots,M$. Red dots denote coefficients whose $95\%$ credible interval does not contain the zero.}
	\label{fig:meanbetagr1t1}
\end{figure}
\begin{figure}[h!]
	\centering
	\includegraphics[width=1\linewidth]{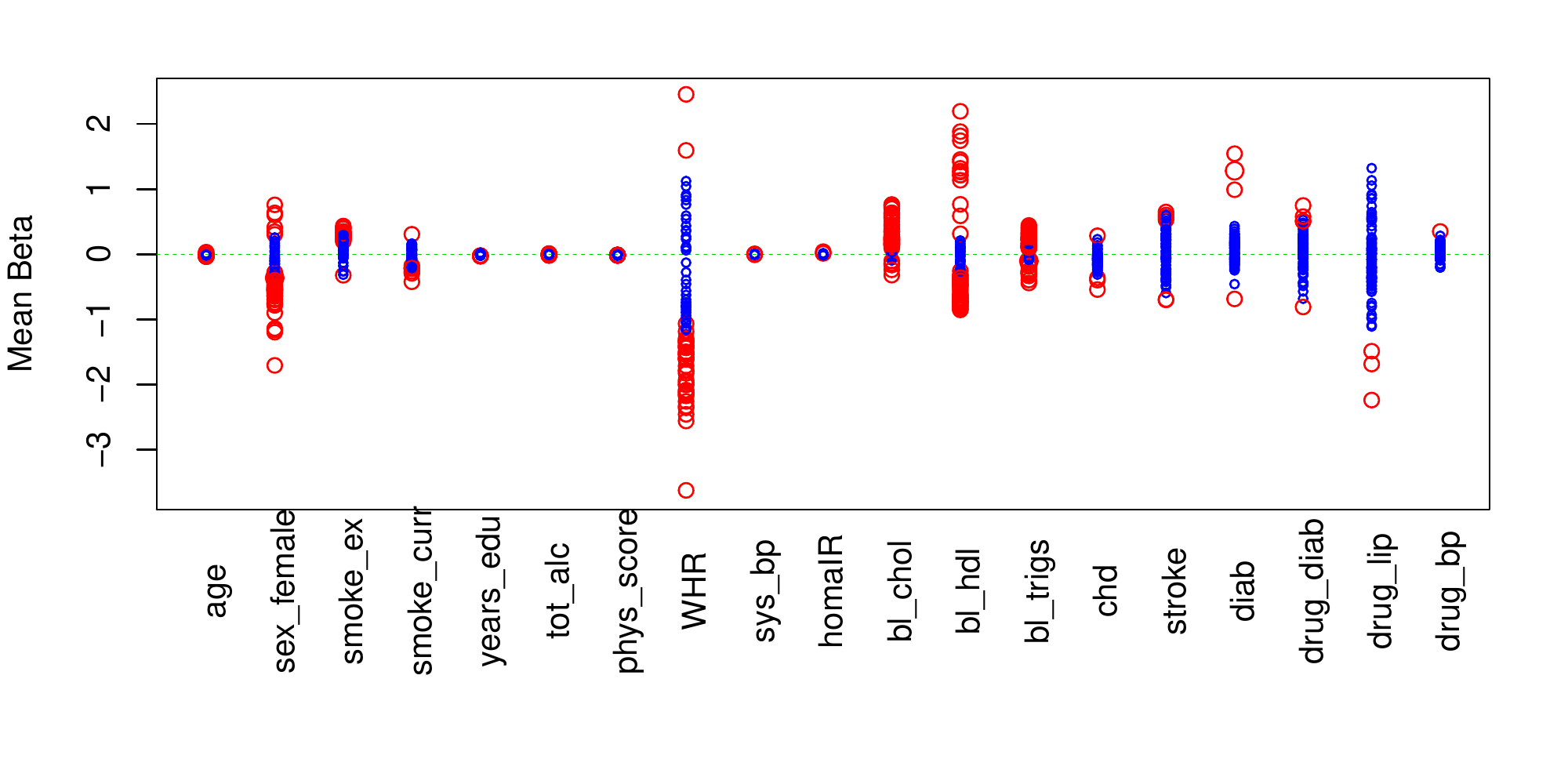}
	\caption{Posterior means of $\eta_{lj}$ for the South-Asians at baseline. Each dot represents the mean of the posterior distribution of a coefficient $\eta_{lj}$, $l=1\ldots,M$.  Red dots denote coefficients whose $95\%$ credible interval does not contain the zero.}
	\label{fig:meanbetagr2t1}
\end{figure}

\begin{figure}[h!]
	\centering
	\includegraphics[width=1\linewidth]{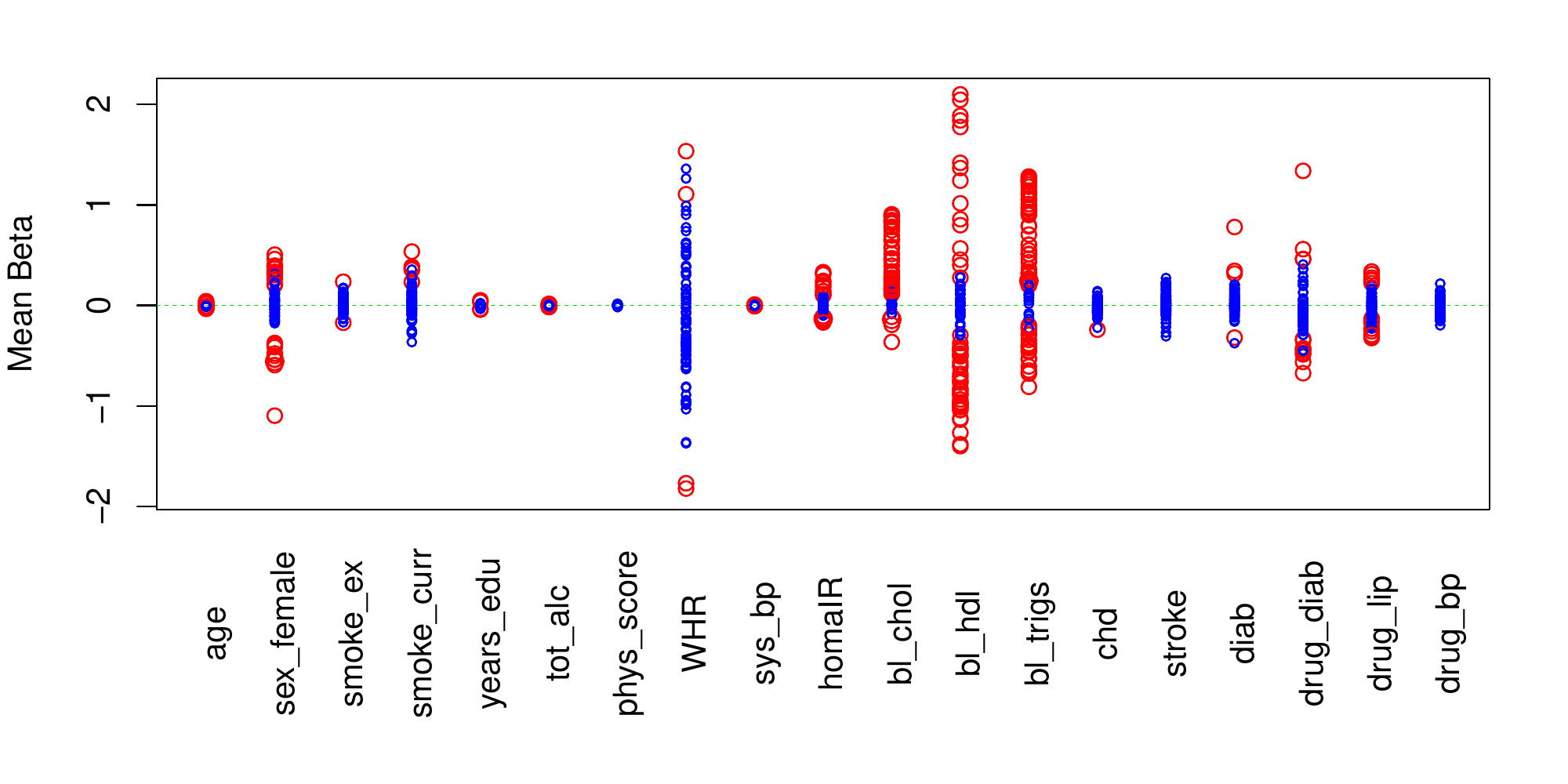}
	\caption{Posterior means of $\eta_{lj}$ for the Europeans at follow-up. Each dot represents the mean of the posterior distribution of a coefficient $\eta_{lj}$, $l=1\ldots,M$.  Red dots denote coefficients whose $95\%$ credible interval does not contain the zero.}
	\label{fig:meanbetagr1t2}
\end{figure}
\begin{figure}[h!]
	\centering
	\includegraphics[width=1\linewidth]{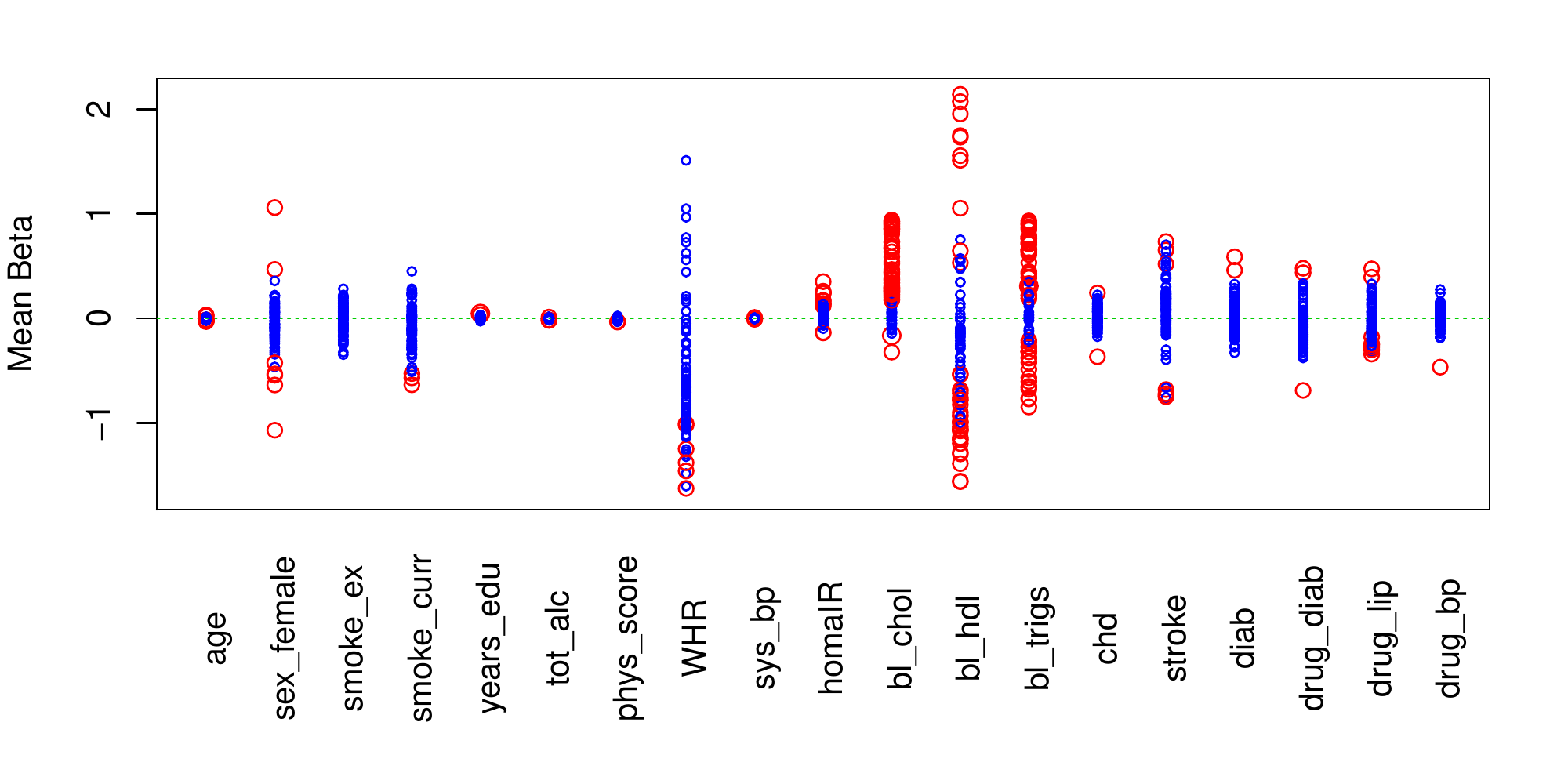}
	\caption{Posterior means of $\eta_{lj}$ for the South-Asians at follow-up. Each dot represents the mean of the posterior distribution of a coefficient $\eta_{lj}$, $l=1\ldots,M$.  Red dots denote coefficients whose $95\%$ credible interval does not contain the zero.}
	\label{fig:meanbetagr2t2}
\end{figure}

\begin{table}
\vspace{-1cm}
\caption{Metabolites extended names. Each lipoprotein is divided into triglycerides, cholesteryl esters, free cholesterol and phospholipids.}
{\begin{tabular}{llc}
		Abbreviation & Full name & Differentially expressed\\
		\hline
		acace & Acetoacetate & \\
		ace & Acetate & \\
		ala & Alanine & \\
		alb & Albumin & \\
		apoa1 & Apolipoprotein A-I & \\ 
		apob & Apolipoprotein B & \ding{51} \\
		bohbut & 3-hydroxybutyrate & \\ 
		ce & Cholesteryl Esters & \ding{51} \\
		cit & Citrate & \\ 
		crea & Creatinine & \\ 
		dha & 22:6, docosahexaenoic acid & \\ 
		faw3 & Omega-3 fatty acids & \\
		faw6 & Omega-6 fatty acids & \ding{51} \\
		fc & Free Cholesterol & \ding{51} \\
		glc & Glucose & \ding{51} \\
		gln & Glutamine & \ding{51} \\ 
		glol & Glycerol & \\
		gly & Glycine & \\
		gp & Glycoprotein acetyls, mainly a1-acid glycoprotein & \ding{51} \\ 
		his & Histidine & \\ 
		ile & Isoleucine & \\ 
		la & 18:2, linoleic acid & \ding{51} \\ 
		lac & Lactate  & \ding{51} \\ 
		leu & Leucine & \\
		mufa & Monounsaturated fatty acids; 16:1, 18:1 & \ding{51} \\ 
		pc & Phosphatidylcholines and other cholines & \\ 
		phe & Phenylalanine & \\ 
		pl & Phospholipids & \ding{51} \\
		pufa & Polyunsaturated fatty acids & \\ 
		pyr & Pyruvate & \\
		sfa & Saturated fatty acids & \\ 
		sm & Sphingomyelins & \ding{51} \\ 
		tg & Triglycerides & \ding{51} \\
		tyr & Tyrosine & \\ 
		unsatdeg & Estimated degree of unsaturation  & \\ 
		val & Valine & \\
		lipids\_s\_hdl & Lipids compounds in small HDL  & \\
		lipids\_m\_hdl & Lipids compounds in medium HDL  & \\	lipids\_l\_hdl & Lipids compounds in large HDL  & \\
		lipids\_xl\_hdl & Lipids compounds in extra large HDL  & \\
		lipids\_s\_ldl & Lipids compounds in small LDL  & \\
		lipids\_m\_ldl & Lipids compounds in medium LDL  & \\
		lipids\_l\_ldl & Lipids compounds in large LDL  & \\
		lipids\_idl & Lipids compounds in IDL  & \\
		lipids\_xs\_vldl & Lipids compounds in extra small VLDL  & \\
		lipids\_s\_vldl & Lipids compounds in small VLDL  & \\
		lipids\_m\_vldl & Lipids compounds in medium VLDL  & \\	
		lipids\_l\_vldl & Lipids compounds in large VLDL  & \\
		lipids\_xl\_vldl & Lipids compounds in extra large VLDL  & \\
		lipids\_xxl\_vldl & Lipids compounds in extra extra large VLDL & \\
		\hline
	\end{tabular} }
	\label{tab:FullNames_Met}
\end{table}

\begin{table}
	\caption{Anthropometric covariates, diabetes indicator, CVD indicators and other control variables. Every variable is used at time 1 $(T_1)$ and time 2 $(T_2)$.}
\resizebox{\linewidth}{!}{
		\begin{tabular}{ll}
		\hline
		Variable name & Label \\
		\hline
		Age & Respondent's age\\
		WHR & Waist to Hip Ratio\\
		Systolic blood pressure & Systolic blood pressure \\
		Homa IR & Homeostasis model assessment Insulin Resistance\\
		Sex female & Dummy variable for female sex (male as reference category)\\
		Smoke\_Ex & Indicator variable for ex smoker category \\
		Smoke\_Current & Indicator variable current smoker category \\
		Pt\_alcohol & Weekly quantity of alcohol consumed \\
		Phys\_score & Physical activity score\\
		Edu\_years & Years of education (assumed constant over time)\\
		bl\_hdl & Concentration of HDL\\
		bl\_trig & Concentration of blood Triglycerides\\
		bl\_chol & Concentration of blood Cholesterol\\
		CHD & Indicator variable for the presence of Coronary Heart Disease up to $T_1$ and $T_2$\\
		stroke & Indicator variable for the presence of stroke up to $T_1$ and $T_2$\\
		Diabetes & Indicator variable for the presence of Diabetes Mellitus (type 2) up to $T_1$ and $T_2$\\
		dm\_treat & Indicator variable for the presence of drug treatment for diabetes up to $T_1$ and $T_2$\\
		bp\_treat & Indicator variable for the presence of blood pressure lowering drugs treatment up to $T_1$ and $T_2$\\
		lipids\_treat & Indicator variable for the presence of blood lipids lowering drugs treatment up to $T_1$ and $T_2$\\
		\hline
	\end{tabular}
}
	\label{tab:AntroCovs}
\end{table}